\documentclass{article}

\usepackage[preprint]{neurips_2022}

\usepackage[utf8]{inputenc} 
\usepackage[T1]{fontenc}    
\usepackage{hyperref}       
\usepackage{url}            
\usepackage{booktabs}       
\usepackage{amsfonts}       
\usepackage{nicefrac}       
\usepackage{microtype}      
\usepackage{xcolor}         

\usepackage{subfigure}
\usepackage{amsmath}
\usepackage{amssymb}
\usepackage{mathtools}
\usepackage{amsthm}
\usepackage{cleveref}

\definecolor{codegreen}{rgb}{0,0.6,0}
\definecolor{codegray}{rgb}{0.5,0.5,0.5}
\definecolor{codepurple}{rgb}{0.58,0,0.82}
\definecolor{backcolour}{rgb}{0.95,0.95,0.92}
\definecolor{lightgreen}{HTML}{A9D18E}
\definecolor{lightblue}{HTML}{9DC2E6}
\definecolor{cite_color}{HTML}{114083}
\definecolor{link_color}{RGB}{153, 0,0}  
\definecolor{url_color}{RGB}{153, 102,  0}
\definecolor{emp_color}{RGB}{0,0,255}

\hypersetup{
 colorlinks=true,
 citecolor=lightblue,
 linkcolor=link_color,
 urlcolor=lightblue}

\usepackage{etoc}
\etocdepthtag.toc{mtchapter}
\etocsettagdepth{mtchapter}{subsection}
\etocsettagdepth{mtappendix}{none}

\title{Can Pre-trained Models Really Learn Better Molecular Representations for AI-aided Drug Discovery?}

\author{%
  Ziqiao Zhang\\
  School of Computer Science\\
  Fudan University\\
  Shanghai, China\\
  \texttt{zqzhang18@fudan.edu.cn}
  \And
  Yatao Bian\textsuperscript{$\ast$}\\
  Tencent AI Lab\\
  Shenzhen, China\\
  \texttt{yatao.bian@gmail.com}\\
  \And
  Ailin Xie\\
  School of Computer Science\\
  Fudan University\\
  Shanghai, China\\
  \texttt{alxie21@m.fudan.edu.cn}\\
  \And
  Pengju Han\\
  School of Computer Science\\
  Fudan University\\
  Shanghai, China\\
  \texttt{pjhan20@fudan.edu.cn}\\
  \And
  Long-Kai Huang\\
  Tencent AI Lab\\
  Shenzhen, China\\
  \texttt{hlogkai@gmail.com}\\
  \And
  Shuigeng Zhou\thanks{Corresponding authors.}\\
  School of Computer Science\\
  Fudan University\\
  Shanghai, China\\
  \texttt{sgzhou@fudan.edu.cn}\\
}

\def \repre{\Phi}
\def \regressor{h}
\def \x{\mathbf{x}}
\def \y{\mathbf{y}}
\def \z{\mathbf{z}}

\def \Z{\mathbf{Z}}
\def \Y{\mathbf{Y}}

\def \R{{\mathbb{R}}}
\def \BA{\mathbf{A}}
\def \chara{\mathbf{e}}

\newtheorem{theorem}{Theorem}
\newtheorem{definition}{Definition}
\newtheorem{proposition}{Proposition}
\newtheorem{lemma}{Lemma}

\usepackage{xspace}
\newcommand{\ours}[0]{\texttt{RePRA}\xspace}

\begin{document}

\maketitle

\begin{abstract}
  Self-supervised pre-training is gaining increasingly more popularity in AI-aided drug discovery, leading to more and more pre-trained models with the promise that they can extract better feature representations for molecules.
  Yet, the quality of learned representations have not been fully explored.
  In this work, 
  inspired by the two phenomena of Activity Cliffs (ACs) and Scaffold Hopping (SH) in traditional Quantitative Structure-Activity Relationship (QSAR) analysis,
  we propose a method named \textbf{Re}presentation-\textbf{P}roperty \textbf{R}elationship \textbf{A}nalysis (\ours) to evaluate 
  the quality of the representations extracted by the pre-trained model and visualize the
  relationship between the representations and properties.
  The concepts of ACs and SH are generalized from the structure-activity context to the representation-property context, and the underlying principles of \ours are analyzed theoretically.
  Two scores are designed to measure the generalized ACs and SH detected by \ours, and therefore the quality of representations can be evaluated.
  In experiments, representations of molecules from 10 target tasks generated by 7 pre-trained models are analyzed.
  The results indicate that the state-of-the-art pre-trained models can overcome some shortcomings of canonical Extended-Connectivity FingerPrints (ECFP), while the correlation between the basis of the representation space and specific molecular substructures are not explicit.
  Thus, some representations could be even worse than the canonical fingerprints.
  Our method enables researchers to evaluate the quality of molecular representations generated by their proposed self-supervised pre-trained models.
  And our findings can guide the community to develop better pre-training techniques to regularize the occurrence of ACs and SH.
\end{abstract}

\section{Introduction}
\label{sec:intro}

The goal of Drug Design and Discovery is to find new molecules with desired properties~\citep{schneider2020rethinking}.
In the process of drug discovery, a key step is to verify whether the properties of a given molecule satisfy specific requirements.
Due to the high cost of wet experiments, molecular property prediction models based on deep learning (DL) are expected to predict the molecular properties fast and precisely,
thereby promoting the research and development of new drugs.

At present, there are a range of molecular property prediction models based on supervised learning~\citep{zhang2021fragat, peng2020top, song2020communicative, xiong2019pushing, coley2017convolutional, ying2021transformers}, which have achieved good performance on some benchmarks~\citep{wu2018moleculenet, hu2021ogb}.
In recent years, self-supervised pre-trained models (PTM) are gaining increasingly more popularity in AI-aided drug discovery.
Many self-supervised pre-trained models have been proposed in the literature~\citep{kim2021merged, xu2021self, rong2020self, maziarka2020molecule, zhang2021motif, hu2019strategies, chen2021extracting, honda2019smiles}.
These models are trained on large-scale unlabeled data by carefully designed pretext tasks to learn representations of molecules from their structures.
Following a transfer learning paradigm, researchers can simply freeze the PTMs as encoders to extract representations and finetune down-stream task layers to make predictions based on the extracted representations. 

Although the authors of current self-supervised PTMs usually claim that their models can learn better representations for molecules, there is no convincing analysis and evidence to support their claims.
And the comparison between the generated representations and canonical representations, e.g. Extended-Connectivity FingerPrints (ECFP)~\citep{rogers2010extended}, which is widely used in computer-aided drug discovery and can be calculated efficiently, is still lacking.

Typically, the prediction accuracy after finetuning on target tasks is used for evaluating the performance of the PTMs.
However, it cannot be leveraged to evaluate the quality of representations.
The first reason is that the prediction accuracy after finetuning is not only related to the quality of representations, but also to the capability of down-stream task layers.
Influenced by the task layers, this evaluation is biased.
On the other hand, a frozen PTM will generate fixed representations, so that the quality of the representations on a specific target task should also be a constant conclusion.
However, the results after fintuning are strongly affected by various factors, so it cannot be an accurate evaluation of the quality.
Therefore, how to evaluate the quality of representations produced by self-supervised pre-trained models, is worthy of further investigation for building more powerful self-supervised pre-trained models to support drug design and discovery.

In this paper, we propose a method named Representation-Property Relationship Analysis (\ours) to evaluate the quality of representations extracted by self-supervised pre-trained models.
This method is inspired by the two phenomena Activity Cliffs ~\citep{perez2015activity} and Scaffold Hopping~\citep{bohm2004scaffold} in drug discovery.
It includes a map to visualize the quality of representations and two scores to quantitatively evaluate the quality. The proposed method has been applied to evaluating 7 pre-trained models on 10 target tasks. 
Our contributions and findings are as follows:
\begin{itemize}
    \item We propose a method 
    called \ours
    with a visualization tool and two quantitative assessment scores for researchers to evaluate the quality of representations generated by pre-trained models.
    \item We generalize the concepts of Activity Cliffs and Scaffold Hopping from structure-activity context to representation-property context, and give an theoretical analysis of the underlying principles of \ours.
    \item We use \ours to evaluate 7 pre-trained models on 10 target tasks and conduct a comprehensive comparison between the representations of these models.
    \item Our analysis indicates that some of these pre-trained models can overcome the shortcomings of ECFP, 
    but the basis of the representation space generated by these models does not explicitly correspond to specific molecular substructures.
    \item Our discovery can guide the community to develop new pre-training techniques to cope with ACs and SH and generate better representations for molecules.
\end{itemize}

\section{Preliminaries}
\subsection{Notations}

Throughout this work we assume
 $\chara_i\in\R^n$ is the $i^\text{th}$ standard basis vector.
We use boldface letter $\x\in \R^n$
 to indicate an $n$-dimensional vector, where $x_i$ is
the $i^\text{th}$ entry of $\x$. We use a boldface capital letter
$\BA\in\R^{m\times n}$ to denote an $m$ by $n$ matrix and use $A_{ij}$
to denote its ${ij}^\text{th}$ entry.
$\|\cdot\|$ means the Euclidean norm, and $<\cdot,\cdot>$ means the inner product of Euclidean vector space by default.

\subsection{Problem Setting}
We consider a transfer learning process as follows:
\[\hat \y = \regressor(\repre(\x))\]
where $\repre(\x)$ is a model pre-trained on a source task $T_s$ and is fixed to be used as an encoder to extract representations of samples for the target task $T_t$.
$\regressor(\cdot)$ denotes additional task layers to be finetuned on the target task.
The dataset of the target task $T_t$ is denoted as $D=\{(\x_i, \y_i)\}_{i=1}^N$, where $N$ is the size of the dataset.
We denote the representations extracted by the pre-trained model as $\z_i=\repre(\x_i)$, and the properties of sample as $\y_i=[y_k]_{k=1}^K,~\y_i \in R^K$, where $K$ is the number of properties.
Here, we use \textit{properties} because $y_k$ can be either label for classification task or value for regression task.
Our problem is to evaluate the quality of extracted representations $\Z = \{\z_i\}_{i=1}^N$ based on the relationship between the representations $\Z$ and the properties $\Y = \{\y_i\}_{i=1}^N$ of samples for a specific target task.

\subsection{Activity Cliffs and Scaffold Hopping}
\label{ACSH}

Activity cliffs (ACs) and scaffold hopping (SH) are two phenomena in drug discovery.
ACs are defined as pairs of structurally similar molecules with largely different potency towards the same target protein~\citep{perez2015activity}.
And SH is a series of methods for discovering new compounds, which attempt to make changes to the central core structure of the molecule to find new molecules with similar or higher activity~\citep{bohm2004scaffold}.
It indicates pairs of molecules sharing similar activity but largely different structural features.

\subsection{Extended-Connectivity FingerPrint}
ECFP~\citep{rogers2010extended} is a circular topological fingerprint method to describe molecular structures.
An ECFP is a binary vector, of which each bit indicates the existence or not of a specific circular substructure, i.e., a substructure consists of one central node and its $R$-hop neighbors.
ECFP can be calculated efficiently by some toolkits such as RDKit.
Thus, it is a traditionally widely-used method to represent structures of molecules for many applications in drug discovery~\citep{li2007large, jeon2019resimnet, rifaioglu2021mdeepred, menke2021using}, e.g. high-throughput screening and ligand-based virtual screening.

\section{Related Work}

\label{sec:relatedworks}

Transferability aims to evaluate the ability of transferring a model pre-trained on a source task $T_s$ to a target task $T_t$~\citep{you2021logme}. 
Some methods have been proposed~\citep{tran2019transferability, you2021logme, huang2021frustratingly, nguyen2020leep}. 
Among them, LogME~\citep{you2021logme} and TransRate~\citep{huang2021frustratingly} are proposed to evaluate transferability by estimating the compatibility between the extracted representations and properties.
These works rank the PTMs by their proposed scores with the same order as the prediction accuracy of these PTMs after finetuning the down-stream task layers on specific target tasks.
Weighted Kendall's coefficient $\tau_\omega$ is proposed to assess the consistency of these two orders to show the precision of their proposed scores of transferability.

Although these works are similar to ours to some extent, the research goals are totally different.
Our work aims to evaluate the quality of representations extracted by PTMs, rather than to fit the order of prediction accuracy after finetuning.
As the accuracy is sensitive and biased, the order is a variable ground-truth and different finetuning results will lead to different $\tau_\omega$.
So, the estimation of $\tau_\omega$ lacks fairness, and it is not appropriate to be used to assess the precision of our method.
In this work, we will not participate the comparison between transferability methods by $\tau_\omega$.
Detailed discussion is shown in Appendix~\ref{sec:finetune}.
And the introduction of the above two methods is in Appendix~\ref{sec:LogMEandTransRate}

\section{Method}
\label{sec:method}
\subsection{Generalized Activity Cliffs and Scaffold Hopping}
In drug discovery, the concepts of Activity Cliffs and Scaffold Hopping generally focus on the relationship between molecular structures and bio-activities.
To generally evaluate the quality of representations extracted by PTMs on arbitrary molecular property prediction tasks, we firstly generalize the concepts of ACs and SH to the relationship between representations and properties.

\begin{definition}[Generalized Activity Cliffs]
Generalized Activity Cliffs are defined as pairs of molecules sharing \textit{similar} representation $\z$ with large difference in properties $\y$.
\end{definition}

\begin{definition}[Generalized Scaffold Hopping]
Generalized Scaffold Hopping is defined as a pair of molecules with large difference in their representations, but their properties are similar.
\end{definition}

With the generalized concepts of ACs and SH, we can give a proposition about ideal representation-property relationship as follows:

\begin{proposition}
\label{prop:1}
In a representation-property relationship, the existence of generalized ACs and SH is unfavorable.
Generalized ACs and SH should not exist in an ideal representation-property relationship.
\end{proposition}

The occurrence of generalized ACs indicates the discontinuous of representation-property relationship, which will burden the down-stream task layers to fit such relationship.
To deal with the samples of generalized ACs, a model will either choose a simple but more generalizable half-plane and ignore the information brought by these samples, or use a complex half-plane to predict these samples correctly, which may face the risk of over-fitting.
And the existence of generalized SH reveals that the samples with similar properties is widely distributed in the representation space.
In such a case, a model may need a higher-dimensional complex half-plane for classification, or a more complex function to fit the representation-property relationship.
To illustrate Prop.~\ref{prop:1}, we present some toysets as examples in Appendix~\ref{sec:toyset}.
By default,  ACs and SH refer to the generalized ACs and SH in the sequel.

\subsection{Ideal Representation-Property Relationship}
Based on Proposition~\ref{prop:1}, a formal definition of ideal representation-property relationship is as follows:

\begin{definition}[$(\delta, \epsilon)$-ideal]
Given representations $\Z = \{\z_i\}_{i=1}^N$ and properties $\Y = \{\y_i\}_{i=1}^N$, for any $\delta, \epsilon \in \R$, if the relationship of $\Z$ and $\Y$ satisfy the following condition:
\begin{align}
\label{assump1}
    \forall \z_i, \z_j \in \Z, d_z(\z_i,\z_j) < \delta \Leftrightarrow d_y(\y_i, \y_j) < \epsilon, 
\end{align}
where $d_z(\cdot)$ and $d_y(\cdot)$ are metric functions on the set $\Z$ and $\Y$, then the relationship is $(\delta, \epsilon)$-ideal.
\end{definition}

\paragraph{Remark.}
This condition can be split into two sub-conditions.
If a representation-property relationship satisfy the sub-condition from left to right, it indicates that for all pairs of molecules with similar representations, their properties are similar.
So that there are no ACs.
And if the relationship satisfy the sub-condition from right to left, for all pairs of molecules with similar properties, their representations are close. 
So that there is no SH.

In another word, if a relationship satisfies Eq.~(\ref{assump1}), there are no ACs and SHs.
According to Prop.~\ref{prop:1}, the relationship is considered to be \textit{ideal}.
As the ACs and SH are defined by two thresholds $\delta$ and $\epsilon$, this \textit{ideal} assessment on the relationship is conditional.
So, we denote the relationship as $(\delta, \epsilon)$-ideal, and the standard to define ACs and SH as \textbf{$(\delta, \epsilon)$-standard}.
If ACs and SH defined by $(\delta, \epsilon)$-standard can be detected from a relationship between representations and properties, the relationship is not ideal.
So, based on detected ACs and SH, we can evaluate the representation-property relationship, and therefore evaluate the quality of representations.

\begin{figure*}[t]

\vskip 0.2in
\begin{center}
\centerline{\includegraphics[width = \linewidth]{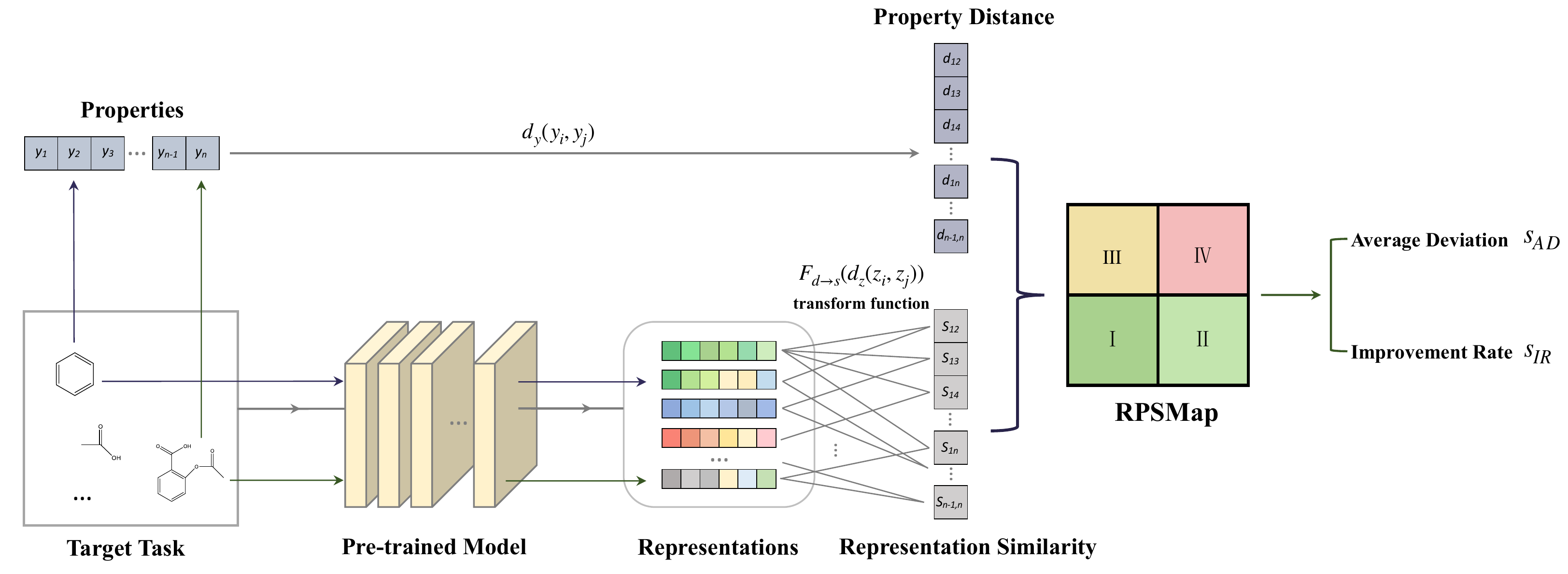}}
\caption{Flow chart of 
\ours  
for evaluating the quality of representations generated by pre-trained models on a target task.}
\label{fig:framework}
\end{center}
\end{figure*}

\subsection{Representation-Property Relationship Analysis (\ours)}

Here we introduce our method called Representation-Property Relationship Analysis (\ours) to detect and measure ACs and SH defined by the $(\delta, \epsilon)$-standard to evaluate the quality of representations extracted by a PTM on a target task.
The flow of \ours is shown in Fig.~\ref{fig:framework}.
With a given PTM and a target task dataset, representations $\Z$ of all molecules in the dataset are firstly calculated.
Then, for each pairs of molecules, the distance between any pair of representations $\Z$ and the distance between any pair of properties $\Y$ are calculated by metric functions $d_z(\cdot)$ and $d_y(\cdot)$, respectively.

With the pairwise distances of representations and properties, a visualization tool is proposed.
Inspired by the Structure-Activity-Similarity Map (SASMap) in drug discovery~\citep{perez2015activity}, here we use Representation-Property-Similarity Map (RPSMap) to name our tool, which is shown in Fig.~\ref{fig:RPSMap}. Following the classical QSAR layout~\citep{perez2015activity}, the x-axis of RPSMap indicates the representation similarity, i.e. $sim(\z_1, \z_2):= F_{d\to s}(d_z(\z_1, \z_2))$.
$F_{d\to s}(\cdot)$ is a function to transform the distances of representations to similarities. A larger distance indicates a smaller similarity.
And the y-axis indicates the distance of properties $d_y(\y_1, \y_2)$.
For a dataset with $N$ samples, $\frac{N(N-1)}{2}$ points will be plot in the RPSMap.
$p_{ij}$ is the point corresponding to sample pair $(\z_i, \y_i)$ and $(\z_j, \y_j)$.
It reflects the relationship between their representations and properties.
Given the $(\delta, \epsilon)$-standard (remember that the threshold $\delta$ should be transformed into similarity by $F_{d\to s}(\cdot)$), an RPSMap can be divided into four regions, as shown in Fig.~\ref{fig:RPSMap}.
If point $p_{ij}$ lies in Region I or IV, then the relationship of sample pair $(\z_i,\y_i)$ and $(\z_j, \y_j)$ is regarded as SH or ACs, respectively.
And points in Regions II and III are normal.
So, RPSMap can serve as a visualization tool for researchers to study the quality of molecular representations.

Finally, two scores, the Average Deviation score $s_{AD}$ and the Improvement Rate score $s_{IR}$ are proposed for quantitative assessment based on the detected ACs and SH, which will be introduced in Sec.~\ref{sec:scores}.
Due to space limit, the choice of metric functions and the corresponding $F_{d\rightarrow s}(\cdot)$ is discussed in Appendix~\ref{sec:metrics}.

\subsection{Choice of Thresholds}
\label{sec:thresholds}

The standard of ACs and SH in drug discovery is based on consensus, but the standard of generalized ACs and SH is totally determined by $\delta$ and $\epsilon$.
So, an important problem is how to choose the thresholds.
In this section, we first study how the samples distribute in the representation space if the relationship is $(\delta, \epsilon)$-ideal.
Then, threshold parameters are introduced for detecting ACs and SH and evaluating the quality of representations.

\begin{theorem}[]
\label{theor:1}
Given representations $\Z=\{\z_i\}_{i=1}^{N}$ and multi-task multi-class classification properties $\Y = \{\y_i\}_{i=1}^{N}$, $\y_i = [y_k]_{k=1}^K$, $y_k \in \{0,1,\dots,c_k\}$, where $K$ is the number of tasks and $c_k$ is the number of classes of task $k$. 
For any $\epsilon^* \in (0,1)$,
if $\Z$ and $\Y$ is $(\delta,\epsilon^*)$-ideal with Euclidean distance metrics $d_z(\cdot)$ and $d_y(\cdot)$, then the samples of the same label are gropued into a cluster of diameter $d<\delta$, and the margin $m$ between any two clusters is greater than $\delta$ ($m >\delta$).
\end{theorem}

An $(\delta,\epsilon^{*})$-ideal example is shown in Fig.~\ref{fig:class_ideal}.
Theorem~\ref{theor:1} reveals that if a relationship between representations and properties is $(\delta, \epsilon^{*})$-ideal, the distribution of such representations is consistent with the case that training an encoder till the Triplet Margin Loss~\citep{schroff2015facenet} $ L = \max(d(a,p) - d(a,n) + \delta, 0) = 0$, where $p$ is a sample having the same label as that of the anchor $a$, and $n$ is a sample of different label.

For regression tasks, our analysis starts from the case that the representations and properties are scalars for convenience. 

\begin{lemma}[]
\label{lemma:1}
Given representations $\Z = \{z_i\}_{i=1}^N, z_i \in \R$ and properties $\Y = \{y_i\}_{i=1}^N, y_i \in \R$.
If the relationship between $\Y$ and $\Z$ is linear, i.e., $y = w \cdot z + b$, where $b$ is a bias, then for any $\delta \in \R$, the relationship is $(\delta, \|w\| \delta)$-ideal with Euclidean distance as metric functions.
\end{lemma}

\begin{lemma}
\label{lemma:2}
Assume an infinite dataset $D = \{(z,y)\,|\,y=w \cdot z + b\}$.
Denote $\Z = \{z_i\}, \Y = \{y_i\}, (z_i, y_i) \in D$.
For any sample $(z_0, y_0)\in D$, if it is disturbed to $(z_0, y^*_0)$ with noise $n$, i.e.,  $y^*_0 = y_0 + n$, then for any $\delta \in \R$, this disturbed point can be detected by $(\delta, \|w\| \delta)$-standard ACs and SH with Euclidean distance as metrics.
\end{lemma}

\begin{theorem}
\label{theor:2}
Given an infinite dataset $D=\{(z_i, y_i)\}$ with representations $\Z = \{z_i\}, z_i \in \R$ and regression properties $\Y = \{y_i\}, y_i \in \R$.
If the relationship between $\Y$ and $\Z$ is $(\delta, \epsilon)$-ideal, then the relationship is linear, i.e. $y = w \cdot z + b$, $\|w\| = \frac{\epsilon}{\delta}$.
\end{theorem}

\begin{theorem}
\label{theor:3}
Given a finite dataset $D=\{(z_i,y_i)\}_{i=1}^N$ with representations $\Z = \{z_i\}_{i=1}^N, z_i \in \R$ and regression properties $\Y = \{y_i\}_{i=1}^N, y_i \in \R$.
If the relationship between $\Z$ and $\Y$ is $(\delta, \epsilon)$-ideal, it is not guaranteed that this relationship is linear.
\end{theorem}

Proofs are given in Appendix~\ref{sec:proof}.
For the case that $\z$ and $\y$ are vectors, 
the discussion is in Appendix~\ref{sec:vectorcase} due to space limit.


Theorem~\ref{theor:1} to~\ref{theor:3} establish the relationship between $(\delta, \epsilon)$-ideal and how the samples are distributed in the representation space.
The detection of ACs and SH by the $(\delta, \epsilon)$-standard is to detect samples violating the corresponding ideal distribution.
However, in practice, it is non-trivial to know the underlying threshold parameters.
An improper $(\delta, \epsilon)$-standard will underestimate the representations even if they are ideal.

To solve this problem, in this work we suggest two groups of $(\delta, \epsilon)$-standards for classification tasks and regression tasks, respectively.

\paragraph{Classification.}
For classification tasks, it is easy to choose an $\epsilon$ to distinguish pairs of samples with the same or different labels.
However, $\delta$ is difficult to be specified.
A larger $\delta$ will weaken the ability to detect SH, since those pairs with $d_z>\delta$ will be fewer.
While more ACs may be detected undesirably if the margin $m< \delta$.
On the other hand, a smaller $\delta$ will detect more pairs of samples as SH undesirably if the diameters of clusters are larger than $\delta$.

So, to avoid false detection and underestimation, we suggest to use a loose standard to detect ACs and SH.
Denote $d_{ij} = \|\z_i - \z_j\|$, let
\begin{equation}
\label{eq:delta1}
    \delta_1 = \frac{1}{|D_{in}|}\sum_{(i, j)\in D_{in}}d_{ij},\quad 
    \delta_2 = \frac{1}{|D_{out}|}\sum_{(i,j)\in D_{out}}d_{ij}, 
\end{equation}
where $D_{in} = \{(i,j) | \y_i = \y_j\}$ and
$D_{out} = \{(i,j) | \y_i \neq \y_j\}$.
Then, our scheme is to use the $(\delta_1, 0.5)$-standard to detect ACs and the $(\delta_2, 0.5)$-standard to detect SH.

\paragraph{Regression.}
Similar to classification tasks, to avoid false detection, two groups of $(\delta, \epsilon)$-standards are introduced to detect ACs and SH for regression tasks, respectively.
Here, we suggest to use the median of $d_{ij}$ to distinguish pairs of samples that are expected to have large representation distance and property differences (corresponding to different labels in classification tasks) and those having small representation distance and property differences (corresponding to similar labels in classification tasks).
Denote $d^{m}$ as the median of $\{d_{ij}\}$, and $\Delta_{ij} = \|\y_i-\y_j\|$, let:
\begin{align}
\label{eq:deltaepsilon1}
    \delta_1 = \frac{1}{|D_{near}|}\sum_{(i,j)\in D_{near}}d_{ij},&\quad
    \epsilon_1 = \frac{1}{|D_{near}|}\sum_{(i,j)\in D_{near}}\Delta_{ij} \\
    \delta_2 = \frac{1}{|D_{far}|}\sum_{(i,j)\in D_{far}}d_{ij},&\quad
    \epsilon_2 = \frac{1}{|D_{far}|}\sum_{(i,j)\in D_{far}}\Delta_{ij},
\end{align}
where $D_{near} = \{(i,j) | d_{ij} < d^{m}\}$ and $D_{far} = \{(i,j) | d_{ij} \geq d^{m}\}$.
Then, our scheme is to use the $(\delta_1, \epsilon_1)$-standard to detect ACs and the $(\delta_2, \epsilon_2)$-standard to detect SH.

The suggested thresholds give relatively loose standards as an alternative to detect ACs and SH by using statistical information of the representations.
There may be better schemes for threshold choice, and we remain this for future work.

\subsection{Scores}
\label{sec:scores}
In this section, we propose two scores to evaluate the representation-property relationship based on the ACs and SH detected by a given $(\delta, \epsilon)$-standard.

\paragraph{Average Deviation ($s_{AD}$).}
According to Prop.~\ref{prop:1}, it would be a better representation-property relationship if fewer
ACs and SH are detected.
So, the ratio of these \textit{noise} points among all pairs of samples should be taken into account.
Denote the set of points located in region $l$ as $R_l$, and the number of points in region $l$ as $C_l = | R_l |$.
The ratio of points located in $R_1$ and $R_4$ is
\begin{equation}
    e = \frac{2(C_1+C_4)}{N(N-1)}.
\end{equation}

In addition, the deviation of the noise points from the ideal regions reflects the extent of the detected points violating the $(\delta, \epsilon)$-standard, which should also be considered.
For a point $p_{ij} \in R_1 \cup R_4$, the deviation of the point from the ideal regions $R_2$ or $R_3$ is
\begin{equation}
    m_{ij} = \min(|d_y(\y_i,\y_j) - \epsilon|,|sim(\z_i,\z_j)- F_{d \to s}(\delta)|).
\end{equation}
The mean of the deviation of all noise points is
\begin{equation}
    \bar{m} = \frac{\sum_{p_{ij} \in R_1 \cup R_4} m_{ij}}{C_1 + C_4}.
\end{equation}

Thus, integrating the ratio of the noise points and their deviations, the average deviation score that could be used to evaluate the relationship is
\begin{align}
    s_{AD} = e \cdot \bar{m},
\end{align} 
which is the lower the better.

\paragraph{Improvement Rate ($s_{IR}$).}
Deep learning models are expected to generate better representations for molecules than traditional methods, e.g. ECFP.
So, the improvement of the representations compared with ECFP will be another score for evaluation.

Denote $\z^{fp}_i$ as the ECFP of molecule $\x_i$.
For each point $p_{ij}$ in the RPSMap of PTM (denoted as $M_{PTM}$), calculate $d_{fp}(\z_i^{fp},\z_j^{fp})$ and draw another RPSMap (denoted as $M_{fp}$).
$C_1^{PTM}$ and $C_4^{PTM}$ indicate the numbers of points located in $R_1$ and $R_4$ of $M_{PTM}$, and $C_1^{fp}$ and $C_4^{fp}$ denote those of $M_{fp}$.
Then we have
\begin{equation}
    s_{IR} = \frac{C_1^{PTM}}{C_1^{fp}} + \frac{C_4^{PTM}}{C_4^{fp}}.
\end{equation}

$s_{IR} \in [0,+\infty)$, the lower the better.
When $s_{IR} = 2$, it indicates that the representations produced by a pre-trained model is not better or worse than ECFP.
When $s_{IR} < 2$, there is an improvement, and when $s_{IR} > 2$, the representations are even worse than ECFP.

\section{Experiments and Analysis}
\label{sec:EXP}
In this section, we use the proposed \ours to evaluate the quality of representations produced by 7 pre-trained models, of which the model files are publicly available, on 10 target tasks.
The information of these PTMs and the downstream tasks are introduced in Appendix~\ref{sec:PTM}.
In what follows, we will analyze the results and show what can be revealed by our method.
Due to space limit, only part of the results are presented for discussion.

\subsection{Analysis by RPSMap}
\label{sec:RPSMapsAnalysis}
Here, we compare the RPSMaps of ECFP and representations extracted by PTMs to discover the advantages and drawbacks of these models.
The radius of ECFP is set to be 2 and the length is 1024.
As discussed in Appendix~\ref{sec:metrics}, we use MinMaxEud as the metric function to draw RPSMap and calculate the scores.
And the distribution of similarities with CosineSim as metric is also drawn for analysis.
Using ESOL as an example, the RPSMap is shown in Fig.~\ref{fig:ESOLMinMaxEudRPSMap}.
And the distribution of cosine similarity is shown in Fig.~\ref{fig:ESOL_Cosine_distribut_all}

\begin{figure}[tbp]
	\centering
	\subfigure[ECFP]{
	\includegraphics[width=3.25cm]{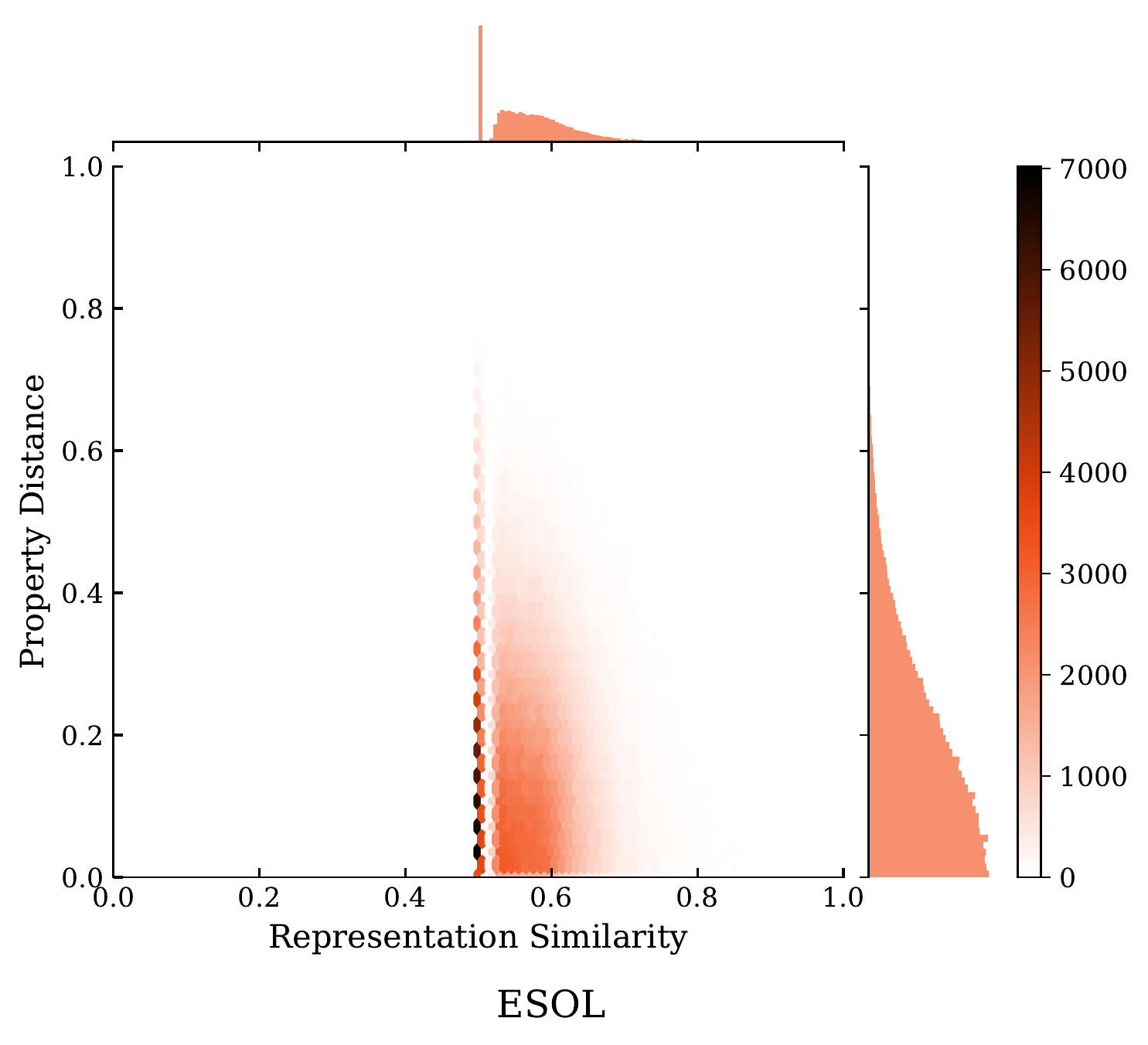}
}
    \subfigure[GROVER]{
    \includegraphics[width=3.25cm]{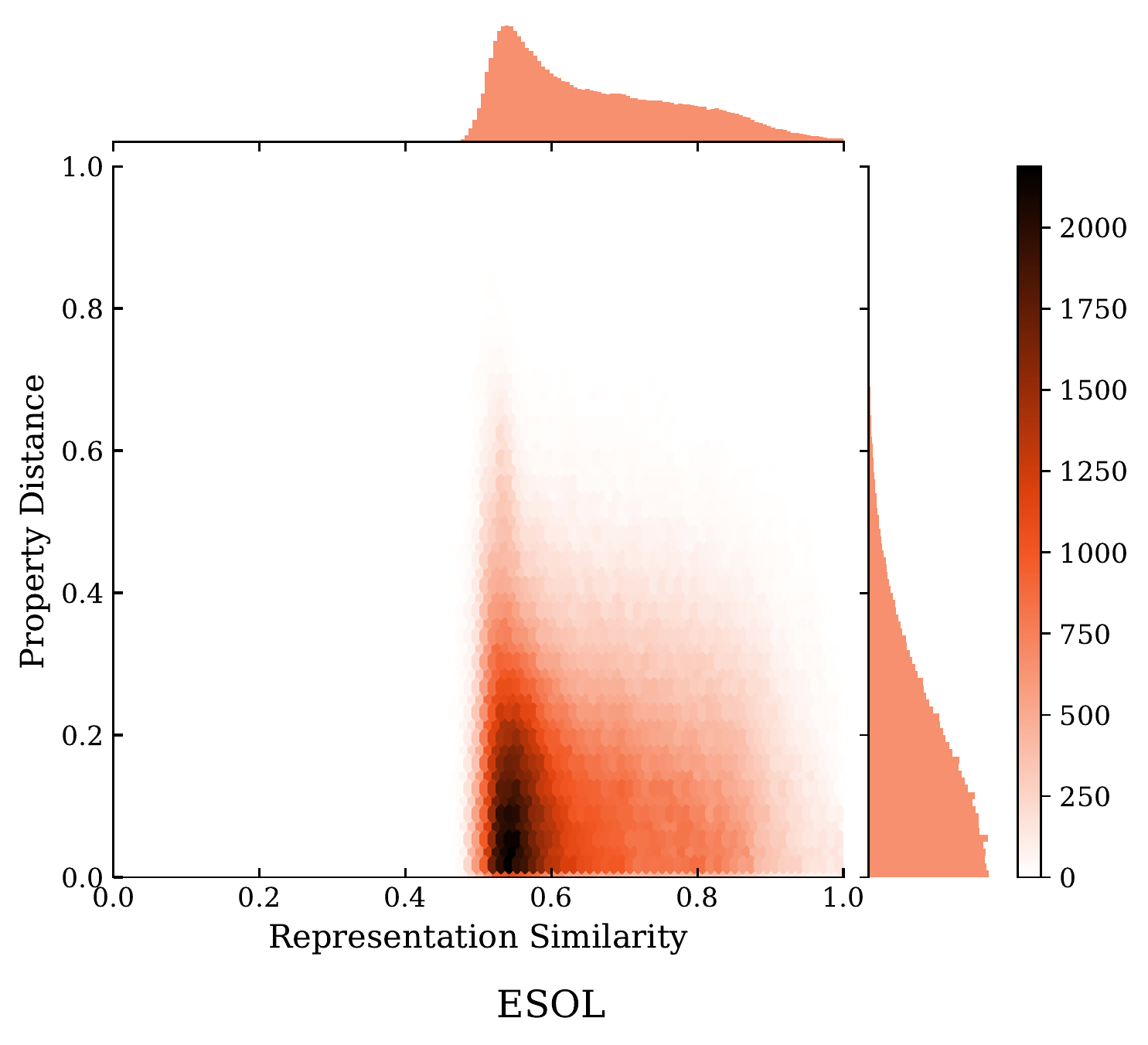}
}
    \subfigure[GraphLoG]{
	\includegraphics[width=3.25cm]{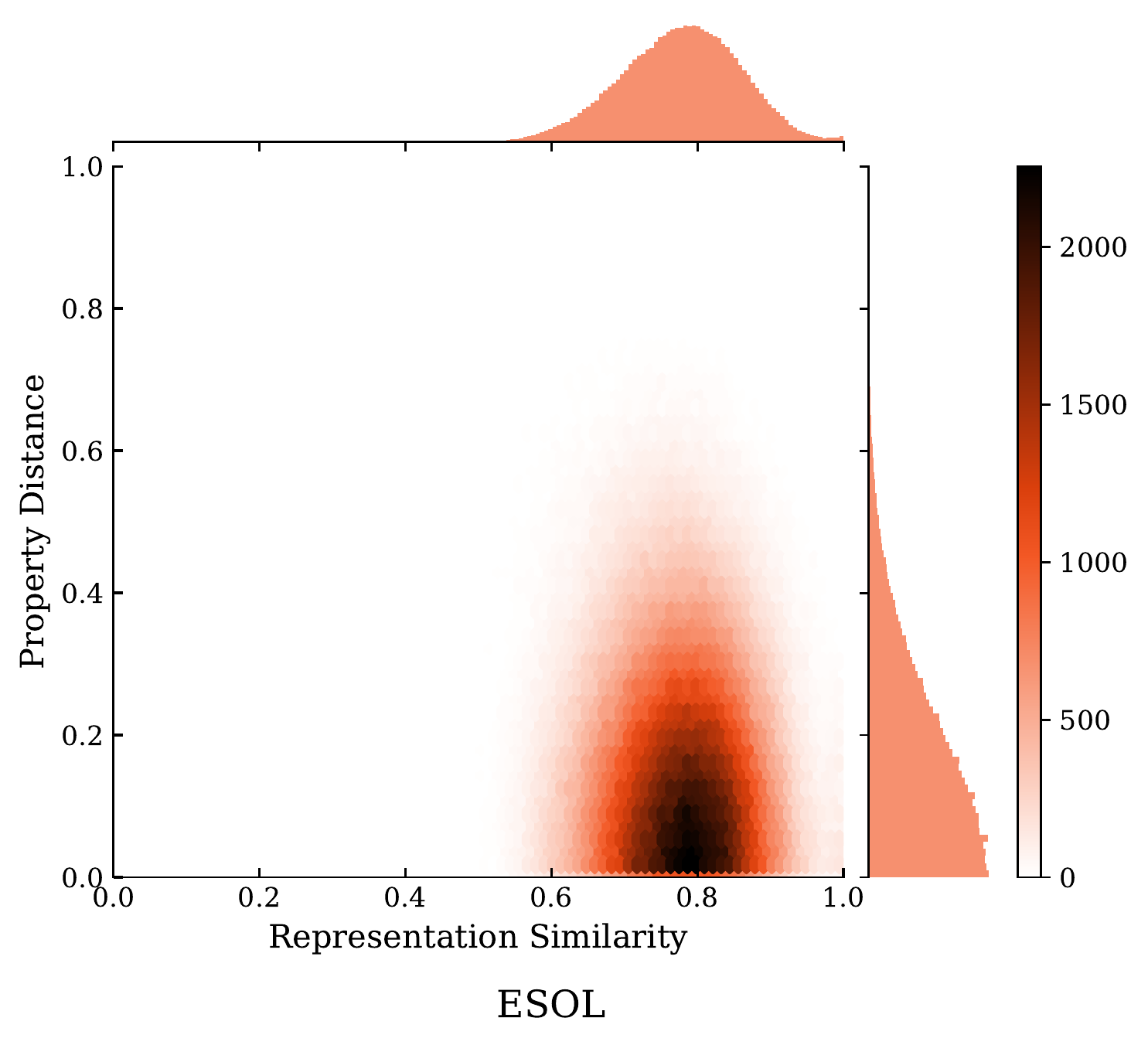}
}
    \subfigure[Pretrain8]{
	\includegraphics[width=3.25cm]{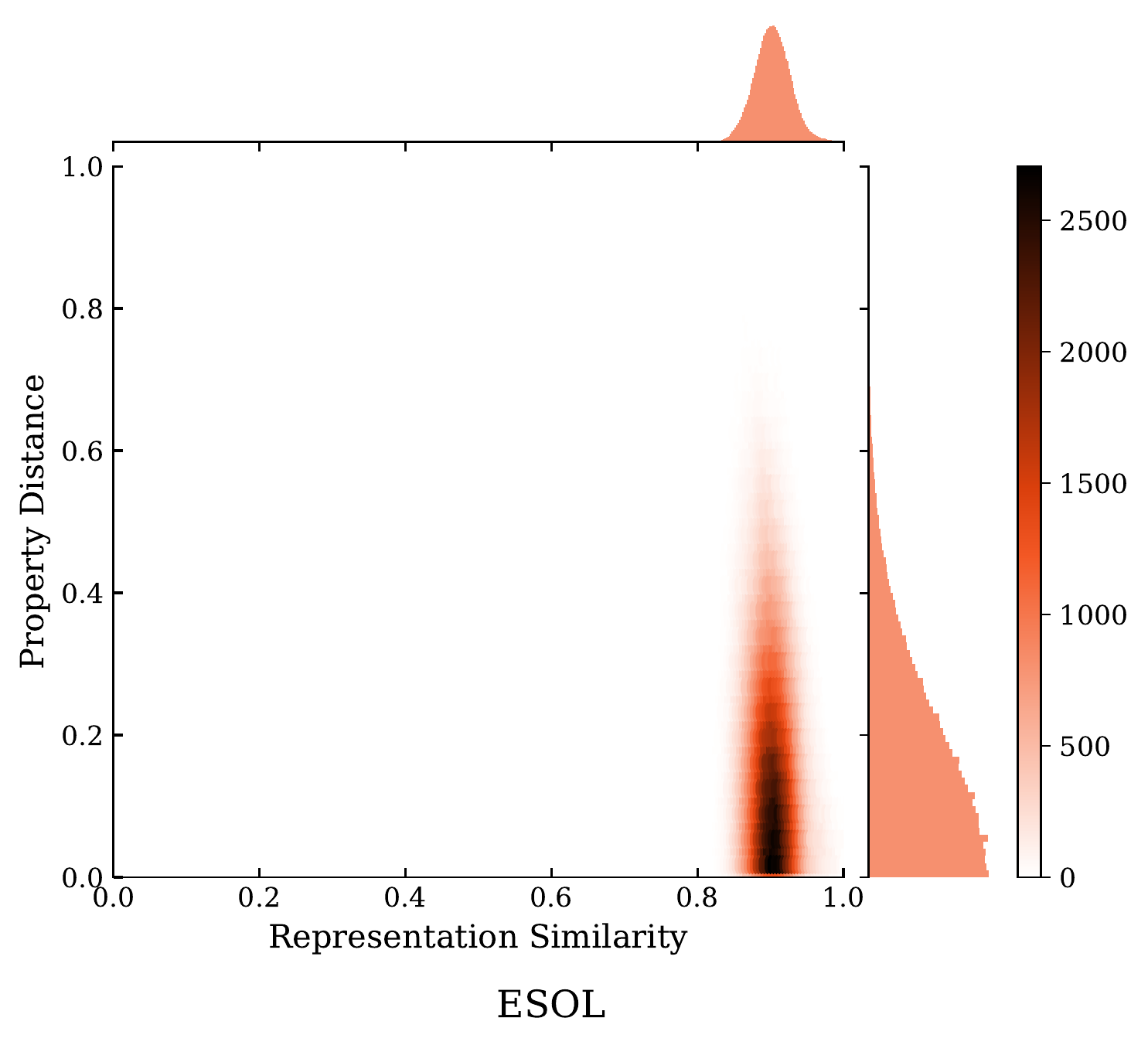}
}
	\caption{Distributions of molecular similarities with CosineSim as metric function.}
	\label{fig:ESOLCosDist}
\end{figure}

\begin{figure}[tbp]
	\centering
	\subfigure[ECFP]{
	\label{fig:RPSMap_ECFP_ESOL}
	\includegraphics[width=3.2cm]{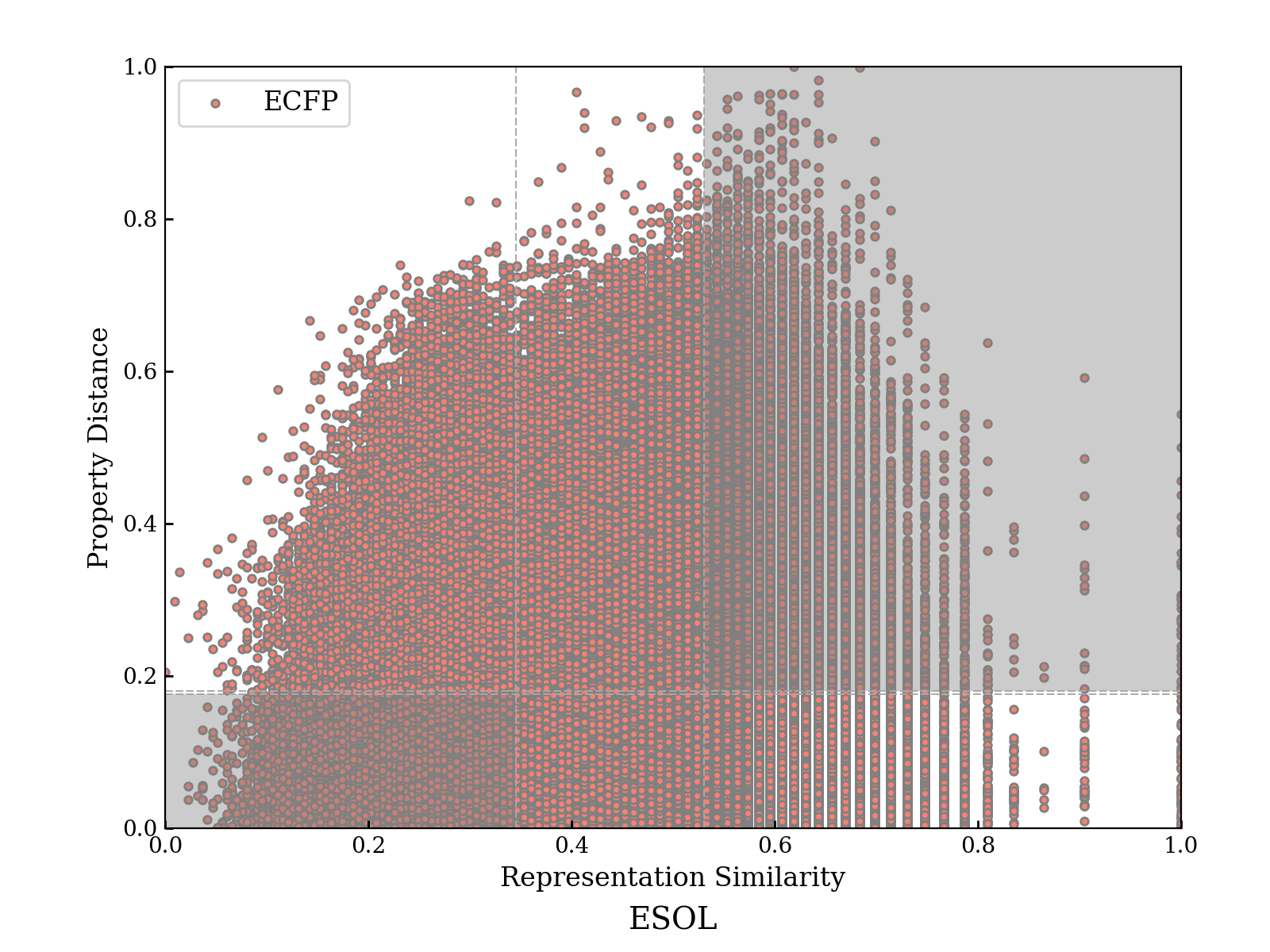}
}
    \subfigure[GROVER]{
    \includegraphics[width=3.2cm]{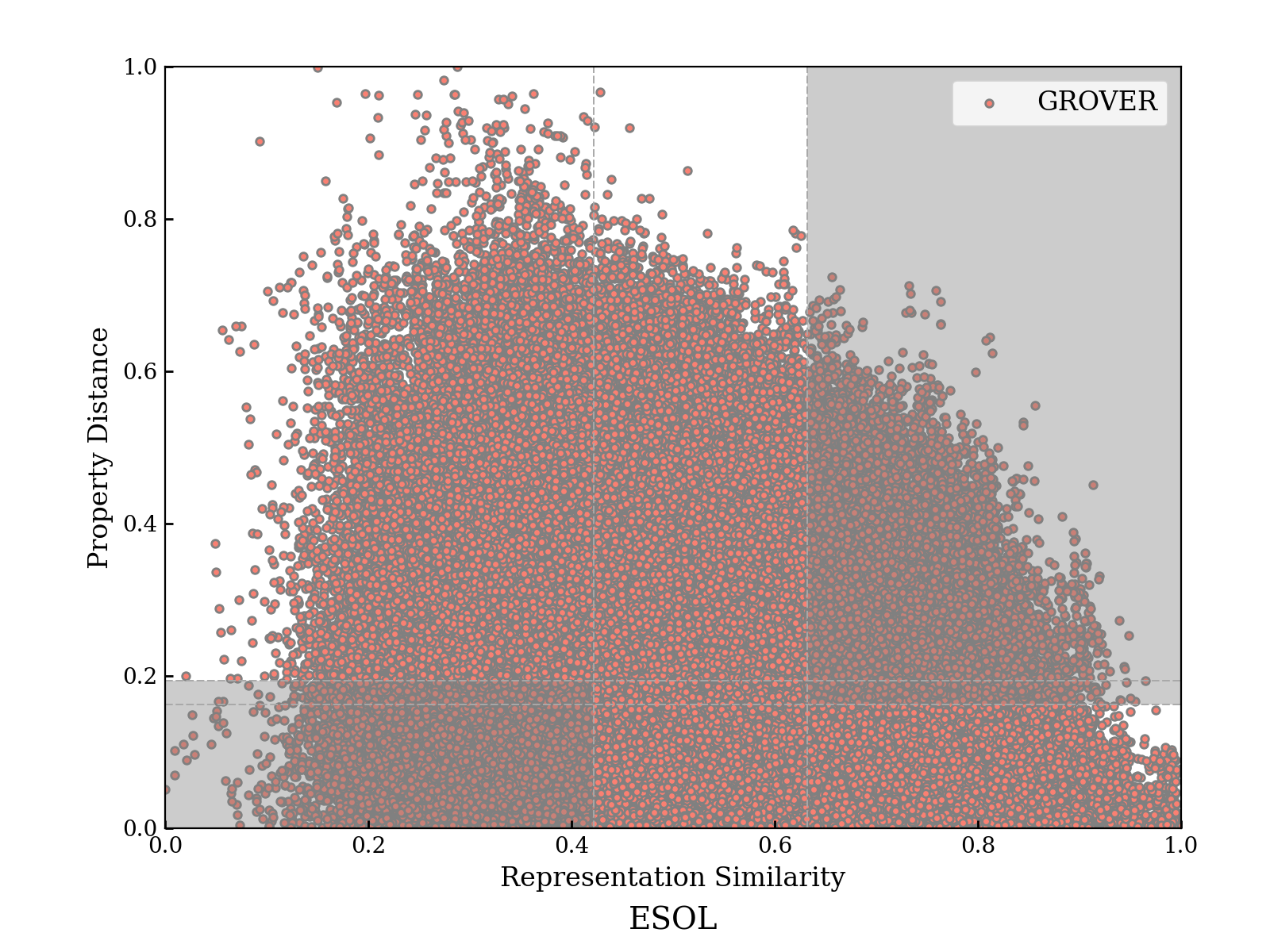}
}
    \subfigure[GraphLoG]{
    \label{GraphLoGCosDistrib}
	\includegraphics[width=3.2cm]{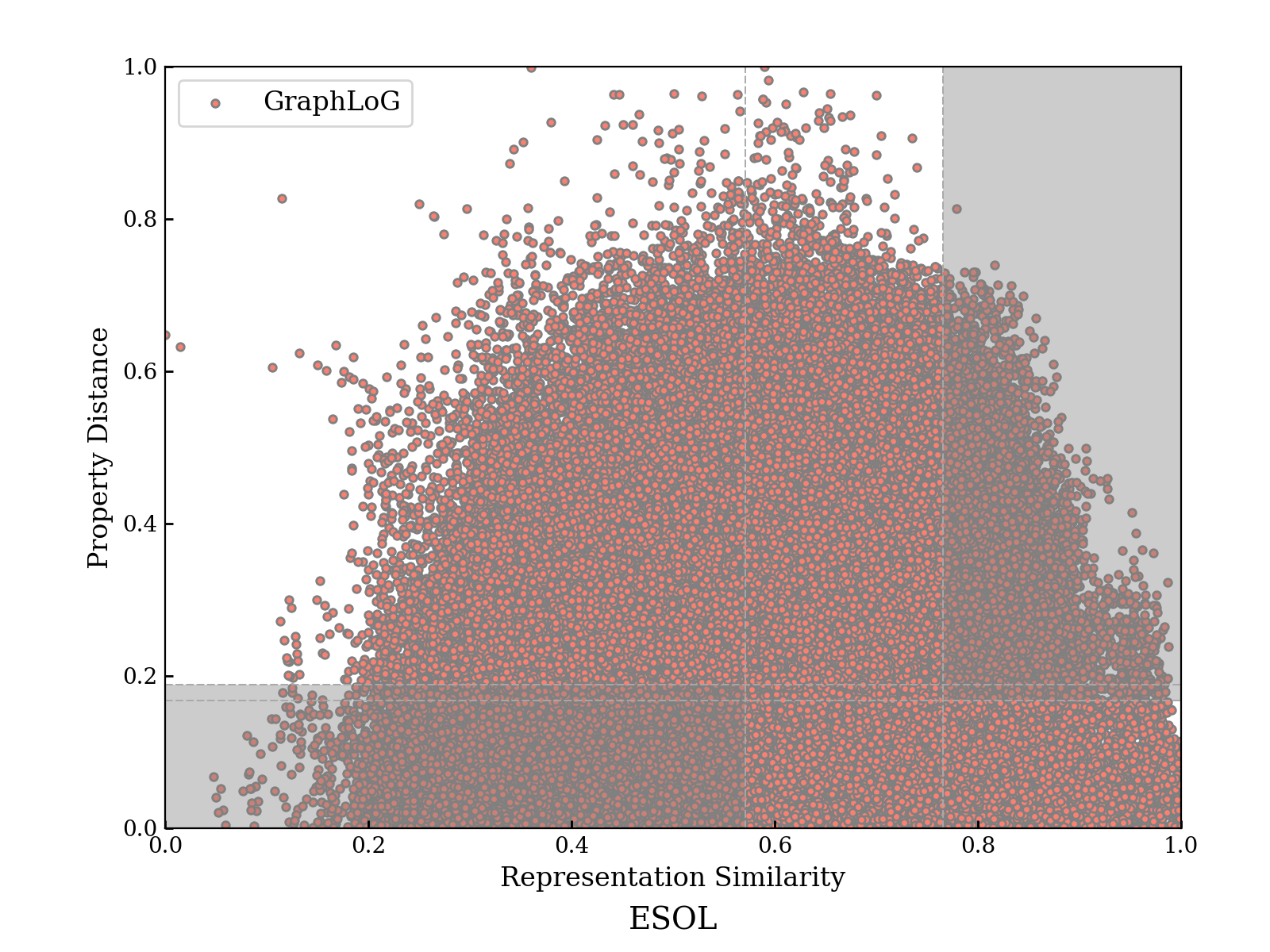}
}
    \subfigure[Pretrain8]{
	\includegraphics[width=3.2cm]{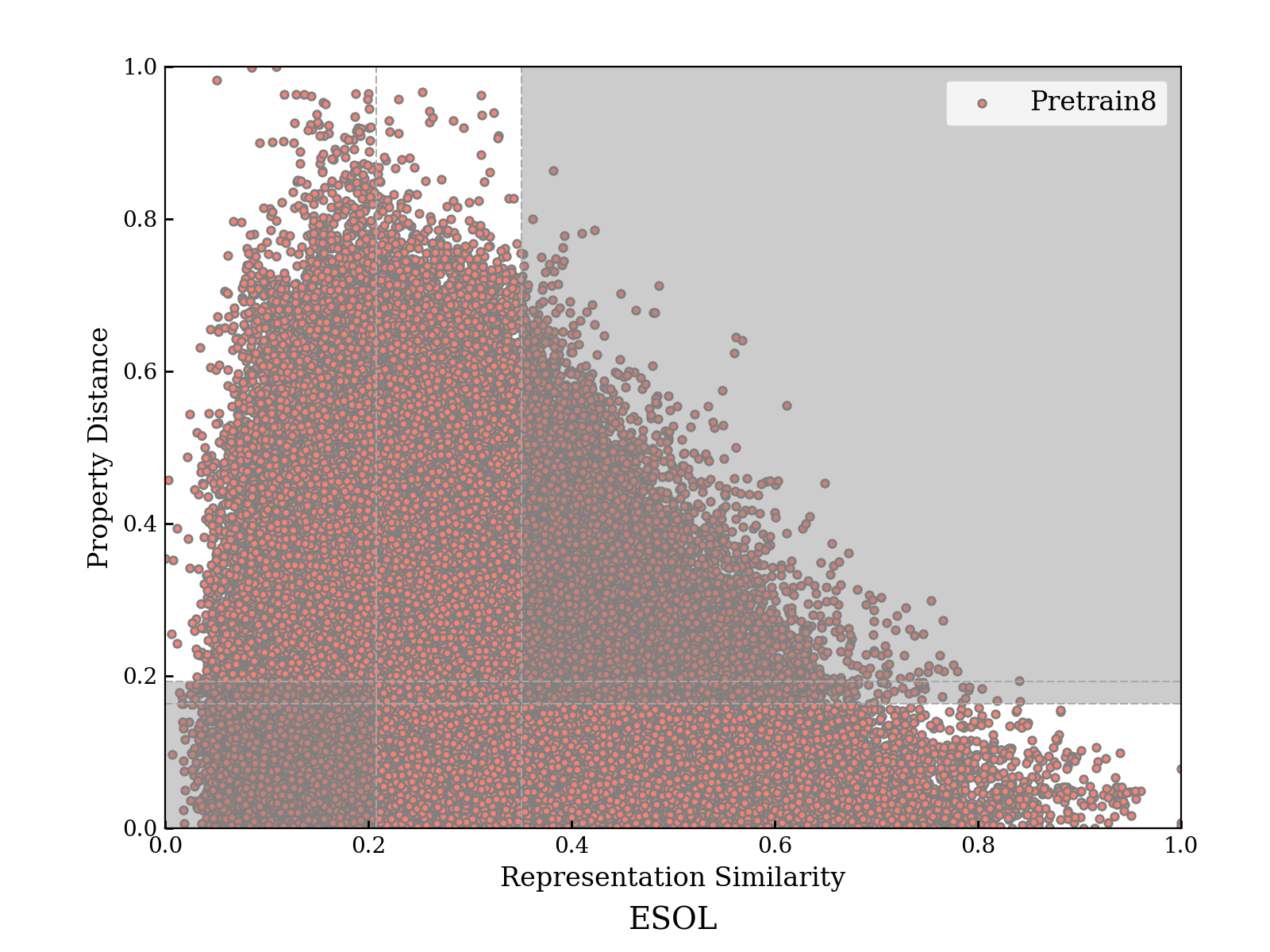}
}
	\caption{RPSMaps of representations with MinMaxEud as the metric function. The shadow areas stand for the region $R_1$ and $R_4$ for detecting ACs and SH.}
	\label{fig:ESOLMinMaxEudRPS}
\end{figure}

The RPSMaps of three pre-trained models and ECFP are selected as examples for comparison here, which are shown in Fig.~\ref{fig:ESOLCosDist} and Fig.~\ref{fig:ESOLMinMaxEudRPS}.
The regions for ACs and SH detection are shadowed in the RPSMaps.
By comparing these maps, the advantages and disadvantages of these models against ECFP can be discovered.

First, as each bit of ECFP corresponds to a circular substructure~\citep{rogers2010extended}, there is no common substructures between two molecules if the CosineSim of two molecules is zero (similarity = 0.5 in the map).
From Fig.~\ref{fig:ESOLCosDist}(a), it can be seen that most pairs of the molecules in ESOL share no common substructures.
While, it is difficult to obtain such finding from the distributions of other PTMs.
This indicates that the relationship between the basis of representation space of PTMs and the existence of substructures is not clear.

Notably, molecular motif prediction is used as a pretext task in GROVER~\citep{rong2020self}.
And it is shown in Fig.~\ref{fig:ESOLCosDist} that the distribution of CosineSim of GROVER is closer to that of ECFP than other PTMs, which indicates a better correlation between representations of GROVER and the substructures.
Similar discoveries can also be found in FreeSolv (see Fig.~\ref{fig:FreeSolv_Cosine_distribut}).

Second, each bit of ECFP can only represents the occurrence of a substructure.
The number of occurrences of each substructure and the similarity between substructures are lost.
This makes ECFP underestimate or overestimate the similarity of some specific pairs of molecules, which consequently causes undesirable ACs and SH, as shown in Fig.~\ref{fig:RPSMap_ECFP_ESOL}.
For instance, ECFP cannot discriminate alkanes with different lengths of carbon chains, e.g. \textit{CCCCCCCCCCCCCCCC} and \textit{CCCCCCCCCCCCCCCCCCCC} in ESOL, as their circular substructures are the same but the number of occurrences are different.
This characteristic of ECFP generates many pairs of samples with 1.0 molecular similarity but different properties, which can be found in the RPSMap.

However, for PTMs, since they typically leverage the entire molecular graph or SMILES to learn representations, they are able to capture the similarity between substructures and discriminate molecules with the same substructures.
As shown in Fig.~\ref{fig:ESOLMinMaxEudRPS}, there are fewer points with 1.0 molecular similarity than ECFP in the RPSMaps of PTMs.
And from Tab.~\ref{tab:fracofc1c4}, it is shown that fewer ACs and SH are detected in most PTMs than in ECFP.

In addition,
there are decreasing trends in the RPSMaps of all the three PTMs when the similarity increases.
And the RPSMap of Pretrain8 is obviously better than that of the other models.       
However, 
it can be figured out that the representation similarity of sample pairs with similar properties may widely distributed, which leads to SH.
Note that these PTMs are pre-trained only based on the structures of molecules without any information of properties, so that it may be difficult for PTMs to cluster the representation of samples with similar properties but different structures.

Our findings can be summarized as follows:

\begin{itemize}
    \item The PTMs can learn the similarity of structures better than ECFP, and discriminate the molecules with the same substructures that ECFP fails to.
    \item The relationship between the basis of the representation space generated by PTMs and the substructures of molecules is not prominent for most PTMs studied in this work.
    \item It is difficult for PTMs to avoid SH if the PTMs are trained only based on the molecular structures.
\end{itemize}

\subsection{Quality of Representations Evaluated by Scores}
The average deviation score values $s_{AD}$ and the improvement rate score values $s_{IR}$ of three selected PTMs on ESOL are computed and presented in Tab.~\ref{tab:esolscores}.
And all results are reported in Tab.~\ref{tab:allscores}

From Tab.~\ref{tab:esolscores}, we can see that both of the two scores tend to favor the representation extracted by Pretrain8.
It is consistent with the observation of the RPSMaps in Fig.~\ref{fig:ESOLMinMaxEudRPS}.
And from $s_{IR}$, we can see that the GROVER and Pretrain8 can generate better representations than ECFP, since their $s_{IR} < 2$.
While the $s_{IR} > 2$ for the GraphLoG model, which indicates that the representations generated by GraphLoG on the ESOL dataset are worse than ECFP.
The finetuning results in Tab.~\ref{tab:finetune_c} conform to this finding.

Among all of the 7 PTMs on 10 datasets, it is shown that Pretrain8 has a better representation space in terms of these two scores, since it achieves the best value on most datasets.
This result is consistent with the observation on the RPSMaps.
While in Tab.~\ref{tab:LogMEValue}, the LogME value favors GROVER on most datasets.
Although it is proved that a $(\delta, \epsilon)$-ideal relationship is consistent with a linear relationship to some extent, the divergence between our method and LogME may attribute to the alternative loose standards.

\begin{table}[htbp]
\caption{Scores for representation quality evaluation on the ESOL dataset.}
\label{tab:esolscores}
\centering
\begin{small}
\begin{tabular}{ccccc}
\toprule
Scores & GROVER & GraLoG &  Pre8 & ECFP\\
\midrule
$s_{AD}$ & 0.019 & 0.021 & \textbf{0.014} & 0.019\\
$s_{IR}$ & 1.923 & 2.047 & \textbf{1.813} & -\\
\bottomrule
\end{tabular}
\end{small}
\end{table}

\section{Discussions and Conclusion}
\label{sec:discussion}
In this paper, we focus on evaluating the quality of representations of drug molecules extracted by self-supervised pre-trained models.
A novel method \ours is proposed, which is inspired by Activity Cliffs and Scaffold Hopping in drug discovery, and mainly concerns the relationship of similarities between molecules.
\ours includes RPSMap for visualization analysis, and two scores for quantitative quality evaluation.

In the experiments, we use \ours to evaluate 7 pre-trained models on 10 target tasks to show what can be discovered by our method.
Analyzed by RPSMap, it can be seen that most of the pre-trained models can learn better structural similarity of molecules than ECFP.
But the basis of the representation space does not  explicitly correspond to the molecular substructures.
In addition, it is difficult for a PTM to avoid SH if the model is pre-trained only based on the molecular structures.
And the proposed scores can describe the quality of representations well.

This study provides an effective method for the drug D\&R community to evaluate the quality of representations generated by their proposed pre-trained models.
In addition, the discoveries may give directions for the community to develop new pre-training techniques.
First, to make the extracted representation more interpretable, molecular substructures should receive more attention by models, and the relationship between the basis of representations and the substructures should be enhanced.
Second, the occurrences of ACs and SH should be regularized, since these phenomena will impact the prediction accuracy of task layers.
However, most pre-trained models are trained only on the molecular structures, so that they can only encode molecules according to their structural similarity.
The relationship between molecular structures and properties is not always ideal.
In this case, it is difficult for pre-trained models to cope with ACs and SH.
So, researchers are encouraged to involve more domain-knowledge in training pre-trained models to realize such regularization.

For future works, more metric functions can be leveraged to provide descriptions of representations from different perspectives, such as manifold learning~\citep{lin2008riemannian}.
Better schemes of thresholds are also required.
In addition, the methods for out-of-distribution (OOD) could also be involved, since samples in most molecular datasets show OOD characteristics due to the distributions of their scaffolds~\citep{kim2021merged} and various other factors~\citep{2022arXiv220109637J}.





\clearpage 
\appendix

\begin{center}
	\LARGE \bf {Appendix}
\end{center}

\etocdepthtag.toc{mtappendix}
\etocsettagdepth{mtchapter}{none}
\etocsettagdepth{mtappendix}{subsection}
\tableofcontents

\setcounter{table}{0}
\renewcommand{\thetable}{S\arabic{table}}
\setcounter{figure}{0}
\renewcommand{\thefigure}{S\arabic{figure}}
\setcounter{equation}{0}
\renewcommand{\theequation}{S\arabic{equation}}


\clearpage

\section{Broader Impacts}
\label{sec:broaderimpacts}
As the labeled data in drug discovery is expensive and time-consuming, self-supervised pre-trained models trained on large amount of unlabeled data are expected to generate better representations than canonical FingerPrint methods and promote the development of new drugs.
Evaluating the quality of representations extracted by these pre-trained models is an important problem, but it is lack of concern.
By introducing the \ours method, our work can serve as an initial step towards tackling the problem of representation evaluation, with the hope to encourage the community to concern and evaluate the quality of representations rather than just reporting the prediction accuracy after finetuning.
The discoveries revealed by analyzing state-of-the-art pre-trained models may give directions for the community to develop new pre-training techniques.
Besides, this paper does not raise any ethical concerns.
This study does not involve any human subjects, practices to data set releases, potentially harmful insights, methodologies and applications, potential conflicts of interest and sponsorship, discrimination/bias/fairness concerns, privacy and security issue, legal compliance, and research integrity issues.

\section{Toysets for Illustration of Proposition~\ref{prop:1}}
\label{sec:toyset}
In this section, toysets are used to explain why the occurrence of generalized ACs and SH is not ideal. 
A classification task is used as an example, and it is easy to draw similar conclusion for the regression tasks.

\begin{figure}[thbp]
	\centering
	\subfigure[]{
	\label{fig:ACandSH(a)}
	\includegraphics[width=6cm]{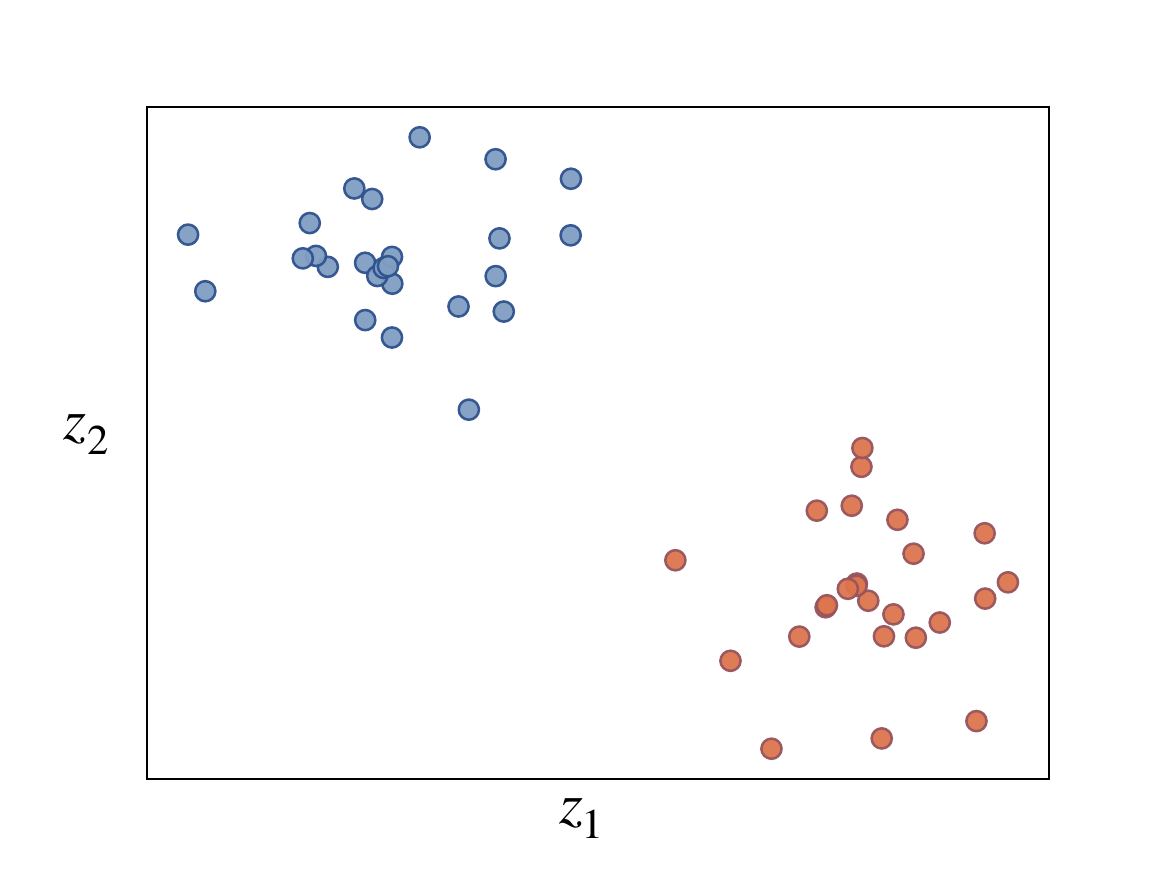}
}
    \subfigure[]{
    \label{fig:ACandSH(b)}
    \includegraphics[width=6cm]{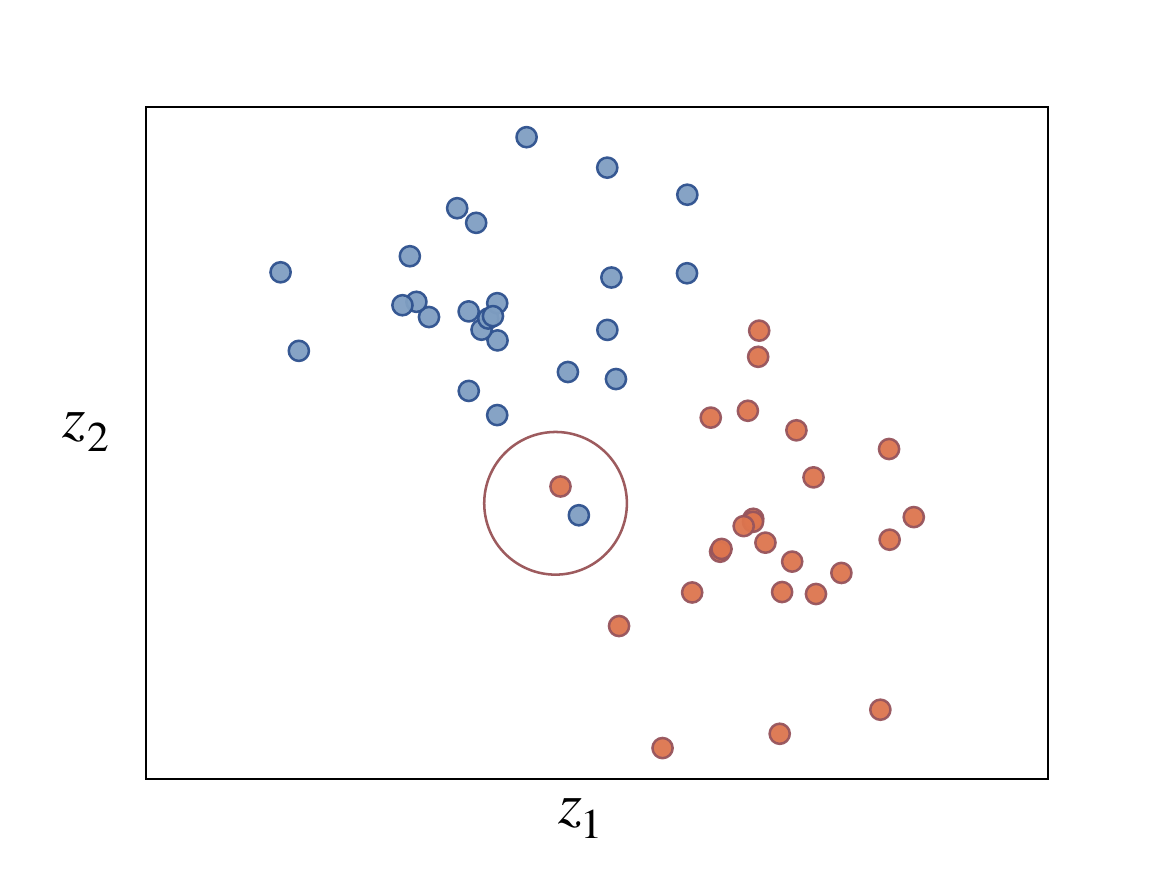}
}
    \subfigure[]{
    \label{fig:ACandSH(c)}
	\includegraphics[width=6cm]{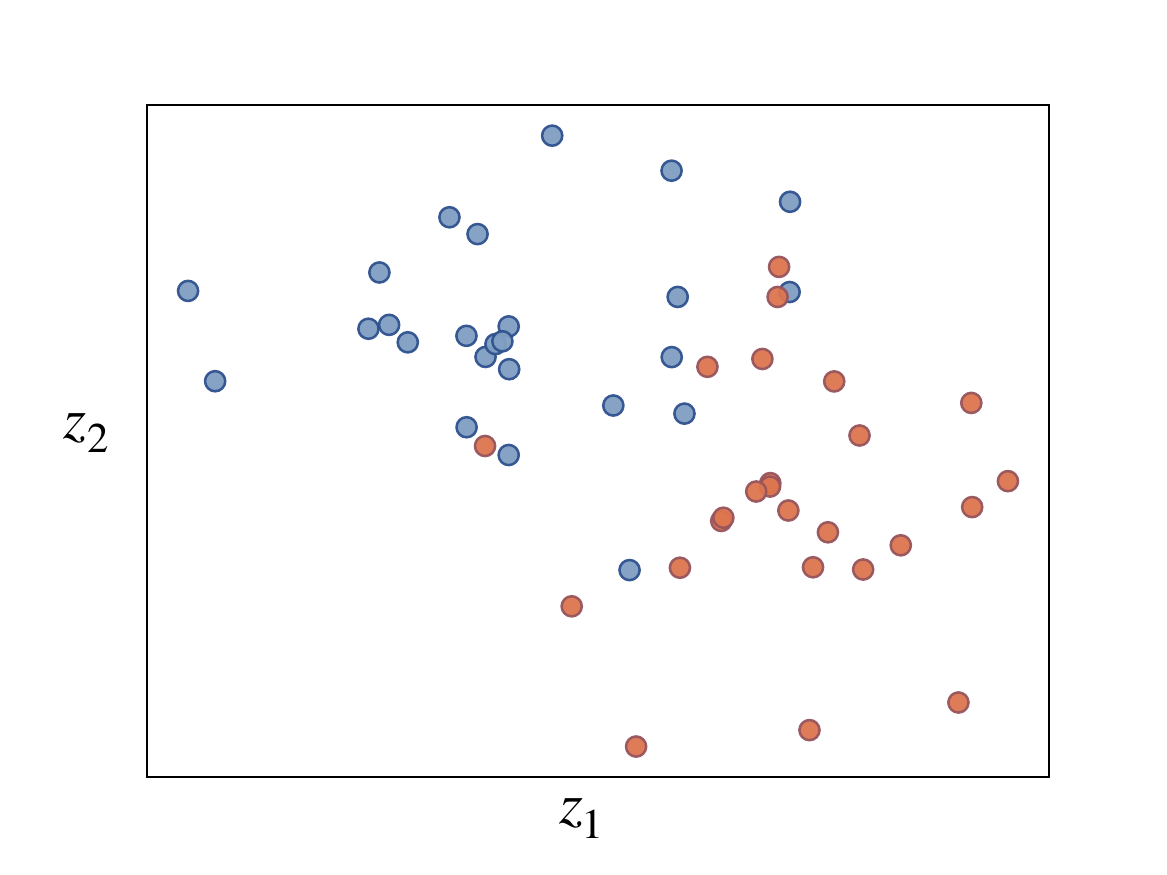}
}
    \subfigure[]{
    \label{fig:ACandSH(d)}
	\includegraphics[width=6cm]{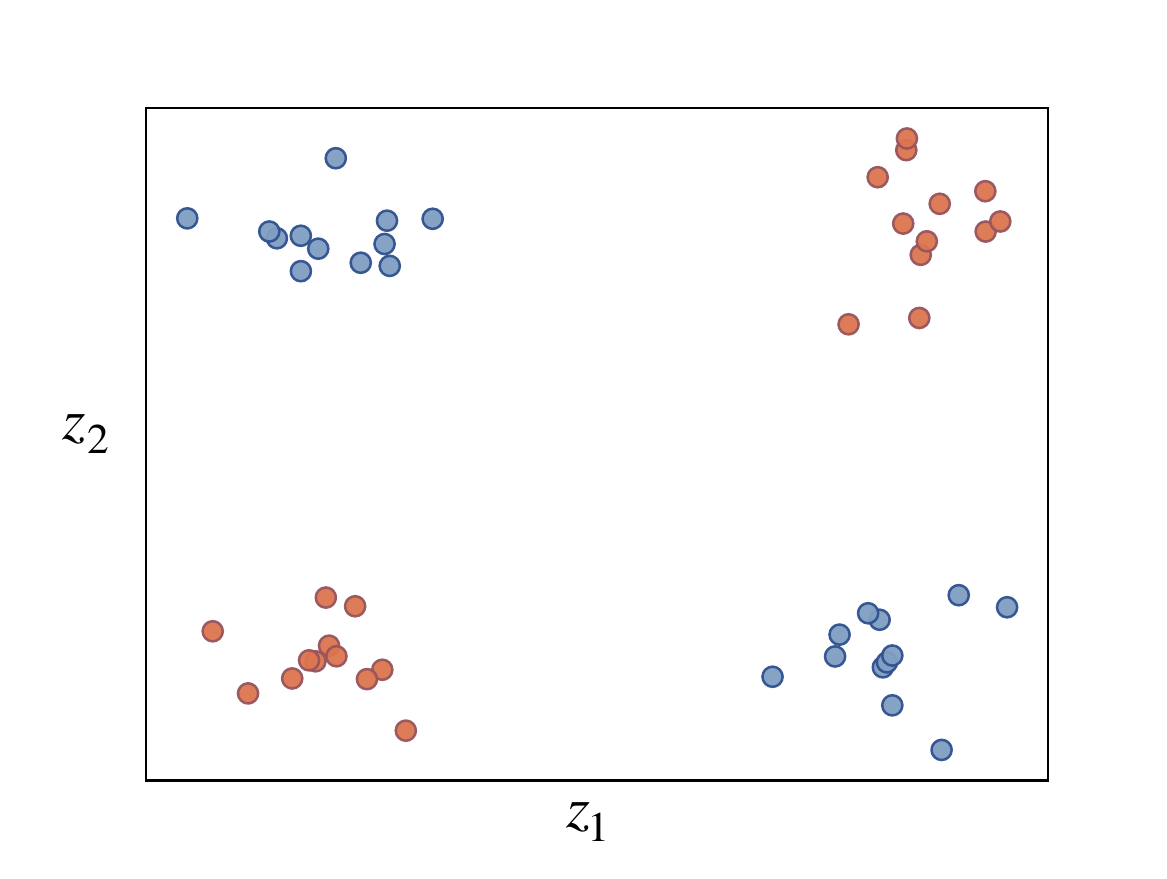}
}
	\caption{Toysets for illustration of ACs and SH. Colors indicate different classes.}
	\label{fig:ACandSH}
\end{figure}

First, as shown in Fig.~\ref{fig:ACandSH}, assuming a binary classification task, the representation-property relationship is ideal if the representations are clustering in two clusters according to their labels with an explicit margin between clusters.
A linear model can easily learn a half-plane to make prediction perfectly.

With diameter of the clusters increasing, ACs will occur.
As shown in Fig.~\ref{fig:ACandSH}(b), the pair points in red circle perform as ACs.
In this case, down-stream task layers can either learn to fit a simple linear half-plane, or to fit a more complex half-plane to make prediction.
For the first choice, although the model can achieve a relatively good prediction accuracy (since only 1 sample is wrongly predicted) and have a good generalization ability, it actually ignores the occurrence of AC.
From a drug discovery aspect, the occurrence of ACs is important source of information to study the QSAR.
And from a general deep learning aspect, the ACs are hard cases that should be focused.
So, it is not proper to ignore the occurence of ACs.
In addition, when the number of ACs is large, as shown in Fig.~\ref{fig:ACandSH}(c), a linear model will no longer lead to a good accuracy.
For the second choice, a complex half-plane can predict the samples of ACs correctly, while the increasing number of parameters requires a large amount of data to train and therefore lead to the risk of over-fitting.
In conclusion, it is better for an encoder to generate representations with no ACs.

To illustrate the influence of SH, considering the case that the representations of a binary classification task are clustering in four clusters, as shown in Fig.~\ref{fig:ACandSH}(d).
In this case, samples in different clusters with the same label perform as SH.
It is obvious that, although there are explicit margins between the clusters, one cannot find a linear half-plane to classify these samples correctly.
The down-stream task layers needs to fit a higher-order half-plane.
And if the number of clusters increases, this half-plane will be more complex and consequently lead to the risk of over-fitting.
So, the occurrence of SH is not ideal as well.

\clearpage
\section{Supplemental Figures for Sec.~\ref{sec:method}}

\begin{figure}[thbp]
    \centering
    \includegraphics[width=8cm]{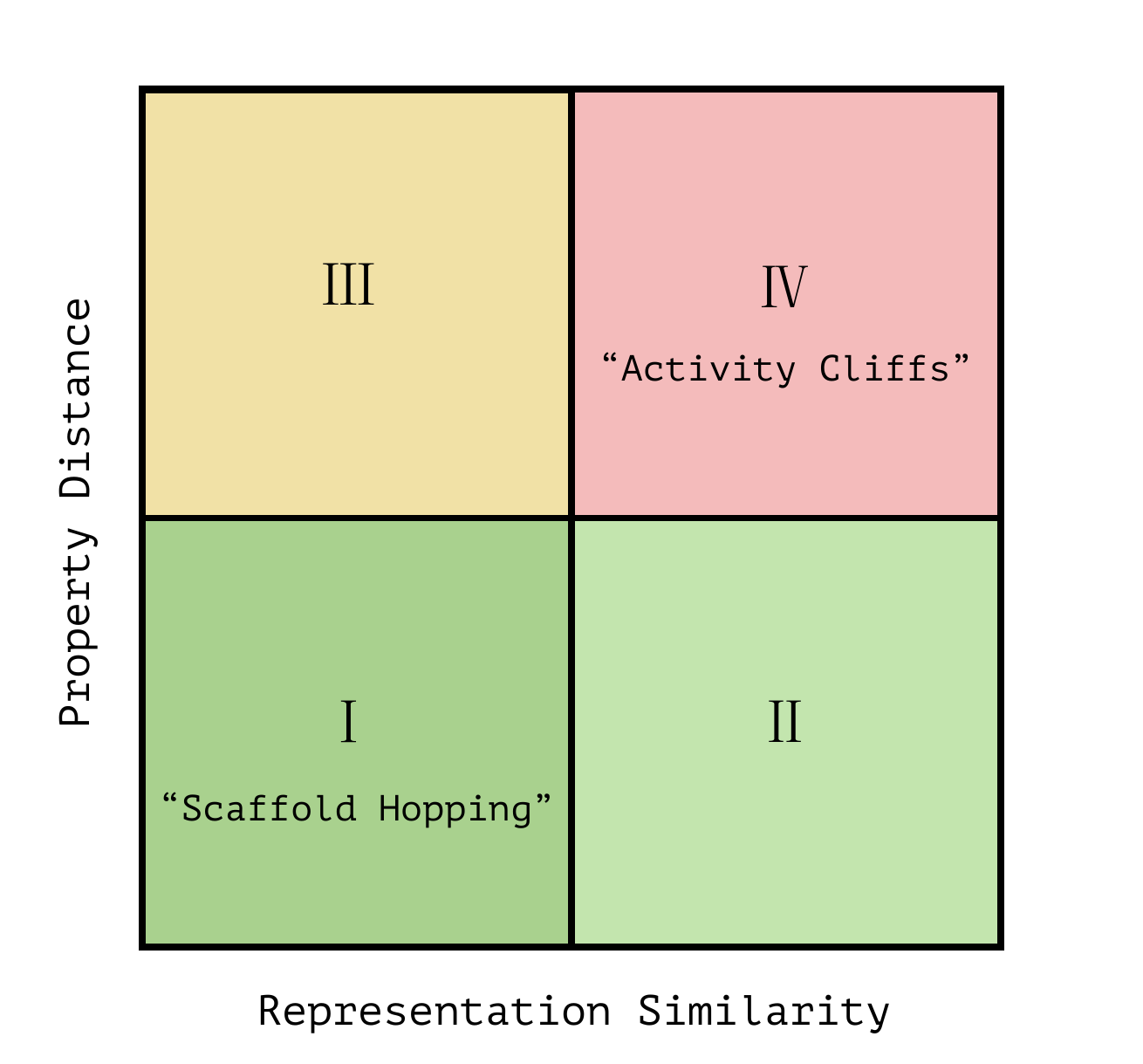}
    \caption{An example of the Representation-Property Similarity Map (RPSMap).}
    \label{fig:RPSMap}
\end{figure}

\begin{figure}[thbp]
    \centering
    \includegraphics[width=10cm]{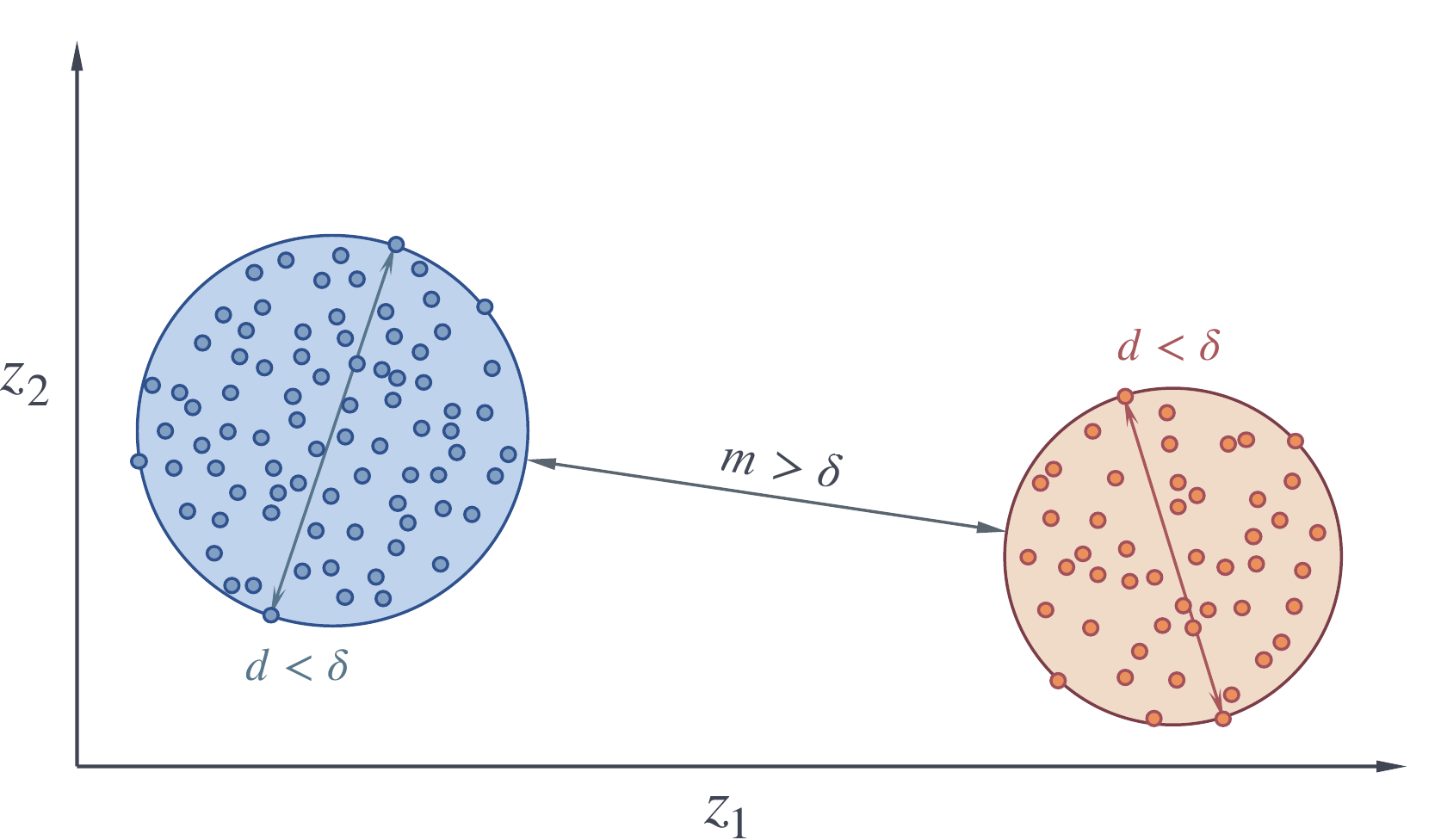}
    \caption{An example of the distribution of samples if the relationship between representations and the properties is $(\delta, \epsilon^*)$-ideal.}
    \label{fig:class_ideal}
\end{figure}

\clearpage
\section{Introduction of LogME and TransRate}
\label{sec:LogMEandTransRate}
As discussed in Sec.~\ref{sec:relatedworks}, there is no work that directly related to the problem we are studying.
LogME~\cite{you2021logme} and TransRate~\cite{huang2021frustratingly} similarly evaluate the relationship between representations and properties, but the fundamental research goal of their works are different from ours.
In this section, we give a brief introduction of these two methods.

\subsection{LogME}
You et~al.~\cite{you2021logme} proposed LogME to evaluate transferability.
In their work, they firstly assumed that an ideal relationship between representations $\Z$ and labels $\Y$ should be linear.
Under this assumption, the compatibility of $\Z$ and $\Y$ is suggested to be calculated by the performance of a simple linear layer used as $\regressor$.
Considering that using the likelihood $p(y|\Z,\omega^*)$ of a linear model with the best $\omega^*$ to estimate the performance is prone to over-fitting, the authors proposed to use the evidence $p(y|\Z)=\int p(\omega)p(y|\Z,\omega)d\omega$ for alternative.

\subsection{TransRate}
Huang et~al.~\cite{huang2021frustratingly} proposed TransRate to measure the transferability. 
In their work, it is argued that the transferability should characterize how well the \textbf{optimal} model performs on the target task.
They actually assumed that the task layer $\regressor$ should be optimized after fine-tuning, and it can extract and leverage all of the information about $\Y$ contained in $\Z$ to make predictions.
Based on this assumption, the authors proposed to use the mutual information $MI(\Z;\Y)$ as a strong indicator to estimate the compatibility of $\Z \rightarrow \Y$.
\begin{align}
    TrR_{T_s \rightarrow T_t}(\repre) = H(\Z) - H(\Z|\Y).
\end{align}

\clearpage
\section{Proof of Theorems.}
\label{sec:proof}
\paragraph{Proof of Theorem~\ref{theor:1}}
\begin{proof}
Multi-task multi-class properties $\Y = \{\y_i\}_{i=1}^{N}$, $\y_i = [y_k]_{k=1}^K$, $y_k \in \{0,1,\dots,c_k\}$, where $K$ is the number of tasks and $c_k$ is the number of classes of task $k$.
With Euclidean distance as metric function, denote $d_{ij} = \|\z_i-\z_j\|$, $\Delta_{ij} = \|\y_i - \y_j\|$.

Given two samples $i,j$, if $\y_i \neq \y_j$, then $\min(\Delta_{ij}) = 1$.
So, for any $\epsilon^* \in (0,1)$, if $\Delta_{ij} < \epsilon^*$, then $\y_i = \y_j$, $i, j$ are samples with the same labels.
And if $\Delta_{ij} > \epsilon^*$, then $i,j $ are samples with different labels.

If the relationship between $\Z$ and $\Y$ is $(\delta, \epsilon^*)$-ideal, then the following sub-conditions holds:
\begin{equation}
\label{eq:sub1}
    \forall \z_i, \z_j \in \Z, d_{ij} < \delta \Rightarrow \Delta_{ij} < \epsilon^*
\end{equation}
and
\begin{equation}
\label{eq:sub2}
    \forall \z_1, \z_2 \in \Z, \Delta_{ij} < \epsilon^* \Rightarrow d_{ij} < \delta.
\end{equation}

If sub-condition~\ref{eq:sub2} holds, it indicates that any samples with the same labels are clustering with diameter $d<\delta$.

And the converse-negative proposition of sub-condition~\ref{eq:sub1} is:
\begin{equation}
    \label{eq:sub3}
    \forall \z_i, \z_j \in \Z, \Delta_{ij} > \epsilon^* \Rightarrow d_{ij} > \delta.
\end{equation}
Recall that the samples with the same labels are clustering.
When $\Delta_{ij} > \epsilon^*$, $\y_i \neq \y_j$, $i,j$ are in different clusters.
For two clusters $p$ and $q$, the margin between clusters is denoted as $m_{pq} = \min(d_{ij})$, where $i,j$ are any points in $p$ and $q$, respectively.
So, if sub-condition~\ref{eq:sub1} holds, it indicates that for any clusters $p$ and $q$, the margin $m_{pq} > \delta$.
\end{proof}

\paragraph{Proof of Lemma~\ref{lemma:1}}
\begin{proof}
As $y = w \cdot z + b$, we have $\|y_i - y_j\| = \|w(z_i-z_j)\| = \|w\| \cdot \|z_i-z_j\|$.
So, $ \forall z_i, z_j \in \Z$, if $d_{ij} < \delta$, $\Delta_{ij} = \|y_i-y_j\| = \|w\| \cdot d_{ij} < \|w\| \delta$, the sub-condition~\ref{eq:sub1} holds.
Similarly, the sub-condition~\ref{eq:sub2} holds.
So, the relationship is $(\delta, \|w\|\delta)$-ideal.
\end{proof}

\begin{figure}[thbp]
    \centering
    \includegraphics[width=8cm]{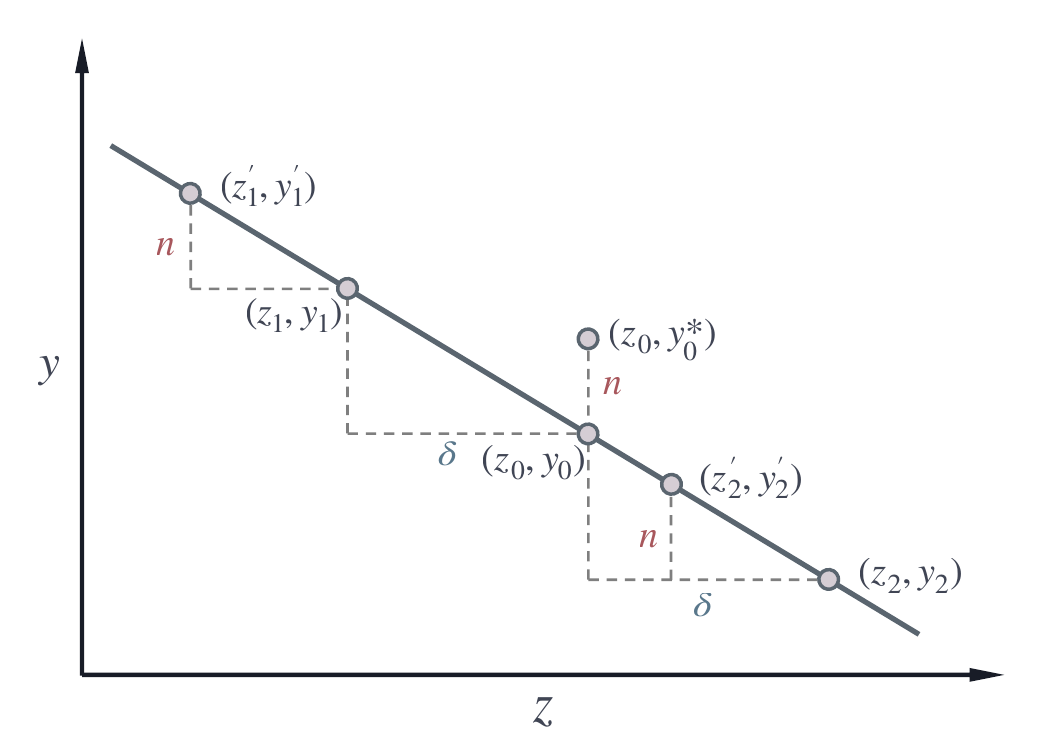}
    \caption{The location of the points in the proof of Lemma~\ref{lemma:2}}
    \label{fig:line}
\end{figure}

\paragraph{Proof of Lemma~\ref{lemma:2}}
\begin{proof}
Assume that $w < 0$ and $n > 0$ which is easy for discussion, and the proof of the opposite case is similar.

With the disturbed point $(z_0,y^*_0)$ given, two points $(z_1, y_1)$ and $(z_2, y_2)$ can be found from $D$, where $z_1 = z_0 - \delta$ and $z_2 = z_0 + \delta$.
And two more points $(z'_1, y'_1)$ and $(z'_2, y'_2)$ can be found, where $y'_1 = y_1 + n$ and $y'_2 = y_2 + n$.

The location of these points is shown in Fig.~\ref{fig:line}.

Denote 
\begin{align}
    D_1 &= \{(z, y) \,|\, y = w \cdot z+b, z \in (z_1, z_2)\},\\
    D_2 &= \{(z, y) \,|\, y = w \cdot z+b, z \in (-\infty, z_1) \cup (z_2, +\infty)\}.
\end{align}
It is obvious that
\begin{align}
    \forall (z_i, y_i) \in D_1, d_z(z_i, z_0) &< \delta, \\
    \forall (z_j, y_j) \in D_2, d_z(z_j, z_0) &> \delta.
\end{align}

Similarly, denote
\begin{align}
    D_3 &= \{(z, y) \,|\, y = w \cdot z+b, z \in (z'_1, z'_2)\}, \\
    D_4 &= \{(z, y) \,|\, y = w \cdot z+b, z \in (-\infty, z'_1) \cup (z'_2, +\infty)\}.
\end{align}
Obviously,
\begin{align}
    \forall (z_i, y_i) \in D_3, d_y(y_i, y^*_0) &< \|w\|\delta, \\
    \forall (z_j, y_j) \in D_4, d_y(y_j, y^*_0) &> \|w\|\delta.
\end{align}

As $z'_2 < z_2$, $D_1 \cap D_4 \neq \varnothing$, we have,
\begin{equation}
    \forall (z_i, y_i) \in D_1 \cap D_4, d_z(z_i, z_0) < \delta \Rightarrow d_y(y_i, y^*_0) > \|w\|\delta
\end{equation}
, which perform as ACs.

And as $z'_1 < z_1$, $D_2 \cap D_3 \neq \varnothing$, we have,
\begin{equation}
    \forall (z_j, y_j) \in D_2 \cap D_3, d_z(z_j, z_0) > \delta \Rightarrow d_y(y_j, y^*_0) < \|w\|\delta
\end{equation}
, which perform as SH.

So, the disturbed point $(z_0, y^*_0)$ can be detect by $(\delta, \|w\|\delta)$-standard ACs and SH.
\end{proof}

\paragraph{Proof of Theorem~\ref{theor:2}}
\begin{proof}
From Lemma~\ref{lemma:2}, for an infinite dataset $D=\{(z,y)\}$, if there is one sample $(z_0, y^*_0)$ disturbed from the linear relationship $y = w \cdot z + b$ with noise $n$, $\|w\| = \frac{\epsilon}{\delta}$,
this disturbed point will cause $(\delta, \epsilon)$-standard ACs with points $(z_i,y_i) \in D_1 \cap D_4$ and SH with points $(z_j, y_j) \in D_2 \cap D_3$.
And therefore the relationship is not $(\delta, \epsilon)$-ideal.

So, if a relationship between $\Y$ and $\Z$ is $(\delta, \epsilon)$-ideal, then there is no sample $(z_0, y^*_0)$ biased from  $y = w \cdot z + b$, so the relationship is linear.
\end{proof}

\paragraph{Proof of Theorem~\ref{theor:3}}
\begin{proof}
Similar to the proof of Theorem~\ref{theor:2}, a disturbed point can be detected by $(\delta, \epsilon)$-standard with points in $D_1 \cap D_4$ and $D_2 \cap D_3$.
However, as $D$ is a finite dataset of which points can be considered as being sampled from a continuous space, those detective points in $D_1 \cap D_4$ and $D_2 \cap D_3$ may not exist in $D$.
In this case, if a disturbed point exists in $D$, it may not be detected.
So, it is not guaranteed that an $(\delta, \epsilon)$-ideal relationship is linear.
\end{proof}

\clearpage
\section{Further Discussion about $(\delta, \epsilon)$-ideal on Regression Tasks}
\label{sec:vectorcase}
In Sec.~\ref{sec:thresholds}, we have established the relationship between $(\delta, \epsilon)$-ideal and the distribution of samples when $z$ and $y$ are scalars on regression tasks.
And when $\z$ and $\y$ are vectors, we have found that for some cases, even a linear relationship cannot satisfy the $(\delta, \epsilon)$-ideal condition.
$(\delta, \epsilon)$-ideal in this case is more strict than linear.

For example, considering a linear relationship between $\Y = \{y_i\}, y_i \in \R$ and $\Z = \{\z_i\}, \z_i \in \R^2$, $y = w_1 z_1 + w_2 z_2 + b$, where $w_1 = w_2$.
For any given $(\delta, \epsilon)$-standard, we can find two samples $\z_i = (-E,E)$ and $\z_j = (E,-E)$, $\z_i = \z_j$, where $E$ denotes a very large value. 
Therefore, we have $d_z(\z_i, \z_j) = 2\sqrt{2}E > \delta $, but $y_i = y_j$, $d_y(y_i, y_j) = 0 < \epsilon$ since the thresholds are typically positive.
So, although the relationship between $\Y$ and $\Z$ is linear, there are still SH detected by $(\delta, \epsilon)$-standard.

This finding indicates that, for the case that the representations and properties are vectors on regression tasks, a $(\delta, \epsilon)$-standard may be too strict.
However, in practice, as introduced in Sec.~\ref{sec:thresholds}, it is suggested to use relatively loose standards to detect ACs and SH by using statistical information of representations.
And the distributions of representations with CosineSim as metric function indicate that almost all of the pairs of samples are sharing positive cosine similarity, as shown in Fig.~\ref{fig:ESOL_Cosine_distribut_all} and Fig.~\ref{fig:FreeSolv_Cosine_distribut}.
The special case that $\z_i = - \z_j$ does not exist.
So, our method will still work in practice.

\section{Discussion of Metric Functions}
\label{sec:metrics}
In Sec.~\ref{sec:thresholds}, the relationship between the $(\delta, \epsilon)$-ideal and the distribution of samples is analyzed based on Euclidean distance as metric functions.
However, it cannot be directly exploited in our problem since the scale of representations extracted by PTMs may be different.
So, the Euclidean distance between samples should first be normalized.
Here we use a MinMax normalization method to reflect the relative size of distances between samples in the space, denoted as MinMaxEud for short.
\paragraph{MinMaxEud:}
Denote $d_{ij} = \|\z_i - \z_j\|$, $\Delta_{ij} = \|\y_i - \y_j\|$ then,
\begin{align}
    d_z(\z_i,\z_j) = d^{MinMax}_{ij} &= \frac{d_{ij} - \min \limits_{i,j}(d_{ij})}{\max \limits_{i,j}(d_{ij}) - \min \limits_{i,j}(d_{ij})}, \\
    d_y(\y_i,\y_j) = \Delta^{MinMax}_{ij} &= \frac{\Delta_{ij}-\min \limits_{i,j}(\Delta_{ij})}{\max \limits_{i,j}(\Delta_{ij}) - \min \limits_{i,j}(\Delta_{ij})}.
\end{align}
Note that the thresholds derived by Euclidean distance should be normalized as well.
And the $F_{d \rightarrow s}(\cdot)$ is a simple linear function $sim_z = 1 - d_z$ to transform the MinMaxEud distance dimension to a similarity dimension and scale the range to $[0,1]$.

Furthermore, to generate more interpretable representations, it is expected to make the basis of a well-trained representation space correspond to different structural characteristics of molecules.
The direction of the representation vector will reflects different combination of these structural characteristics.
Thus, the cosine similarity (ConsineSim for short), which embed the angle of two representation vectors, also provide important information.

\paragraph{CosineSim:}
\begin{equation}
    sim_z(\z_i,\z_j) = \cos \theta_{ij} = \frac{<\z_i,\z_j>}{\|\z_i\|\cdot\|\z_j\|}.
\end{equation}
As the CosineSim is a similarity dimension, $F_{d \rightarrow s}(\cdot)$ is not needed.
However, since the range of CosineSim is $[-1,1]$, it needs a transform function $sim'_z = \frac{1}{2}(sim_z+1)$ to scale the range to $[0,1]$.

Based on the discussion, in the experiments of this work, MinMaxEud is used as metric function to draw RPSMap and compute scores, and CosineSim is used to draw distribution for analysis.

\clearpage
\section{Information of Pre-trained Models and Target Tasks Experiments}
\label{sec:PTM}
In this work, we use 7 pre-trained models, including GROVER~\cite{rong2020self}, ChemBERT~\cite{kim2021merged}, GraphLoG~\cite{xu2021self}, MAT~\cite{maziarka2020molecule}, SMILESTransformer~\cite{honda2019smiles}, PretrainGNNs~\cite{hu2019strategies} and Pretrain8~\cite{chen2021extracting}.
In this section, we will give an introduction of these related PTMs for better understanding how these PTMs work.

GROVER~\cite{rong2020self} is proposed to train a Transformer-based backbone model by two pretext tasks, including node/edge-level contextual property prediction and graph-level motif prediction.
For the context prediction task, models are trained to predict a masked neighborhood based on the embedding of the central node.
And for the motif prediction task, models are trained to predict the motifs (calculated by functions in RDKit package) of a given molecular graph, which can provide supervised information about the substructures which construct the molecules.
Two Transformer models are exploited to encode the node and edge embedding, respectively.
A dynamic message passing network (dyMPNN) is also proposed to be served as the input of the Transformer backbones to embed the local neighborhood information into the initial embedding of each token.
The number of layers for message passing is variational during training, provides a dynamic receptive field for each node.
This design intuitively inspired by the idea that the interactions between different atoms in a molecule are not the same, which may require different size of receptive field.

ChemBERT~\cite{kim2021merged} is pre-trained to learn merged molecular representations using the BERT encoder.
Briefly, two pretext tasks are designed.
One is to predict randomly masked atoms, and the other one is to predict the QED values, which incorporate latent information of descriptors and several drug-like molecular properties.
SMILES are used as the input of the BERT encoder.
An additional matrix embedding layer is used to receive the adjacency matrix of the molecular graph and generate the embedding vectors associated with the atomic tokens, which is expected to compensate the lost connectivity of SMILES.
As a result, the input of each token includes token embedding, position embedding and matrix embedding. 

Xu \textit{et~al.}~\cite{xu2021self} proposed GraphLoG model to learn useful molecular representations for downstream tasks.
In their work, graph embedding in the latent space are refined from local and global structure aspects.
Specifically, for local structure perspective, the GraphLoG aims to make structural similar samples closer in the latent space, and enlarge the distance of negative pairs which are formed by random replacement of subgraph pairs.
Thus, GraphLoG is expected to construct a locally smooth latent space by aligning the embedding of correlation graphs/subgraphs.
And for the global structure perspective, graph embedding is modeled by hierarchical prototypes with maximum likelihood estimation by EM algorithm.

The innovation of MAT~\cite{maziarka2020molecule} is the integration of inter-atomic distance (distance matrix) and molecular structures (adjacency matrix) leveraged as biases to enhance the self-attention mechanism in Transformer.
Predicting features of masked atoms is selected as the pretext task.
MAT takes the feature of atoms as input, and randomly masked 15\% of them for each graph and a inear model is applied to predict features of masked atoms based on the embedding after N blocks of Transformer-encoder.
To avoid changing the internal representation in a given layer, MAT also adds a dummy node which is not connected to any other atoms.
In addition, the experiments reveal that the selected self-attention weights from the first layer of MAT concern about six chemical patterns, while weights of the subsequent layers capture more complex arrangements of atoms.

SMILESTransformer~\cite{honda2019smiles} pre-trains a vanilla Transformer with encoder-decoder architecture in an unsupervised reconstruction way by a huge corpus of SMILES. 
During pre-training, cross entropy between the input SMILES and the output probabilities of decoder is leveraged as loss function of pretext.
Transformation from the atom-level representations to molecule-level representations is implemented through concatenating four vectors from the last two layers of the Transformer.

PretrainGNNs~\cite{hu2019strategies} implements three pretext tasks to learn local and global molecular representations to improve out-of-distribution generalization ability on downstream target tasks.
The pretext tasks include node-level context prediction, attribute masking, and graph-level multi-task supervised property prediction.
For context prediction, a main GNN is used to generate embedding of a central node with k-hop neighborhood, and an auxiliary GNN is used to generate a context embedding for the context graph of this central node.
And the pretext task is to predict whether these two embeddings belong to the same node.
For attribute masking, the authors try to pre-train GNN to learn some underlying chemical rules and phenomena by masking atomic types. 
As for multi-task property prediction task at graph level, the PretrainGNNs uses a diverse set of down-stream molecular property labels to construct a supervised task for pre-training.
Although the multi-task property prediction task aims to alleviate the problem of negative transfer, the limited amount of supervised data may also lead to negative transfer on downstream predictions.
In addition, the context prediction and atomic type prediction are closely related, but pre-trained separately in PretrainGNNs, which may cause the loss of domain relevance~\cite{rong2020self}.

Chen \textit{et~al.}~\cite{chen2021extracting} constructed three datasets with increasing scale by sampling from ChEMBL, PubChem and ZINC.
Pairs of real SMILES and masked SMILES are produced by hiding a certain percentage of symbols with specific physio-chemical meaning. 
Bidirectional encoders of Transformer are selected as backbones, and three models are pre-trained by predicting the masked symbols on these three datasets respectively. 
When coping with downstream target tasks, the Wassertein Distance metric between the specific downstream dataset and the pre-training dataset is calculated.
And the corresponding optimal pre-training model is selected for fine-tune to alleviate the impact of negative transfer. 
Certainly, this will bring a large training cost overhead.
In our experiment, representations generated by PTM trained on ChEMBL(C) dataset is evaluated.

The detailed information of pre-training these PTMs are shown in Tab.~\ref{tab:PTMpretraindetail}.

\begin{table}[h]
\caption{Pre-training details of PTMs used in the experiments.}
\label{tab:PTMpretraindetail}
\centering
\begin{tabular}{lcr}
\toprule
Pre-trained Models & dataset & pre-training num \\
\midrule
GROVER & ZINC15 and ChEMBL & 11 million\\
ChemBERT & ZINC & 9 million\\
GraphLoG & ZINC15 & 2 million\\
MAT & ZINC15 & 2 million\\
SMILESTransformer & ChEMBL24 & 861K\\
PretrainGNNs & ZINC15 (self-supervised) & 2 million \\
& ChEMBL (supervised) & 456K\\
Pretrain8 & ChEMBL(C) & 2 million \\
& ChEMBL+PubChem(CP) & 100 million\\
& ChEMBL+PubChem+ZINC(CPZ) & 700 million\\
\bottomrule
\end{tabular}
\end{table}

Due to the limited spaces, we will use abbreviations in the tables later.
The relationship of abbreviations and pre-trained models are shown in Tab.~\ref{tab:pretrainedmodels}.
And the information of 10 target tasks are shown in Tab.~\ref{tab:target_task}.

\begin{table}[h]
\caption{Abbreviation of pre-trained models}
\label{tab:pretrainedmodels}
\centering
\begin{tabular}{lr}
\toprule
Pre-trained Models & abbreviation \\
\midrule
GROVER & GROVER\\
ChemBERT & ChemBERT\\
GraphLoG & GraLoG\\
MAT & MAT\\
SMILESTransformer & S.T.\\
PretrainGNNs & PreGNNs \\
Pretrain8 & Pre8\\
\bottomrule
\end{tabular}
\end{table}

\begin{table}[h]
\caption{Information of Target tasks}
\label{tab:target_task}
\centering
\begin{tabular}{lcr}
\toprule
Data set & tasks & \# sample \\
\midrule
BACE & 1 & 1513 \\
BBBP & 1 & 2039 \\
ClinTox & 2 & 1478 \\
ESOL    & 1 & 1128 \\
FreeSolv & 1 & 642 \\
Lipo      & 1 & 4200\\
QM7   & 1 & 6834 \\
QM8 & 12 & 21786 \\
SIDER & 27 & 1427 \\
Tox21 & 12 & 7831 \\
\bottomrule
\end{tabular}
\end{table}

\clearpage

\section{Prediction Performance of Pre-trained Models Reported in the Literature}
Tab.~\ref{tab:reported_c} and Tab.~\ref{tab:reported_r} shows the prediction performance of 7 pre-trained models on target task after finetuning reported in the literature.
These results are cited from the articles of these models.
For classification tasks, ROC-AUC is used as metric, of which higher score is better.
And for regression tasks, RMSE is used, of which lower score is better.
Missing values indicates that the model is not tested on the target task in the experiments of the article.
Result with the best score of each dataset is bold.

\begin{table}[h]
\caption{Prediction performance of pre-trained models on classification tasks. Using ROC-AUC as performance metric.}
\label{tab:reported_c}
\centering
\begin{tabular}{cccccc}
\toprule
PTMs & BACE & BBBP & ClinTox & SIDER & Tox21\\
\midrule
ChemBERT & 0.820$\pm$0.017 & 0.724$\pm$0.009 & \textbf{0.990$\pm$0.003} & 0.631$\pm$0.006 & 0.774$\pm$0.005 \\
GROVER & \textbf{0.894$\pm$0.028} & 0.940$\pm$0.019 & 0.944$\pm$0.021 & \textbf{0.658$\pm$0.023} & \textbf{0.831$\pm$0.025} \\
GraLoG & 0.835$\pm$0.012 & 0.725$\pm$0.008 & 0.767$\pm$0.033 & 0.612$\pm$0.011 & 0.757$\pm$0.005\\
MAT & - & 0.737$\pm$0.009 & - & - & -\\
PreGNNs & 0.803$\pm$0.009 & 0.665$\pm$0.025 & 0.737$\pm$0.028 & 0.639$\pm$0.009 & 0.779$\pm$0.004 \\
Pre8 & 0.872$\pm$0.036 & \textbf{0.949$\pm$0.016} & 0.963$\pm$0.044 & - & - \\
S.T. & 0.701 & 0.704 & 0.954 & - & 0.802 \\
\bottomrule
\end{tabular}
\end{table}

\begin{table}[h]
\caption{Prediction performance of pre-trained models on regression tasks. Using RMSE as performance metric.}
\label{tab:reported_r}
\centering
\begin{tabular}{lcccccr}
\toprule
pre-trainModels & ESOL & FreeSolv & Lipo & QM7 & QM8 \\
\midrule
ChemBERT & 0.451$\pm$0.033 & 0.755$\pm$0.078 & - & - & - \\
GROVER & 0.831$\pm$0.120 & 1.544$\pm$0.397 & \textbf{0.560$\pm$0.035} & \textbf{72.6$\pm$3.8} & \textbf{0.0125$\pm$0.002} \\
GraLoG & - & - & - & - & - \\
MAT & \textbf{0.278$\pm$0.020} & \textbf{0.265$\pm$0.042} & - & - & -\\
PreGNNs & - & - & - & - & - \\
Pre8 & 0.925$\pm$0.001 & 0.912$\pm$0.069 & 0.755$\pm$0.022 & - & - \\
S.T. & 0.72 & 1.65 & 0.921 & - & - \\
\bottomrule
\end{tabular}
\end{table}

\clearpage

\section{Experiments for Finetuning Pre-trained Models on Target Tasks}
\label{sec:finetune}
As discussed in Sec.~\ref{sec:intro} and Sec.~\ref{sec:relatedworks}, we have claimed that the prediction accuracy after finetuning the down-stream task layers on a specific target task cannot be leveraged to evaluate the quality of representations.
And the $\tau_\omega$ that assess the consistency of the orders of PTMs ranked by accuracy and by the proposed scores is also lack of fairness.
Here, finetuning experiments are conducted to support our claim.

In this experiment, we use MLPs with 1, 2 and 3 layers as down-stream task layers to be finetuned.
As shown in Tab.~\ref{tab:hyperparameters}, the size of hidden layers of these MLP structures are $None$, $[128]$ and $[128,64]$, respectively.
For each choice of MLP structure, four random seeds are tried for data splitting.
Here we use random splitting method, since the out-of-distribution (OOD) phenomenon reflected by scaffold splitting is not considered in this work.
We will remain this for future works.
Other training parameters remain unchanged.
Models are trained 20 times for each group of hyper-parameters, and the average performance are reported.

The experiment is conducted on one Titan RTX GPU with CPU E5-2678 and 128 GB memory.
All of the finetuning results are shown in Tab.~\ref{tab:finetune_c} and Tab.~\ref{tab:finetune_r}.
Results with the best score of each dataset is bold.

As shown in the tables, it can be figured out that the prediction accuracy after finetuning the task layers is very sensitive to many factors of the hyper-parameters.
The prediction accuracy of one PTM on one target task may change dramatically by changing the number of layers of MLP, and even by changing the random seeds for data splitting.
Obviously, a sensitive and variable prediction accuracy is not applicable for evaluating the quality of fixed representations, since the evaluation should also be constant.
In addition, it is obvious that for most of the cases, using a MLP with more layers can lead to better prediction accuracy.
It reveals that the prediction accuracy is biased by the capability of task layers, and it cannot give an accurate evaluation for the quality of representations.

Furthermore, we have computed the LogME of the PTMs on these target tasks and the order of the PTMs ranked by LogME scores is compared with the order ranked by finetuning results in Tab.~\ref{tab:finetune_c} and Tab.~\ref{tab:finetune_r}.
$\tau_{\omega}$ is calculated to assess the consistency between these two orders.
Tab.~\ref{tab:LogMEValue} shows the LogME scores, and Tab.~\ref{tab:LogMEtauomega} shows the  $\tau_{\omega}$ values with different finetuning results as ground-truth.
From the tables, it can be figured out that due to the sensitivity of finetuning results, although the scores of LogME is constant, there will be a great change in $\tau_{\omega}$ when using different ground-truth.
For instance, on QM8, the standard deviation of $\tau_{\omega}$ achieves 0.565 (note that the range of $\tau_{\omega}$ is $[-1,1]$).
The results reveal that $\tau_\omega$ cannot accurately assess the scores for representation quality evaluation, since the ground-truth, i.e. the order of accuracy after finetuning, is variable.
In addition, it would be a trick for the authors to choose finetuning results corresponding to the largest $\tau_\omega$ as ground-truth to overestimate the precision of the proposed evaluation scores in their article.
So, the estimation of $\tau_{\omega}$ is lack of fairness and we will not use the $\tau_{\omega}$ as a leaderboard to compare our scores with others.

\begin{table}[htpb]
\caption{The hyper-parameters for finetuning experiments.}
    \label{tab:hyperparameters}
    \centering
    \begin{tabular}{cc}
    \toprule
         Parameters & Values \\
    \midrule
         BatchSize & 200  \\
         DropRate & 0.2 \\
         lr & 1e-2.5 \\
         WeightDecay & 5 \\
         MLPLayers & [] \\ 
         & [128] \\
         & [128,64] \\
    \bottomrule
    \end{tabular}
    \label{tab:hyperparameters}
\end{table}

\begin{table}[htbp]
\caption{Finetuning results on classification tasks. Using ROC-AUC as performance metric.}
\label{tab:finetune_c}
\centering
\resizebox{\linewidth}{!}{
\begin{tabular}{lccccccccccr}
\toprule
Tasks & Layers & GROVER & ChemBERT & GraLoG & MAT & S.T. & PreGNNs & Pre8 & ECFP\\
\midrule
BACE & one & 0.882 & 0.854 & 0.819 & 0.780 & 0.761 & 0.754 & 0.877 & \textbf{0.901}\\
& & \textbf{0.867} & 0.836 & 0.703 & 0.807 & 0.808 & 0.697 & 0.864 & 0.862\\
& & 0.823 & 0.809 & 0.767 & 0.755 & 0.791 & 0.661 & 0.852 & \textbf{0.860}\\
& & 0.818 & 0.812 & 0.721 & 0.744 & 0.789 & 0.698 & 0.864 & \textbf{0.880}\\
&two & 0.889 & 0.858 & \textbf{0.914} & 0.726 & 0.804 & 0.845 & 0.872 & 0.880\\
& & \textbf{0.887} & 0.847 & 0.857 & 0.731 & 0.832 & 0.802 & 0.882 & 0.854\\
& & 0.871 & 0.814 & 0.847 & 0.702 & 0.837 & 0.815 & \textbf{0.872} & 0.834\\
& & 0.839 & 0.826 & 0.857 & 0.667 & 0.808 & 0.779 & 0.869 & \textbf{0.900}\\
&three & \textbf{0.904} & 0.858 & 0.904 & 0.749 & 0.807 & 0.837 & 0.876 & 0.886\\
& & \textbf{0.889} & 0.848 & 0.844 & 0.784 & 0.839 & 0.792 & 0.868 & 0.859\\
& & \textbf{0.875} & 0.811 & 0.837 & 0.725 & 0.829 & 0.787 & 0.867 & 0.839\\
& & 0.837 & 0.824 & 0.840 & 0.729 & 0.825 & 0.776 & 0.866 & \textbf{0.888}\\
\midrule
BBBP & one & 0.929 & 0.954 & 0.760 & 0.836 & \textbf{0.965} & 0.832 & 0.959 & 0.905 \\
& & 0.883 & 0.949 & 0.761 & 0.822 & \textbf{0.957} & 0.822 & 0.940 & 0.841 \\
& & 0.941 & 0.971 & 0.855 & 0.906 & \textbf{0.980} & 0.884 & 0.969 & 0.923 \\
& & 0.915 & 0.944 & 0.739 & 0.850 & 0.953 & 0.798 & \textbf{0.963} & 0.884 \\
 & two & 0.932 & 0.959 & 0.852 & 0.849 & \textbf{0.971} & 0.887 & 0.954 & 0.906 \\
& & 0.892 & 0.953 & 0.777 & 0.828 & \textbf{0.963} & 0.846 & 0.950 & 0.813 \\
& & 0.947 & 0.973 & 0.877 & 0.901 & \textbf{0.981} & 0.924 & 0.969 & 0.908 \\
& & 0.917 & 0.954 & 0.824 & 0.855 & 0.948 & 0.871 & \textbf{0.957} & 0.879 \\
 & three & 0.934 & 0.960 & 0.842 & 0.853 & \textbf{0.973} & 0.870 & 0.953 & 0.904 \\
& & 0.889 & 0.955 & 0.780 & 0.839 & \textbf{0.964} & 0.854 & 0.942 & 0.822 \\
& & 0.963 & 0.974 & 0.882 & 0.899 & \textbf{0.980} & 0.924 & 0.963 & 0.915 \\
& & 0.909 & 0.953 & 0.825 & 0.865 & 0.956 & 0.856 & \textbf{0.962} & 0.883 \\
\midrule
ClinTox & one & 0.846 & 0.951 & 0.703 & 0.734 & 0.936 & 0.617 & \textbf{0.983} & 0.747 \\
& & 0.892 & 0.991 & 0.618 & 0.870 & \textbf{0.992} & 0.684 & 0.992 & 0.837 \\
& & 0.861 & 0.982 & 0.704 & 0.924 & 0.985 & 0.773 & \textbf{0.993} & 0.734 \\
& & 0.846 & 0.966 & 0.751 & 0.839 & 0.954 & 0.758 & \textbf{0.997} & 0.780 \\
 & two & 0.796 & 0.928 & 0.722 & 0.743 & 0.925 & 0.718 & \textbf{0.967} & 0.697 \\
& & 0.873 & 0.989 & 0.700 & 0.877 & \textbf{0.995} & 0.744 & 0.975 & 0.831 \\
& & 0.862 & 0.977 & 0.693 & 0.916 & 0.990 & 0.803 & \textbf{0.995} & 0.715 \\
& & 0.911 & 0.947 & 0.824 & 0.834 & 0.960 & 0.816 & \textbf{0.993} & 0.719 \\
 & three & 0.825 & 0.938 & 0.728 & 0.758 & 0.931 & 0.693 & \textbf{0.989} & 0.696 \\
& & 0.884 & 0.987 & 0.701 & 0.867 & 0.990 & 0.696 & \textbf{0.996} & 0.811 \\
& & 0.876 & 0.983 & 0.707 & 0.921 & 0.985 & 0.790 & \textbf{0.994} & 0.698 \\
& & 0.925 & 0.951 & 0.832 & 0.832 & 0.949 & 0.800 & \textbf{0.999} & 0.698 \\
\midrule
SIDER & one & 0.537 & 0.574& 0.569 & 0.530 & 0.565 & 0.568 & \textbf{0.591} & 0.573 \\
& & 0.642 & 0.642 & 0.635 & 0.617 & 0.644 & 0.633 & \textbf{0.654} & 0.606\\
& & 0.588 & 0.617 & 0.590 & 0.606 & 0.614 & 0.611 & \textbf{0.649} & 0.560\\
& & 0.618 & 0.637 & 0.621 & 0.604 & 0.625 & 0.632 & \textbf{0.650} & 0.577\\
&two & 0.559 & 0.539 & \textbf{0.607} & 0.512 & 0.569 & 0.567 & 0.601 & 0.568\\
& & 0.622 & 0.609 & \textbf{0.660} & 0.592 & 0.610 & 0.638 & 0.653 & 0.574\\
& & 0.595 & 0.598 & 0.624 & 0.560 & 0.590 & 0.615 & \textbf{0.636} & 0.547\\
& & 0.614 & 0.608 & 0.656 & 0.575 & 0.611 & \textbf{0.668} & 0.655 & 0.552\\
&three & \textbf{0.583} & 0.547 & 0.567 & 0.518 & 0.568 & 0.543 & 0.562 & 0.571\\
& & \textbf{0.662} & 0.641 & 0.649 & 0.607 & 0.651 & 0.637 & 0.632 & 0.592\\
& & 0.630 & 0.604 & 0.620 & 0.591 & 0.626 & 0.631 & \textbf{0.658} & 0.562\\
& & \textbf{0.648} & 0.625 & 0.625 & 0.591 & 0.634 & 0.647 & 0.648 & 0.593\\
\midrule
Tox21 & one & 0.766 & 0.779 & 0.713 & 0.792 & 0.787 & 0.740 & \textbf{0.819} & 0.696 \\
& & 0.771 & 0.778 & 0.703 & 0.777 & 0.773 & 0.743 & \textbf{0.811} & 0.676 \\
& & 0.808 & 0.805 & 0.756 & 0.814 & 0.801 & 0.753 & \textbf{0.839} & 0.712 \\
& & 0.764 & 0.783 & 0.712 & 0.797 & 0.785 & 0.746 & \textbf{0.841} & 0.740 \\
 & two & 0.772 & 0.773 & 0.700 & 0.768 & 0.776 & 0.773 & \textbf{0.822} & 0.741 \\
& & 0.748 & 0.754 & 0.702 & 0.748 & 0.766 & 0.757 & \textbf{0.825} & 0.707 \\
& & 0.785 & 0.777 & 0.753 & 0.790 & 0.789 & 0.774 & \textbf{0.848} & 0.741 \\
& & 0.740 & 0.769 & 0.707 & 0.763 & 0.766 & 0.771 & \textbf{0.847} & 0.780 \\
 & three & 0.832 & 0.800 & 0.750 & 0.796 & 0.806 & 0.788 & \textbf{0.842} & 0.740 \\
& & 0.816 & 0.786 & 0.732 & 0.788 & 0.800 & 0.760 & \textbf{0.838} & 0.711 \\
& & 0.840 & 0.817 & 0.786 & 0.817 & 0.827 & 0.782 & \textbf{0.858} & 0.750 \\
& & 0.831 & 0.797 & 0.749 & 0.806 & 0.809 & 0.796 & \textbf{0.856} & 0.779 \\
\bottomrule
\end{tabular}
}
\end{table}

\begin{table}[htbp]
\caption{Finetuning results on regression tasks. Using RMSE as performance metric.}
\label{tab:finetune_r}
\centering
\resizebox{\linewidth}{!}{
\begin{tabular}{lccccccccccr}
\toprule
Tasks & Layers & GROVER & ChemBert & GraLoG & MAT & S.T. & PreGNNs & Pre8 & ECFP\\
\midrule
ESOL & one & 0.933 & 1.005 & 1.688 & 1.110 & 0.936 & 1.512 & \textbf{0.923} & 1.475 \\
& & 0.917 & 1.017 & 1.822 & 0.982 & 0.937 & 1.396 & \textbf{0.897} & 1.460 \\
& & \textbf{0.763} & 0.818 & 1.867 & 0.907 & 0.837 & 1.583 & 0.950 & 1.681 \\
& & 0.779 & 0.848 & 1.711 & 0.981 & \textbf{0.752} & 1.418 & 0.859 & 1.300 \\
 & two & 0.757 & 0.909 & 1.455 & 0.899 & 0.823 & 1.345 & \textbf{0.699} & 1.442 \\
& & 0.874 & 0.929 & 1.667 & 0.902 & 0.866 & 1.178 & \textbf{0.775} & 1.304 \\
& & \textbf{0.708} & 0.753 & 1.409 & 0.745 & 0.774 & 1.112 & 0.755 & 1.493 \\
& & 0.703 & 0.793 & 1.515 & 0.877 & 0.725 & 1.120 & \textbf{0.680} & 1.366 \\
 & three & \textbf{0.656} & 0.952 & 1.503 & 1.035 & 0.915 & 1.378 & 0.830 & 1.352 \\
& & \textbf{0.796} & 0.987 & 1.721 & 0.965 & 0.919 & 1.309 & 0.832 & 1.256 \\
& & \textbf{0.600} & 0.798 & 1.527 & 0.881 & 0.834 & 1.367 & 0.856 & 1.466 \\
& & \textbf{0.677} & 0.829 & 1.505 & 0.987 & 0.722 & 1.229 & 0.775 & 1.158 \\
\midrule
FreeSolv & one & 0.439 & 0.594 & 0.998 & 0.428 & \textbf{0.410} & 0.747 & 0.417 & 0.680\\
 & & 0.358 & 0.468 & 0.713 & 0.380 & 0.346 & 0.622 & \textbf{0.340} & 0.454\\
 & & \textbf{0.406} & 0.648 & 0.950 & 0.451 & 0.446 & 0.726 & 0.426 & 0.533\\
 & & 0.386 & 0.572 & 0.913 & 0.484 & \textbf{0.318} & 0.722 & 0.335 & 0.569\\
& two & 0.755 & 0.536 & 0.828 & 0.403 & 0.375 & 0.526 & \textbf{0.315} & 0.594\\
& & 0.283 & 0.402 & 0.765 & 0.326 & 0.312 & 0.699 & \textbf{0.230} & 0.419\\
& & \textbf{0.404} & 0.601 & 0.828 & 0.460 & 0.450 & 0.635 & 0.409 & 0.485\\
& & 0.410 & 0.514 & 0.931 & 0.444 & 0.339 & 0.621 & \textbf{0.311} & 0.517\\
& three & \textbf{0.292} & 0.526 & 0.953 & 0.365 & 0.373 & 0.647 & 0.396 & 0.607\\
& & \textbf{0.207} & 0.406 & 0.716 & 0.306 & 0.321 & 0.651 & 0.306 & 0.433\\
& & \textbf{0.317} & 0.595 & 0.873 & 0.431 & 0.444 & 0.690 & 0.425 & 0.502\\
& & \textbf{0.304} & 0.507 & 0.906 & 0.438 & 0.312 & 0.660 & 0.312 & 0.552\\
\midrule
Lipo & one & \textbf{0.821} & 1.056 & 1.151 & 1.059 & 1.040 & 1.114 & 0.901 & 0.997 \\
& & \textbf{0.844} & 1.020 & 1.153 & 1.055 & 1.015 & 1.073 & 0.909 & 0.995 \\
& & \textbf{0.796} & 0.967 & 1.111 & 0.969 & 0.952 & 1.042 & 0.809 & 0.895 \\
& & \textbf{0.758} & 1.046 & 1.133 & 1.042 & 1.008 & 1.110 & 0.831 & 0.945 \\
 & two & 1.050 & 1.064 & 1.09 & 1.054 & 1.017 & 1.033 & \textbf{0.843} & 0.983 \\
& & 0.993 & 1.020 & 1.120 & 1.056 & 1.014 & 0.993 & \textbf{0.852} & 0.980 \\
& & 0.953 & 0.967 & 1.061 & 0.964 & 0.940 & 0.956 & \textbf{0.749} & 0.894 \\
& & 0.957 & 1.040 & 1.068 & 1.039 & 0.985 & 1.049 & \textbf{0.773} & 0.946 \\
 & three & \textbf{0.721} & 1.059 & 1.069 & 1.050 & 0.991 & 1.009 & 0.828 & 0.921 \\
& & \textbf{0.700} & 1.004 & 1.096 & 1.039 & 0.985 & 1.004 & 0.855 & 0.884 \\
& & \textbf{0.648} & 0.956 & 1.031 & 0.971 & 0.927 & 0.942 & 0.763 & 0.813 \\
& & \textbf{0.641} & 1.025 & 1.035 & 1.041 & 0.959 & 1.023 & 0.760 & 0.857 \\
\midrule
QM7 & one & 128.983 & \textbf{124.305} & 502.077 & 140.745 & 146.106 & 531.277 & 201.606 & 1201.089 \\
& & 112.260 & \textbf{109.689} & 497.383 & 128.273 & 133.171 & 553.648 & 187.535 & 1200.918 \\
& & 139.541 & \textbf{133.513} & 503.053 & 142.620 & 155.683 & 553.069 & 200.199 & 1216.250 \\
& & 122.591 & \textbf{121.477} & 501.804 & 133.781 & 148.696 & 574.396 & 201.104 & 1203.450 \\
 & two & 115.997 & 115.441 & 193.193 & \textbf{110.864} & 113.664 & 156.067 & 119.295 & 166.282 \\
& & \textbf{99.470} & 100.614 & 185.654 & 104.233 & 100.947 & 144.781 & 106.794 & 161.940 \\
& & 127.228 & 122.205 & 197.780 & \textbf{122.026} & 123.805 & 162.147 & 128.927 & 183.488 \\
& & 114.682 & 110.494 & 194.981 & \textbf{105.362} & 108.982 & 160.320 & 112.886 & 163.041 \\
 & three & 118.661 & \textbf{115.777} & 193.212 & 116.589 & 120.917 & 162.565 & 124.597 & 179.525 \\
& & \textbf{100.169} & 105.245 & 181.504 & 109.578 & 117.891 & 158.126 & 110.397 & 181.538 \\
& & 131.099 & \textbf{125.116} & 196.39 & 127.591 & 131.451 & 167.397 & 132.692 & 200.146 \\
& & 115.216 & 113.638 & 191.700 & \textbf{112.369} & 122.304 & 170.047 & 119.739 & 179.545 \\
\midrule
QM8 & one & 0.125 & 0.073 & 0.043 & 0.065 & 0.063 & 0.040 & \textbf{0.035} & 0.040\\
& & 0.134 & 0.067 & 0.043 & 0.064 & 0.060 & 0.040 & \textbf{0.036} & 0.041\\
& & 0.150 & 0.074 & 0.042 & 0.069 & 0.062 & 0.039 & \textbf{0.035} & 0.039\\
& & 0.144 & 0.069 & 0.044 & 0.062 & 0.061 & 0.040 & \textbf{0.036} & 0.040\\
&two & 0.048 & 0.047 & 0.039 & 0.047 & 0.043 & 0.036 & 0.034 & \textbf{0.030}\\
& & 0.049 & 0.047 & 0.039 & 0.047 & 0.043 & 0.036 & 0.034 & \textbf{0.031}\\
& & 0.047 & 0.046 & 0.038 & 0.043 & 0.042 & 0.034 & 0.033 & \textbf{0.029}\\
& & 0.049 & 0.048 & 0.039 & 0.047 & 0.046 & 0.036 & 0.035 & \textbf{0.030}\\
&three & \textbf{0.024} & 0.035 & 0.035 & 0.032 & 0.035 & 0.033 & 0.029 & 0.029\\
& & \textbf{0.024} & 0.035 & 0.035 & 0.032 & 0.035 & 0.033 & 0.029 & 0.030\\
& & \textbf{0.023} & 0.034 & 0.034 & 0.031 & 0.033 & 0.031 & 0.028 & 0.028\\
& & \textbf{0.024} & 0.035 & 0.035 & 0.032 & 0.035 & 0.032 & 0.029 & 0.029\\
\bottomrule
\end{tabular}
}
\end{table}

\clearpage

\begin{table}[htbp]
\caption{LogME value of pretrained models on different target tasks.}
\label{tab:LogMEValue}
\centering
\begin{tabular}{lcccccccr}
\toprule
Tasks & GROVER & ChemBERT & GraLoG & MAT & S.T. & PreGNNs & Pre8 & ECFP\\
\midrule
BACE & \textbf{5.415} & -0.537 & -0.622 & -0.596 & -0.599 & -0.617 & -0.503 & -0.481 \\
BBBP & \textbf{4.645} & -0.19 & -0.467 & -0.369 & -0.197 & -0.405 & -0.156 & -0.36 \\
ClinTox & \textbf{9.928} & 0.436 & -0.057 & -0.021 & 0.474 & -0.071 & 0.383 & -0.186 \\
SIDER & \textbf{11.165} & -0.469 & -0.475 & -0.471 & -0.469 & -0.482 & -0.463 & -0.566 \\
Tox21 & -0.175 & -0.17 & -0.197 & -0.178 & -0.168 & -0.19 & \textbf{-0.16} & -0.257 \\
ESOL & \textbf{5.565} & -1.229 & -1.924 & -1.159 & -1.214 & -1.723 & -1.215 & -1.785 \\
FreeSolv & \textbf{17.989} & -0.022 & -1.301 & 0.374 & 0.616 & -1.005 & -0.433 & 0.014 \\
Lipo & \textbf{-1.021} & -1.392 & -1.509 & -1.304 & -1.421 & -1.435 & -1.297 & -1.357 \\
QM7 & -6.219 & -6.096 & -6.594 & \textbf{-6.094} & -6.097 & -6.524 & -6.126 & -6.932 \\
QM8 & \textbf{2.459} & 2.121 & 1.928 & 2.272 & 2.014 & 1.965 & 2.158 & 1.671 \\
\bottomrule
\end{tabular}
\end{table}

\begin{table}[htbp]
\caption{$\tau_\omega$ scores of LogME with different finetuning results as ground-truth.}
\label{tab:LogMEtauomega}
\centering
\resizebox{\linewidth}{!}{
\begin{tabular}{ccccccccccc}
\toprule
Layers & BACE & BBBP & ClinTox & SIDER & Tox21 & ESOL & FreeSolv & Lipo & QM7 & QM8 \\
\midrule
one & 0.741 & 0.431 & 0.242 & 0.019 & 0.792 & 0.629 & 0.552 & 0.787 & 0.676 & -0.618 \\
& 0.846 & 0.31 & 0.484 & 0.578 & 0.798 & 0.755 & 0.651 & 0.787 & 0.676 & -0.618 \\
& 0.637 & 0.343 & 0.291 & 0.273 & 0.563 & 0.794 & 0.818 & 0.787 & 0.676 & -0.618 \\
& 0.67 & 0.575 & 0.252 & 0.25 & 0.763 & 0.628 & 0.697 & 0.85 & 0.676 & -0.648 \\
\midrule
two & 0.26 & 0.282 & 0.343 & -0.053 & 0.819 & 0.755 & 0.059 & 0.378 & 0.8 & -0.774 \\
& 0.617 & 0.236 & 0.479 & 0.155 & 0.819 & 0.67 & 0.692 & 0.424 & 0.503 & -0.682 \\
& 0.301 & 0.271 & 0.291 & 0.209 & 0.737 & 0.76 & 0.818 & 0.464 & 0.86 & -0.682 \\
& 0.42 & 0.617 & 0.431 & -0.035 & 0.335 & 0.689 & 0.596 & 0.675 & 0.847 & -0.682 \\
\midrule
three & 0.556 & 0.31 & 0.315 & 0.41 & 0.847 & 0.707 & 0.793 & 0.72 & 0.721 & 0.564 \\
& 0.745 & 0.236 & 0.37 & 0.613 & 0.798 & 0.772 & 0.7 & 0.692 & 0.633 & 0.564 \\
& 0.651 & 0.271 & 0.319 & 0.444 & 0.814 & 0.728 & 0.754 & 0.639 & 0.814 & 0.497 \\
& 0.377 & 0.575 & 0.304 & 0.745 & 0.769 & 0.754 & 0.763 & 0.431 & 0.76 & 0.497 \\
\midrule
Std & $\pm$ 0.18 & $\pm$ 0.135 & $\pm$ 0.078 & $\pm$ 0.251 & $\pm$ 0.14 & $\pm$ 0.053 & $\pm$ 0.197 & $\pm$ 0.161 & $\pm$ 0.098 & $\pm$ 0.565 \\
\bottomrule
\end{tabular}
}
\end{table}

\clearpage

\section{RPSMaps of Pretrined Models on Target Tasks}
In our experiments, we used RPSMap to analyze 7 pre-trained models and ECFP on 10 target tasks.
We have drawn 80 RPSMaps, which are too many to be presented completely in this paper.
Here we only use 4 datasets as examples, 2 classification tasks and 2 regression tasks.
Other RPSMaps are provided in supplemental materials.

\begin{figure}[thbp]
	\centering
	\subfigure[ECFP]{
	\includegraphics[width=3.2cm]{RPSMaps/MorganFP_ESOL_MinMaxEud_Linear_FPSMap.png}
	\label{subfig:ECFPESOLMinMaxEudRPSMap}
}
    \subfigure[GROVER]{
    \includegraphics[width=3.2cm]{RPSMaps/GROVER_ESOL_MinMaxEud_Linear_FPSMap.png}
}
    \subfigure[ChemBert]{
	\includegraphics[width=3.2cm]{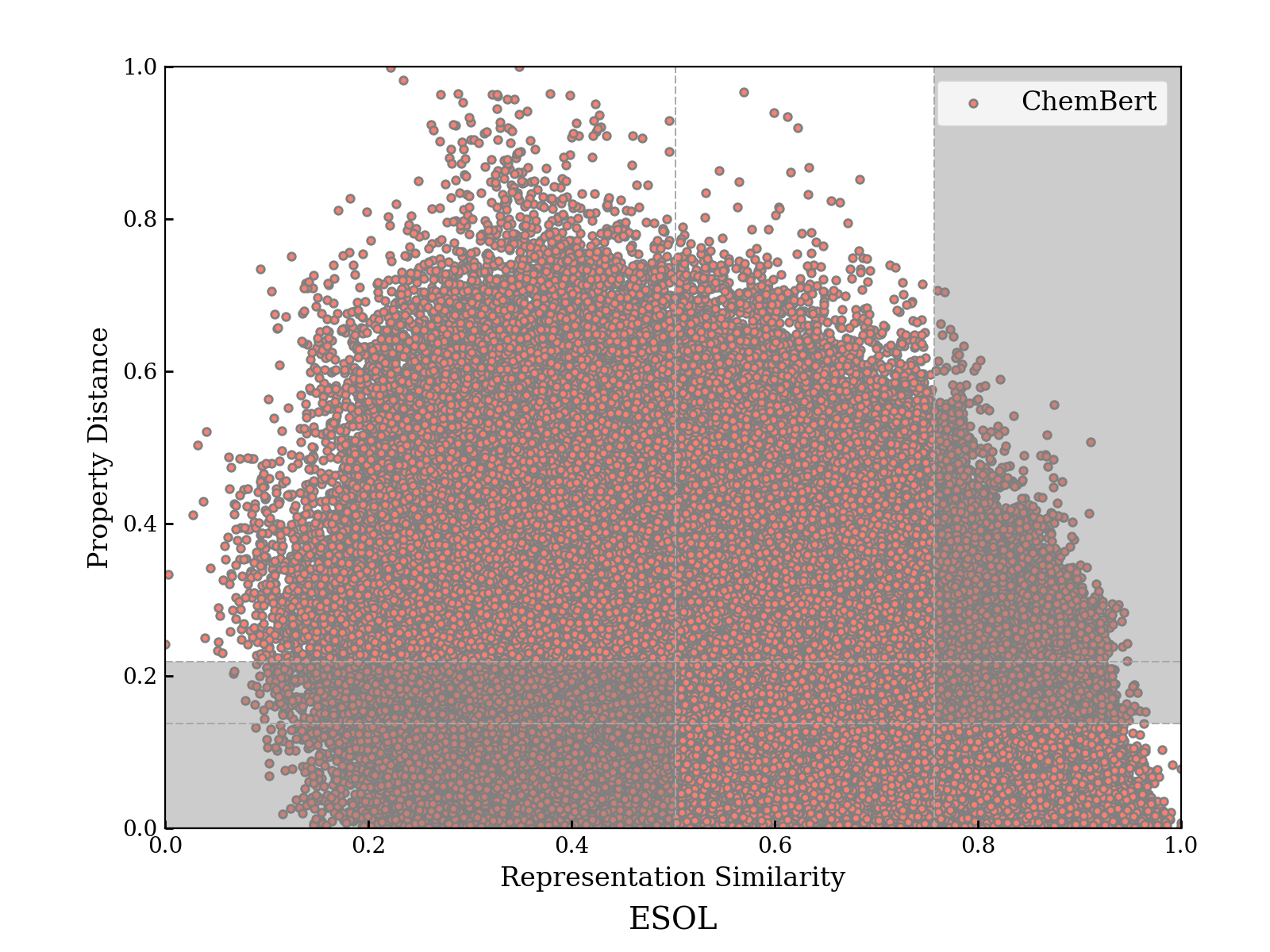}
}
    \subfigure[GraphLoG]{
	\includegraphics[width=3.2cm]{RPSMaps/GraphLoG_ESOL_MinMaxEud_Linear_FPSMap.png}
}
	\subfigure[MAT]{
	\includegraphics[width=3.2cm]{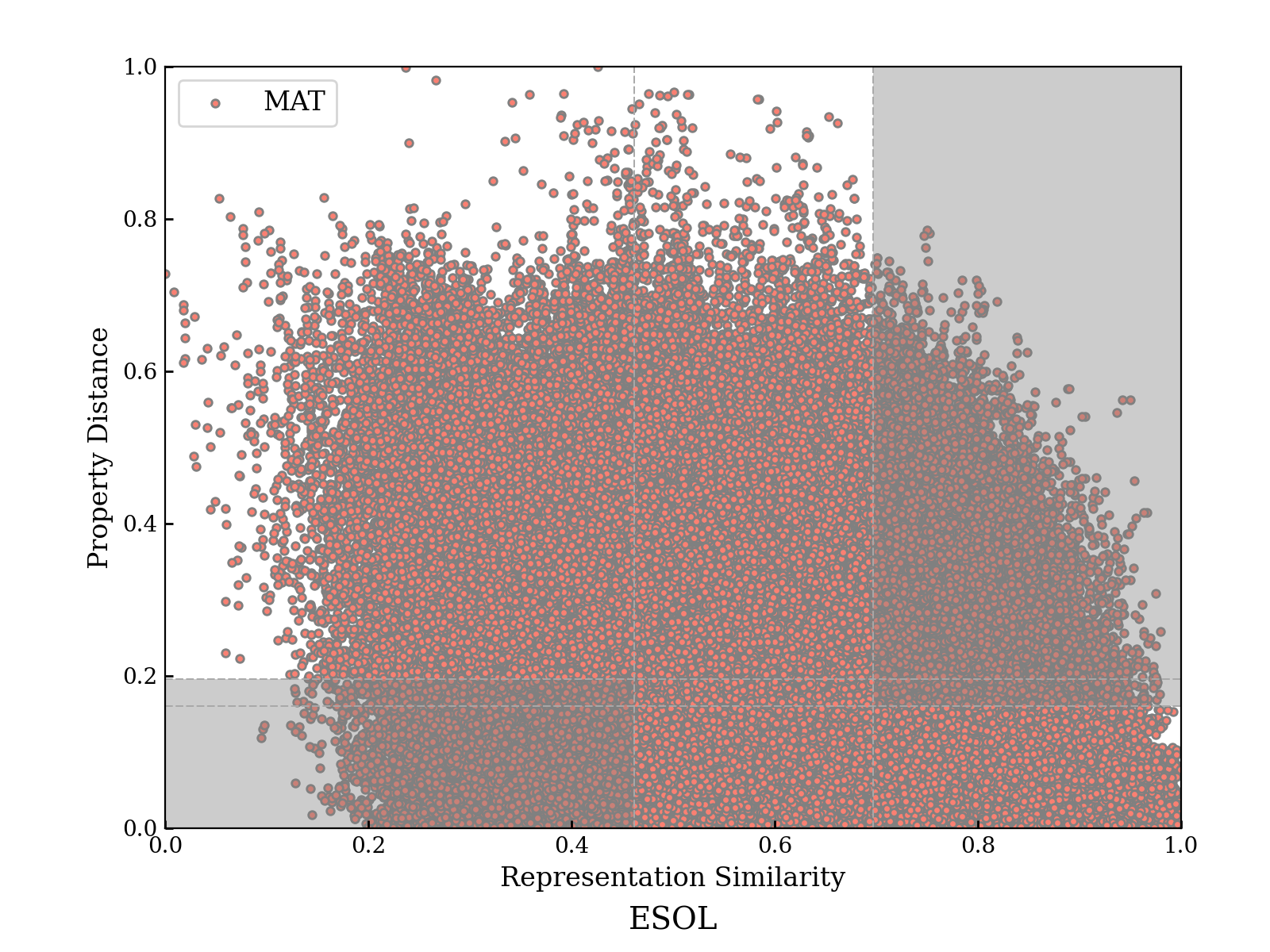}
}
    \subfigure[SMILESTransformer]{
    \includegraphics[width=3.2cm]{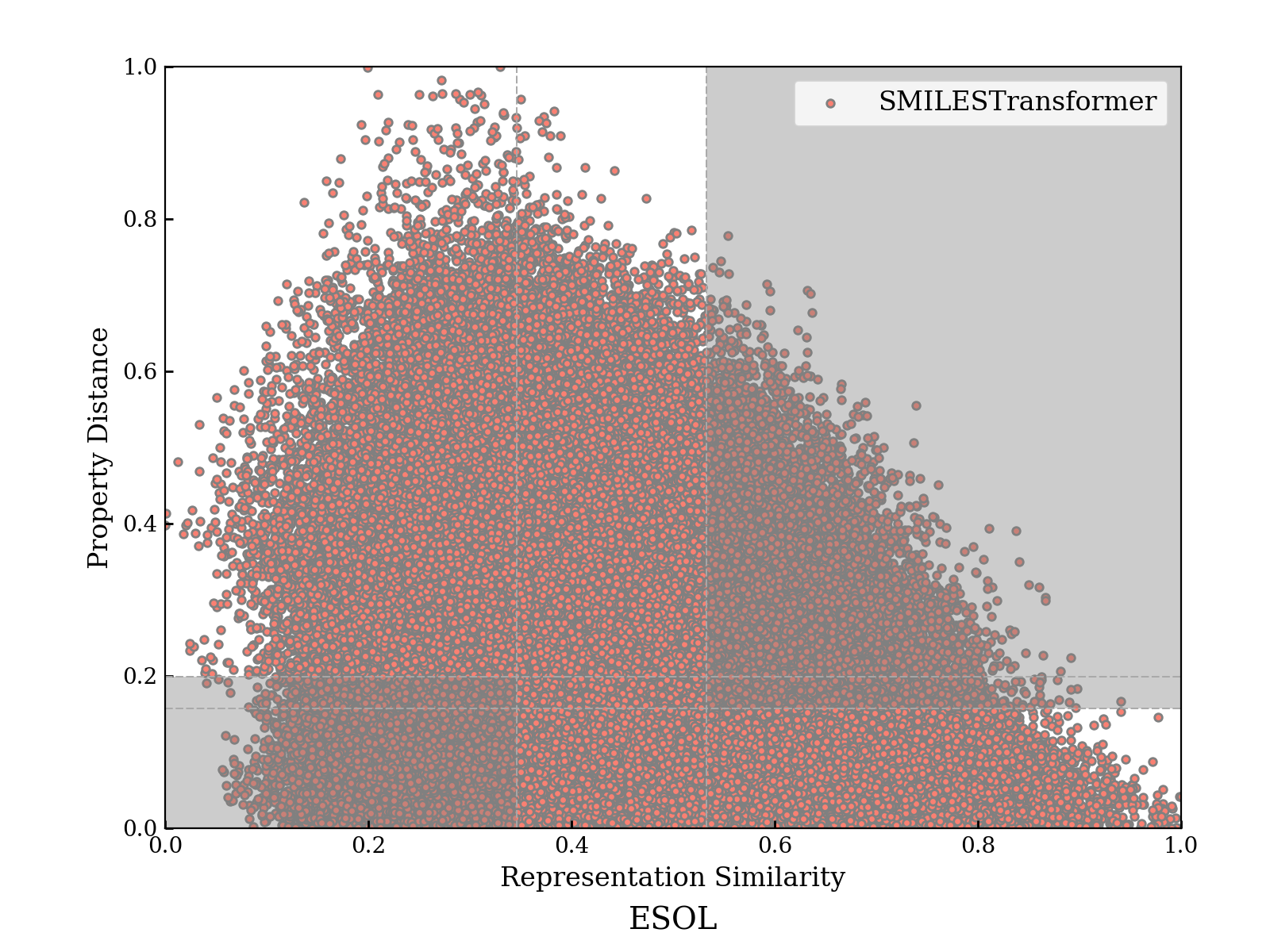}
}
    \subfigure[PretrainGNNs]{
	\includegraphics[width=3.2cm]{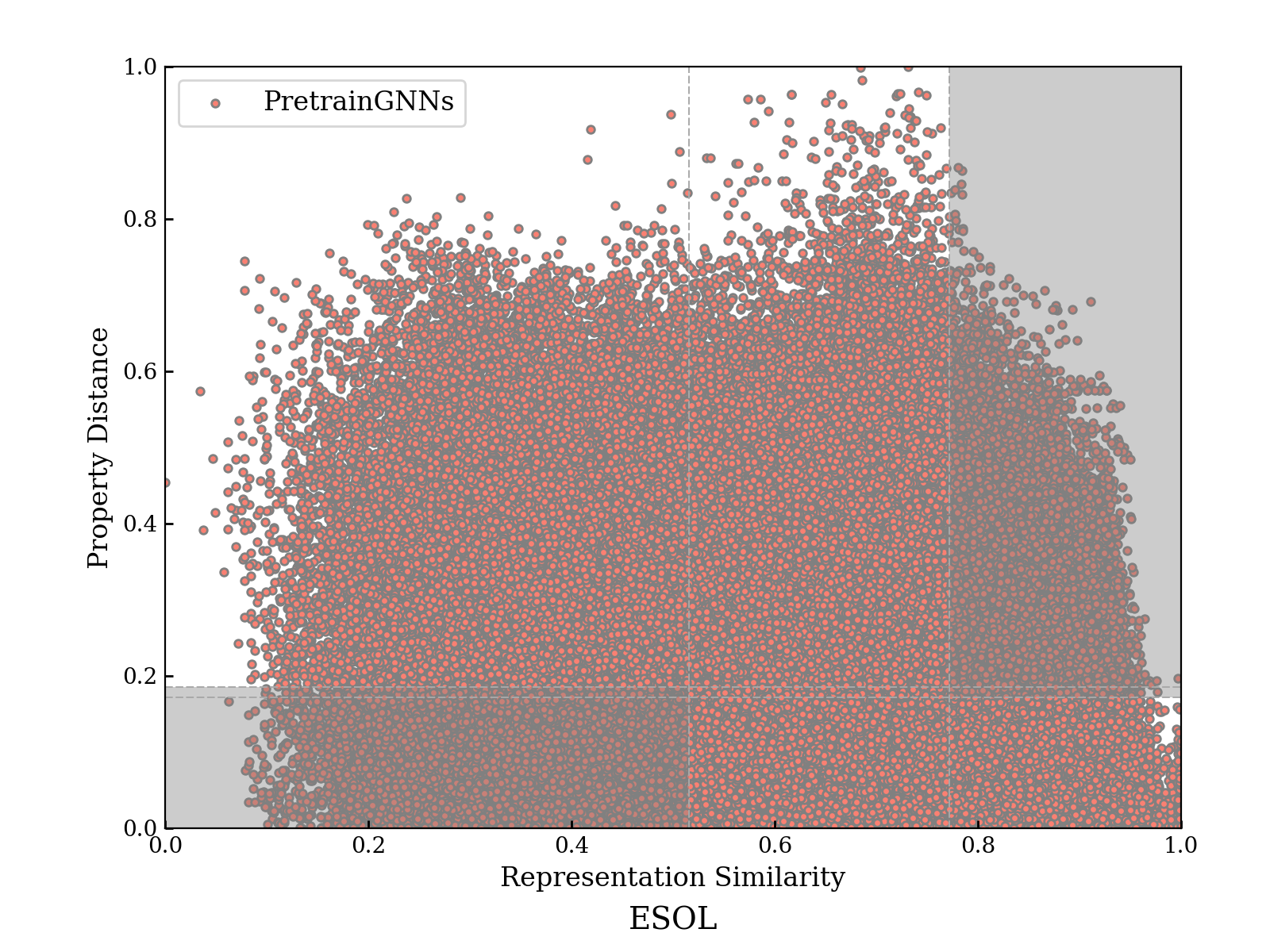}
}
    \subfigure[Pretrain8]{
	\includegraphics[width=3.2cm]{RPSMaps/Pretrain8_ESOL_MinMaxEud_Linear_FPSMap.png}
}

	\caption{RPSMaps with MinMaxEud as metric function on ESOL dataset. }
	\label{fig:ESOLMinMaxEudRPSMap}
\end{figure}

\begin{figure}[thbp]
	\centering
	\subfigure[ECFP]{
	\includegraphics[width=3.2cm]{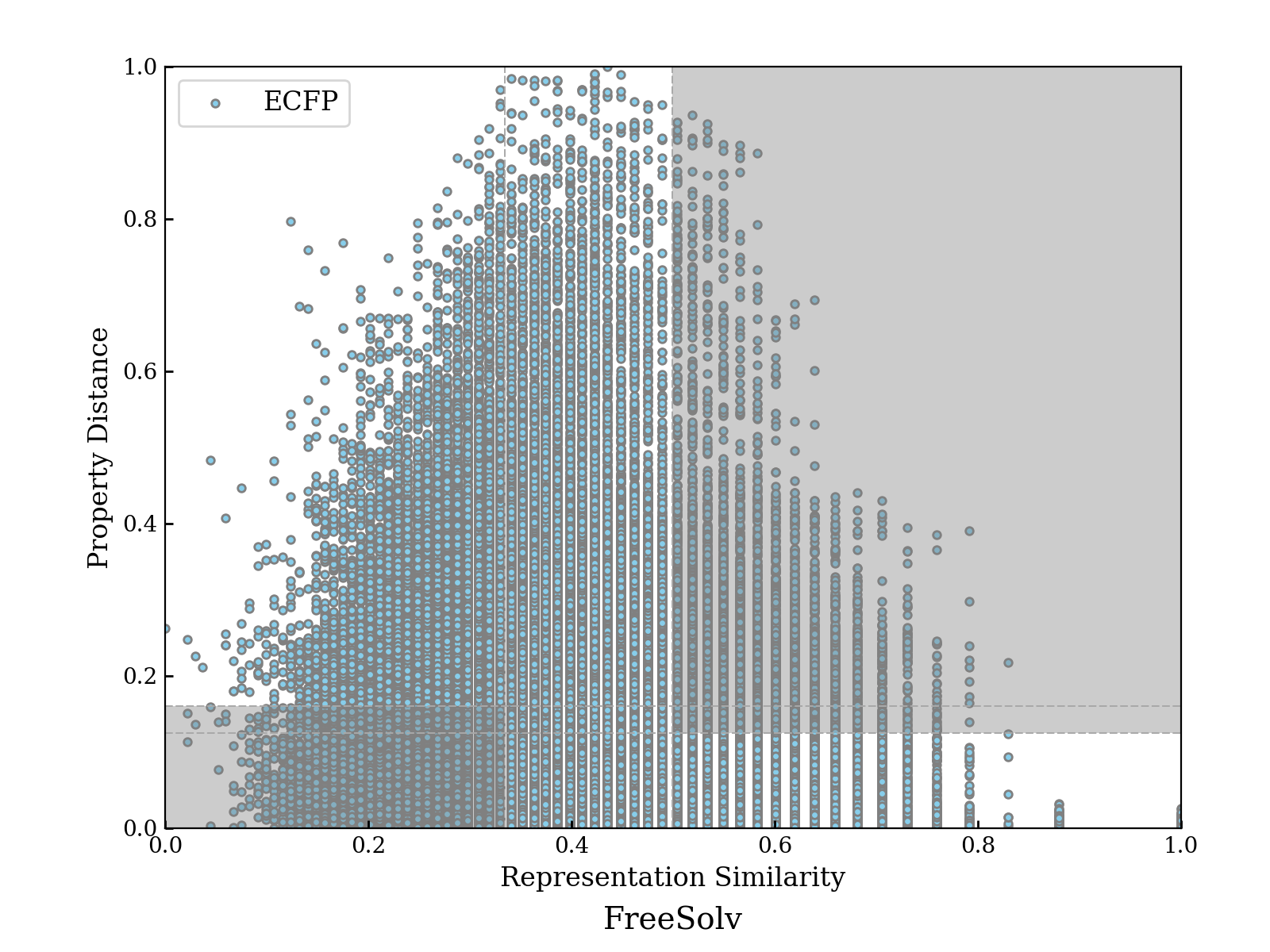}
}
    \subfigure[GROVER]{
    \includegraphics[width=3.2cm]{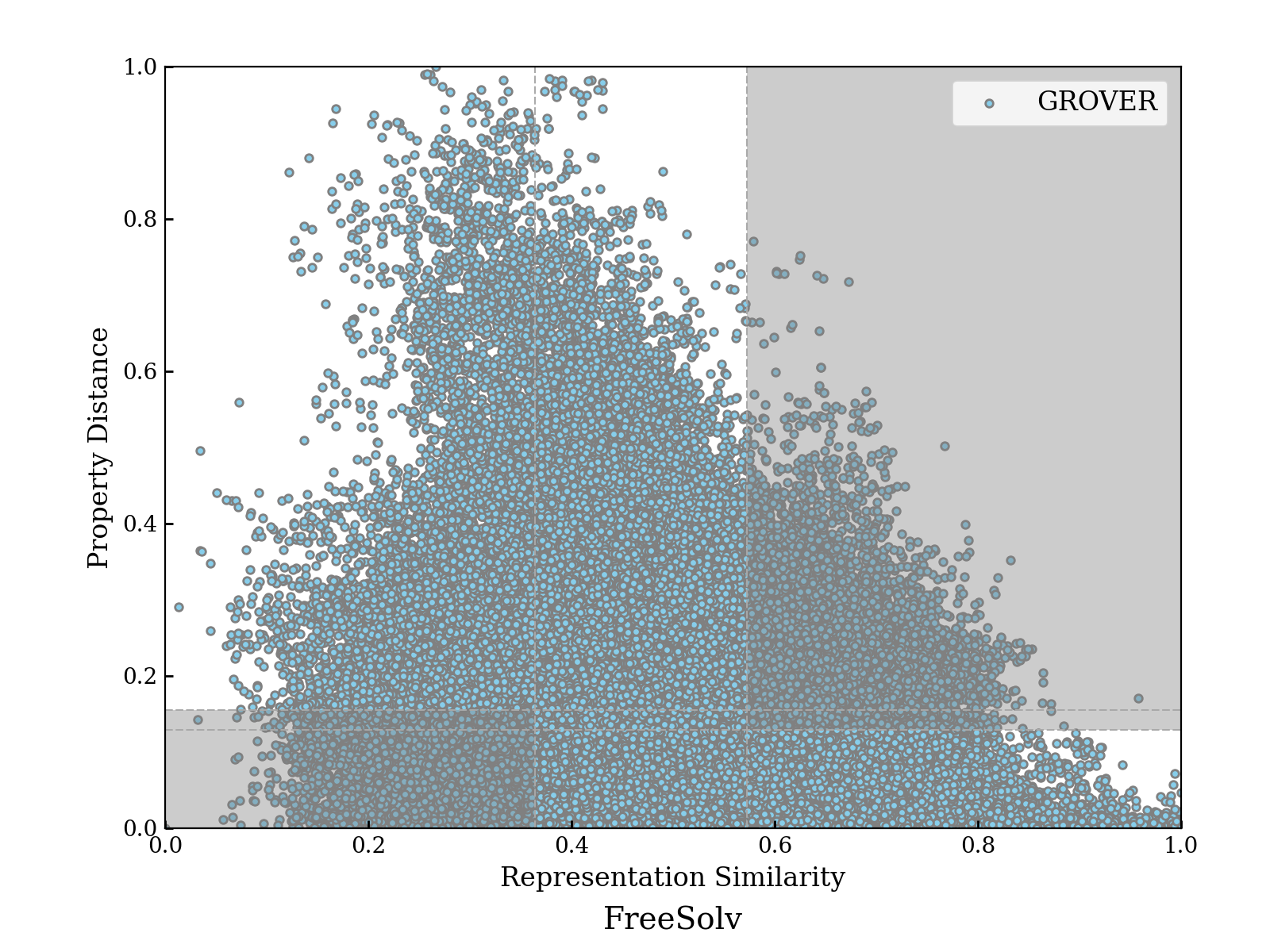}
}
    \subfigure[ChemBert]{
	\includegraphics[width=3.2cm]{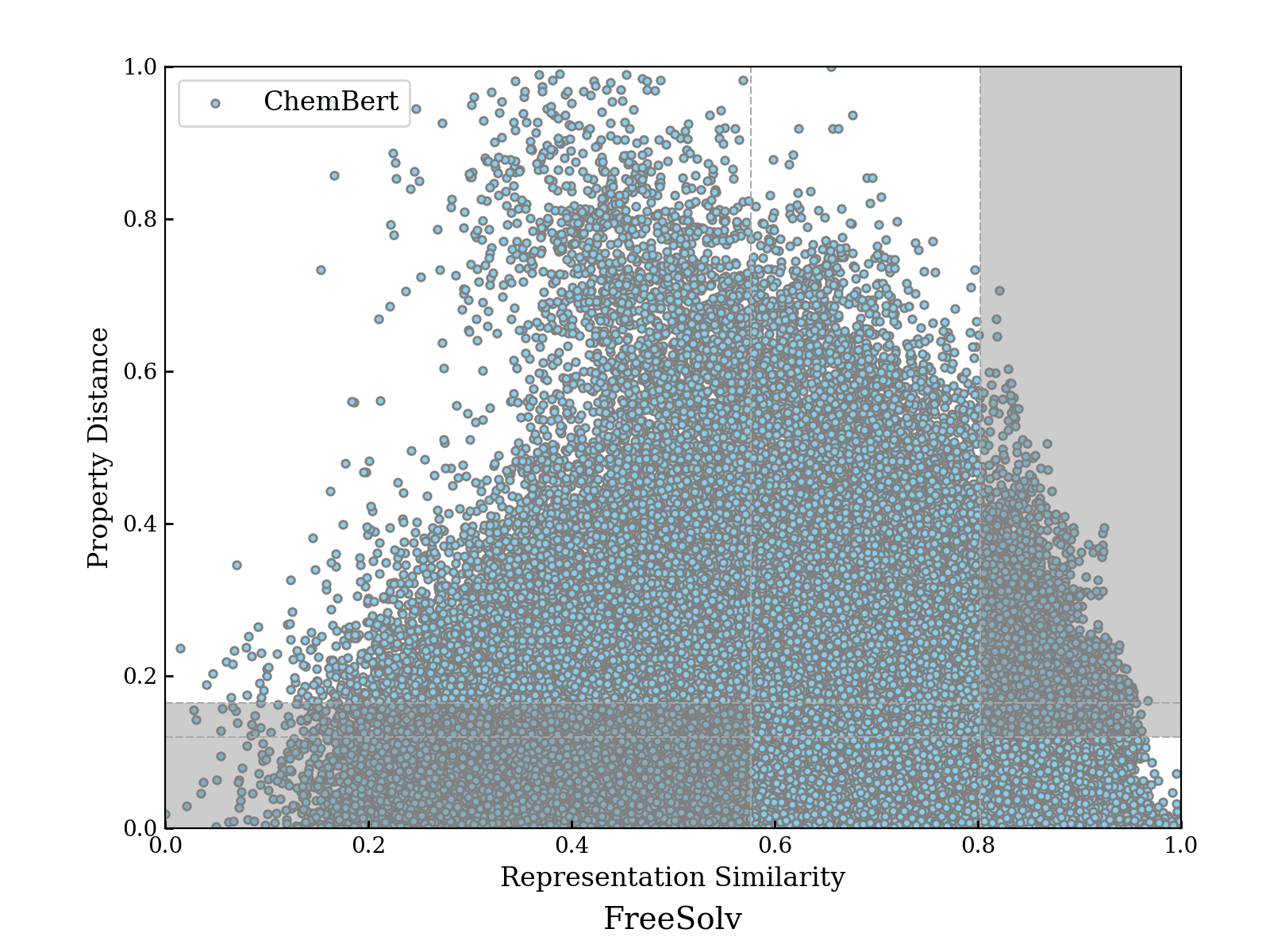}
}
    \subfigure[GraphLoG]{
	\includegraphics[width=3.2cm]{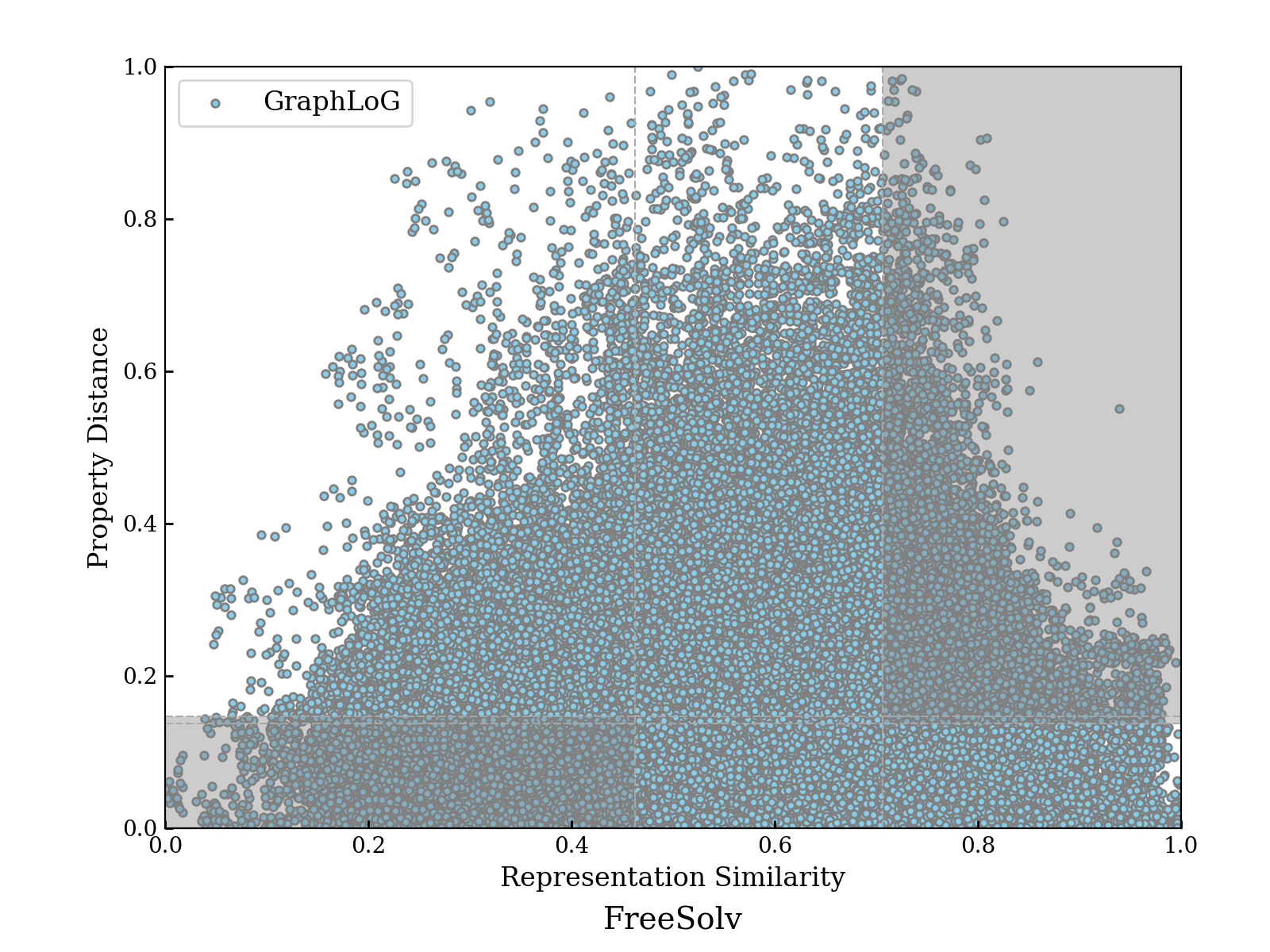}
}
	\subfigure[MAT]{
	\includegraphics[width=3.2cm]{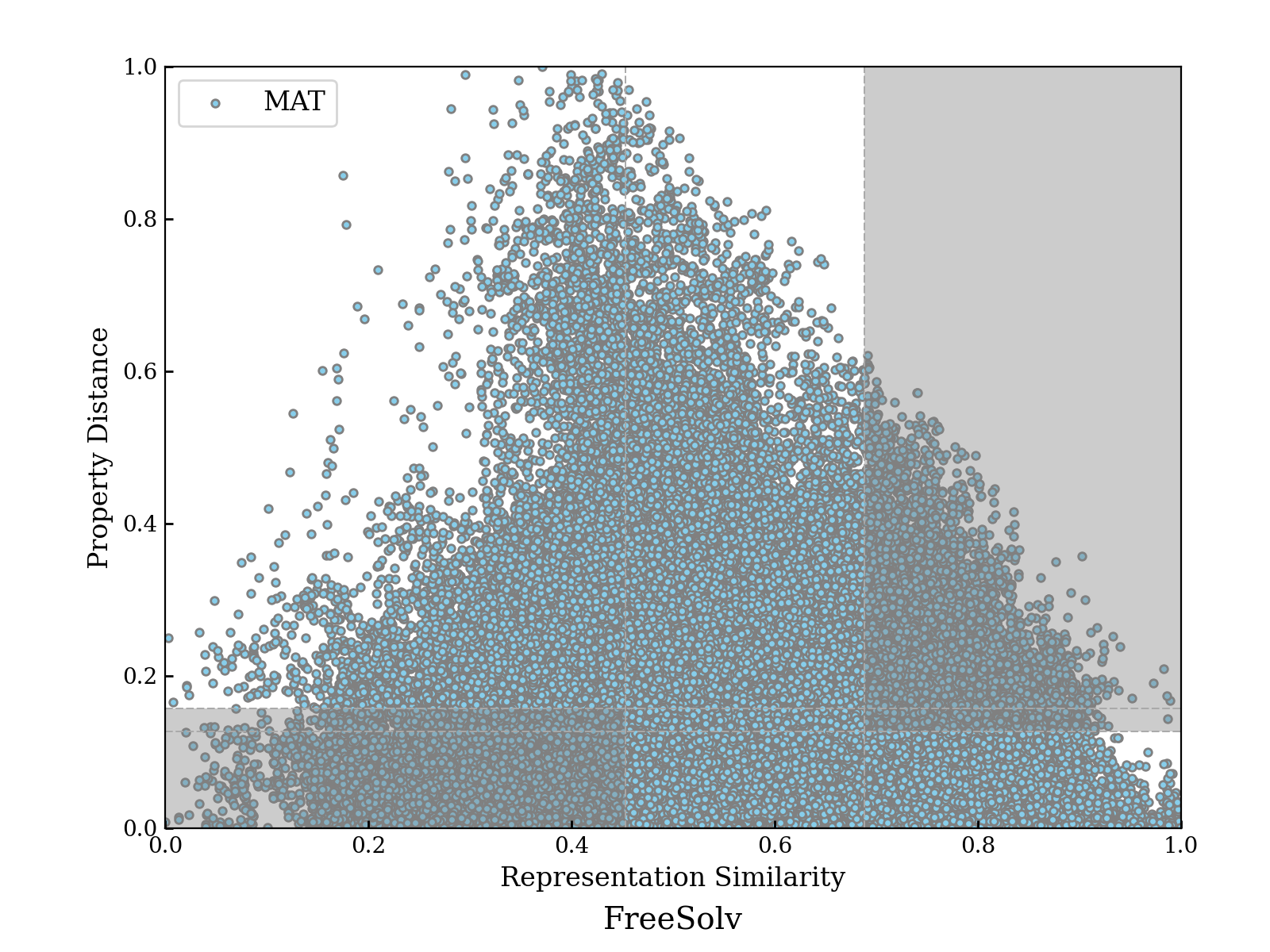}
}
    \subfigure[SMILESTransformer]{
    \includegraphics[width=3.2cm]{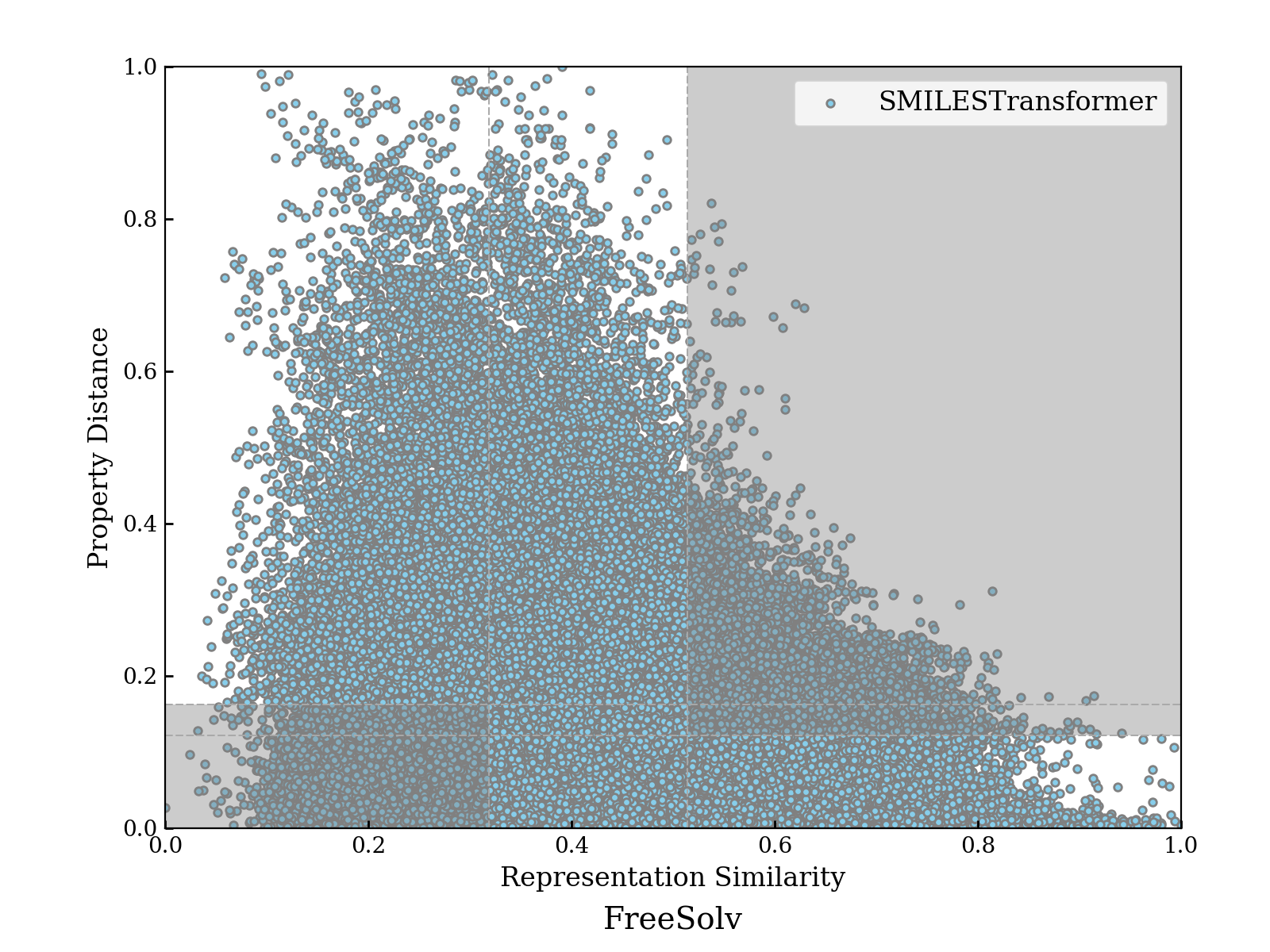}
}
    \subfigure[PretrainGNNs]{
	\includegraphics[width=3.2cm]{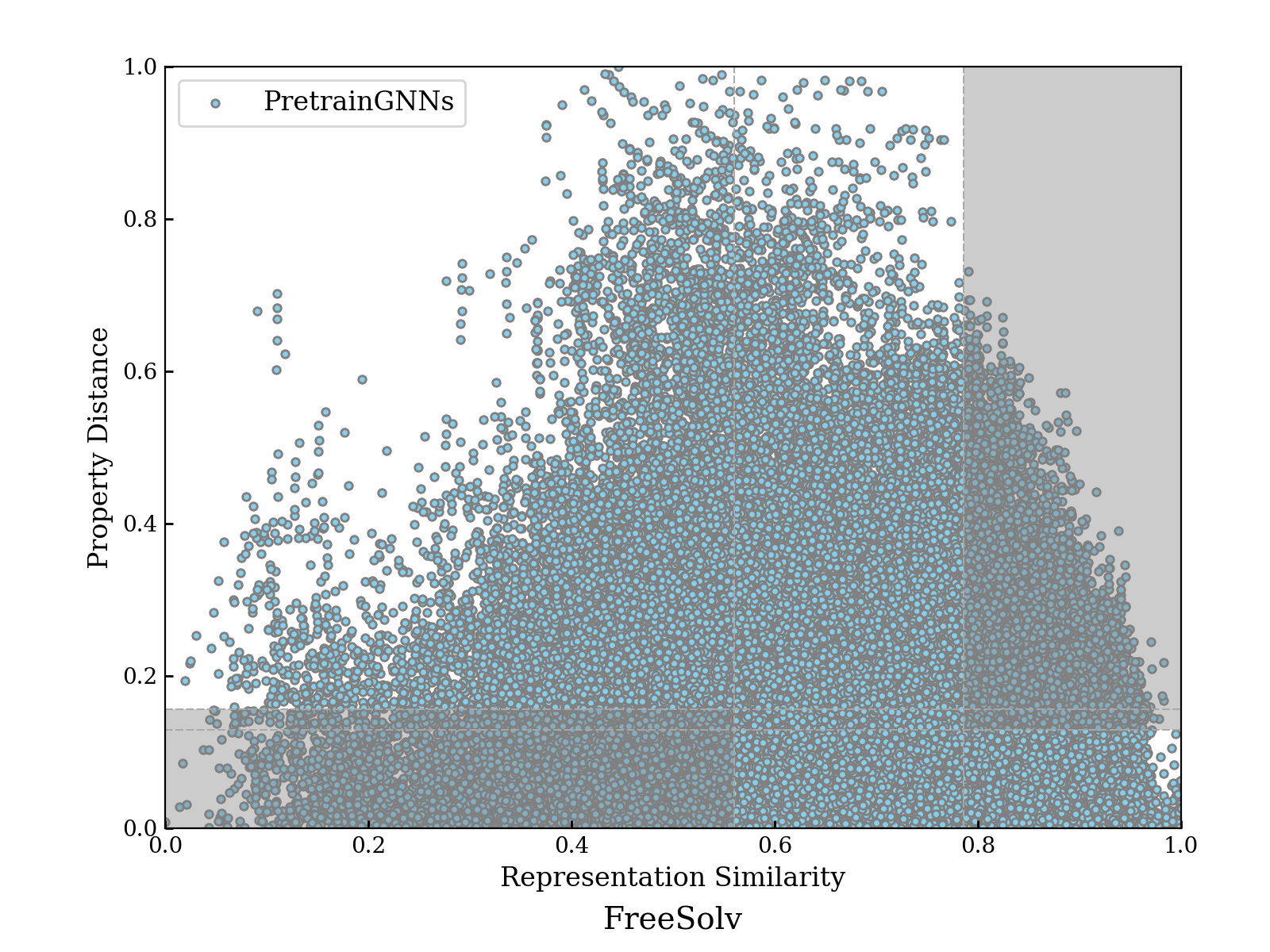}
}
    \subfigure[Pretrain8]{
	\includegraphics[width=3.2cm]{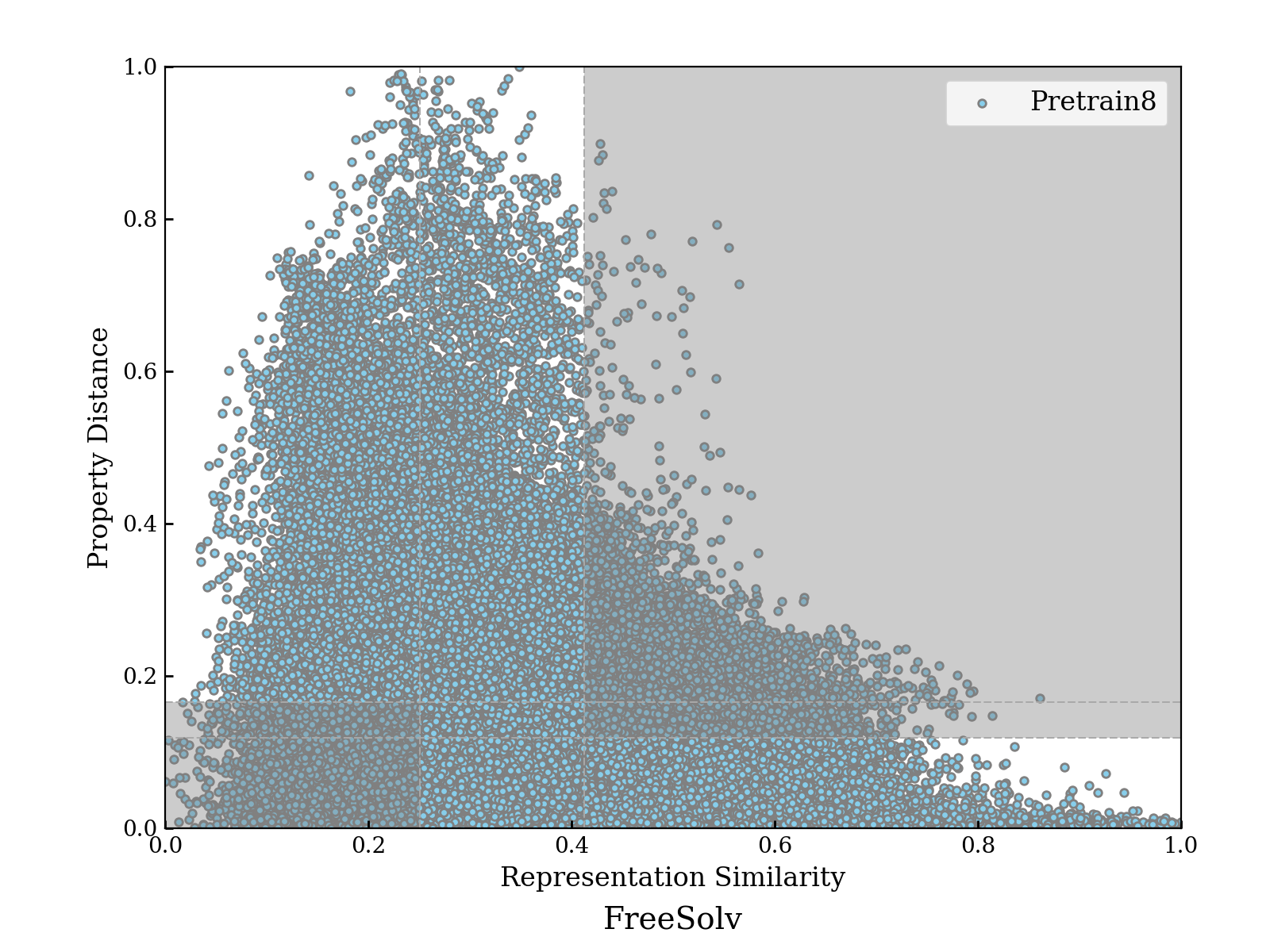}
}

	\caption{RPSMaps with MinMaxEud as metric function on FreeSolv Dataset. }
	\label{fig:FreeSolvMinMaxEudRPSMap}
\end{figure}

\begin{figure}[thbp]
	\centering
	\subfigure[ECFP]{
	\includegraphics[width=3.2cm]{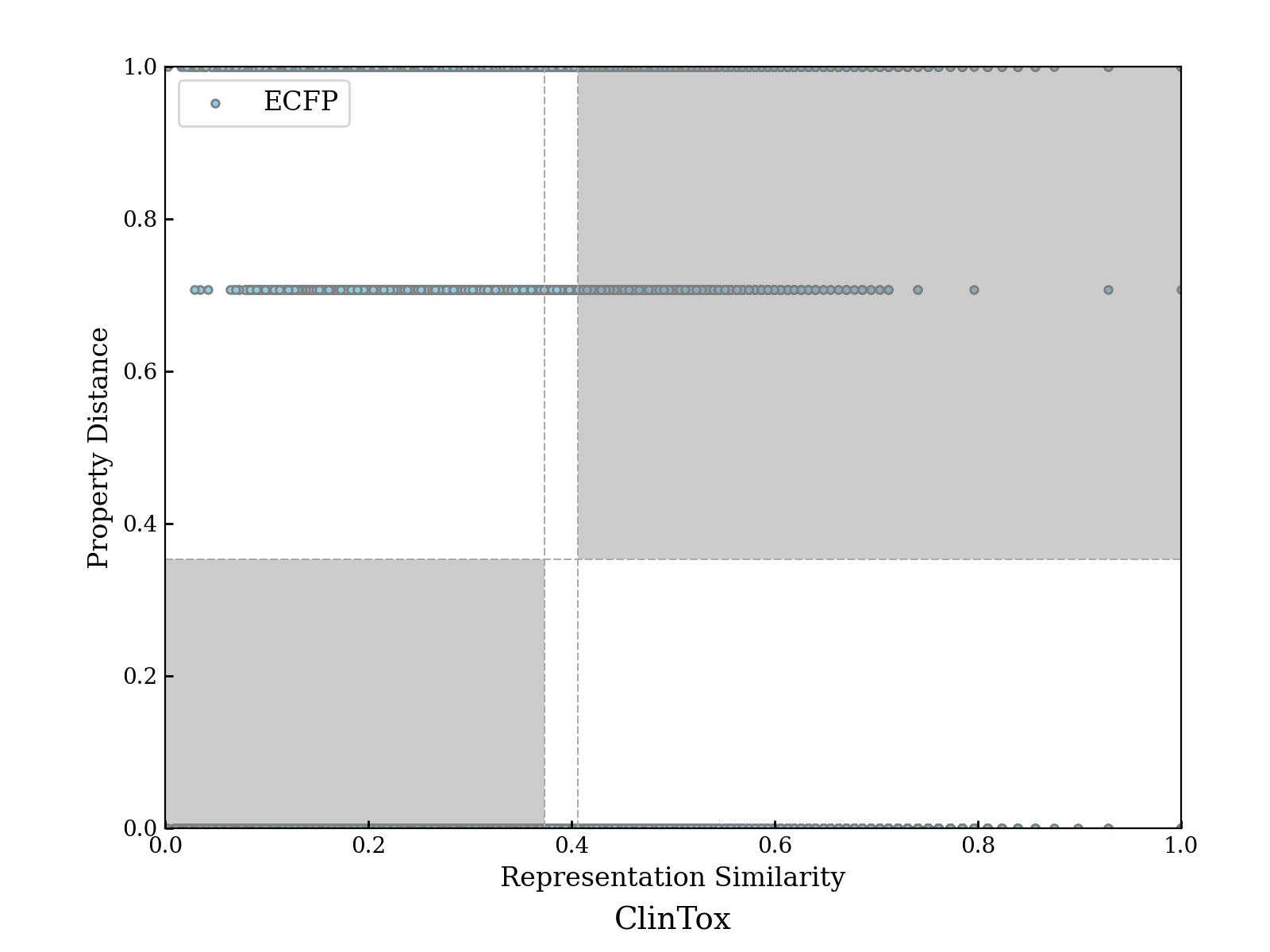}
}
    \subfigure[GROVER]{
    \includegraphics[width=3.2cm]{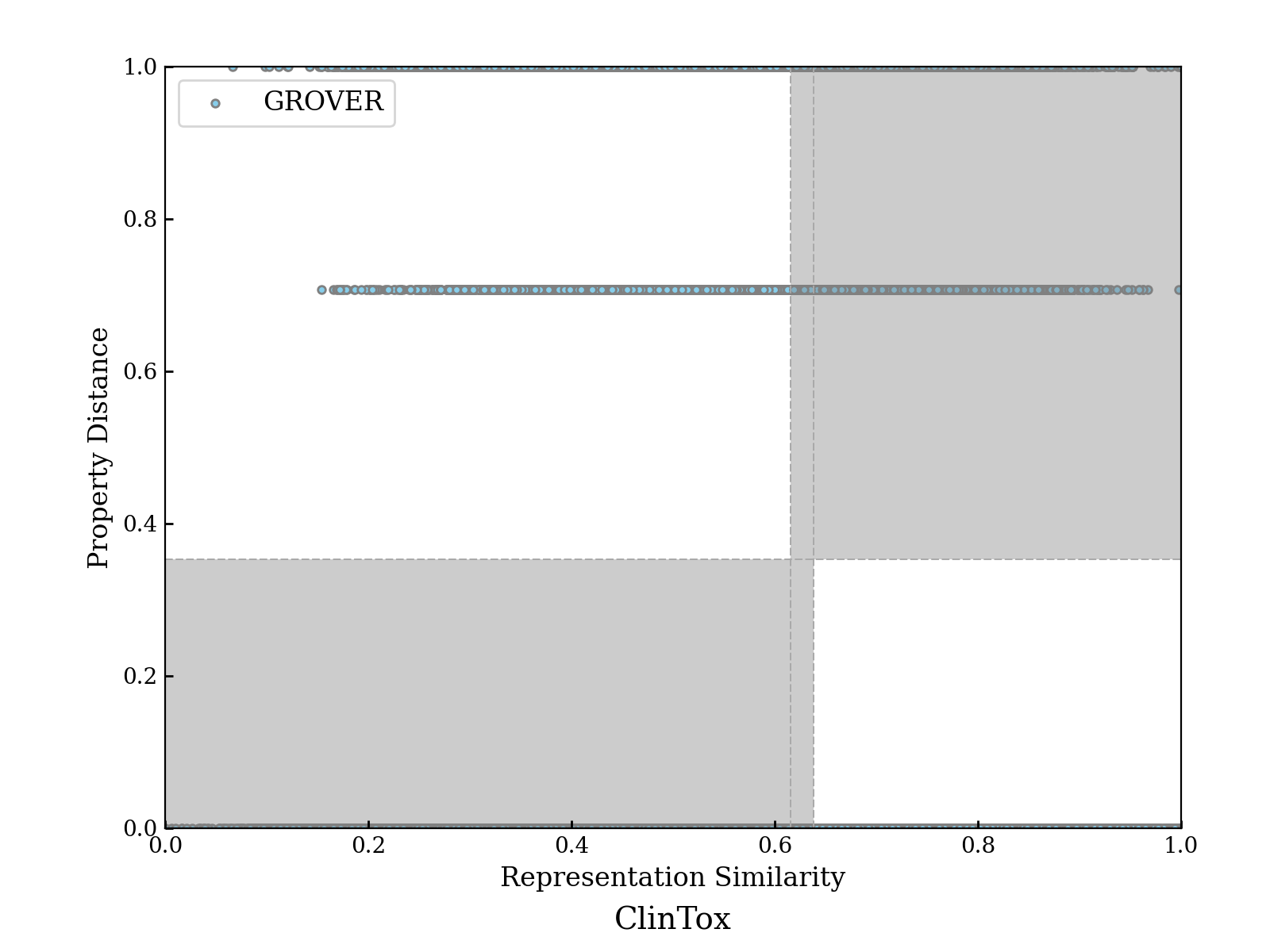}
}
    \subfigure[ChemBert]{
	\includegraphics[width=3.2cm]{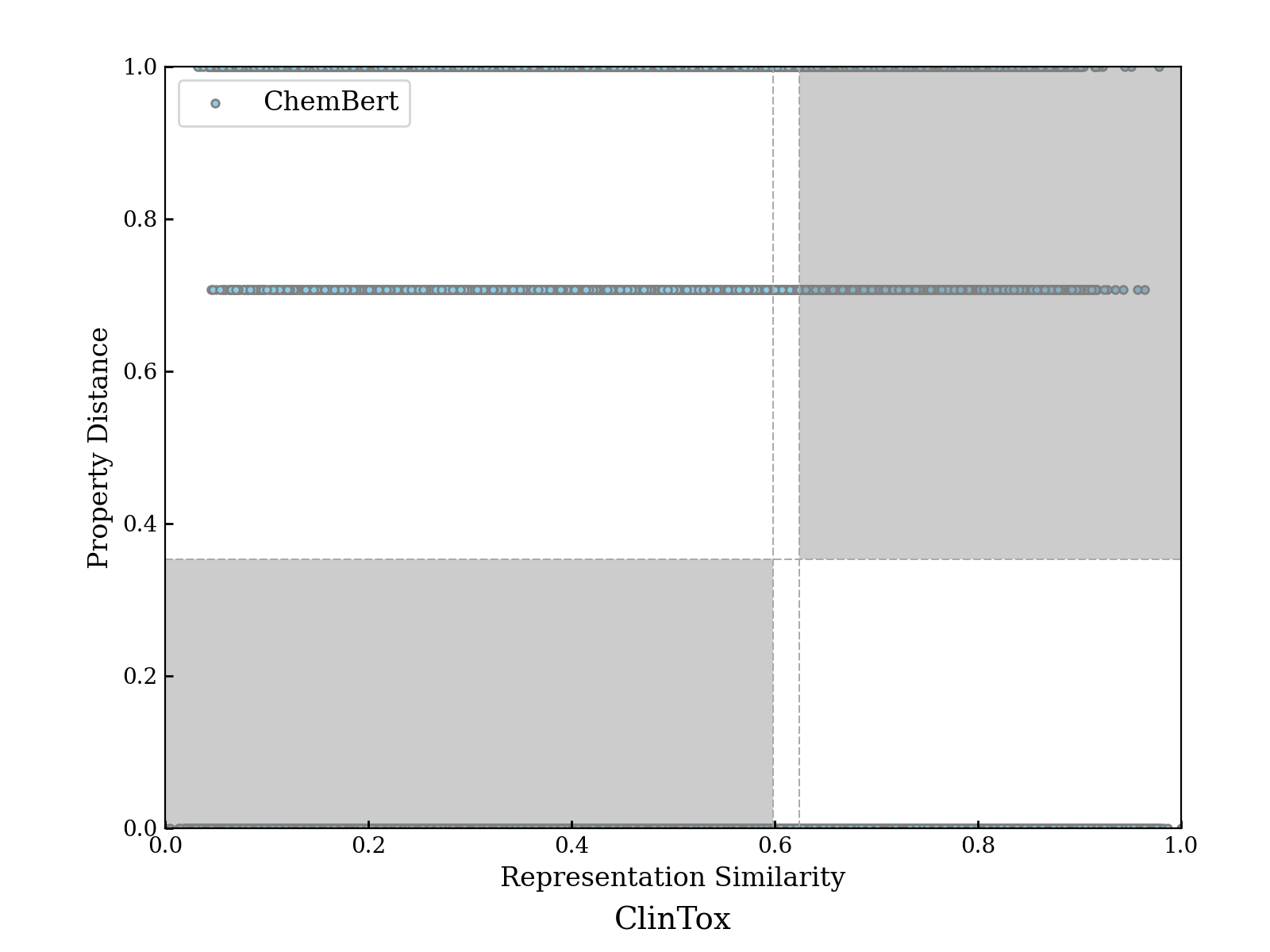}
}
    \subfigure[GraphLoG]{
	\includegraphics[width=3.2cm]{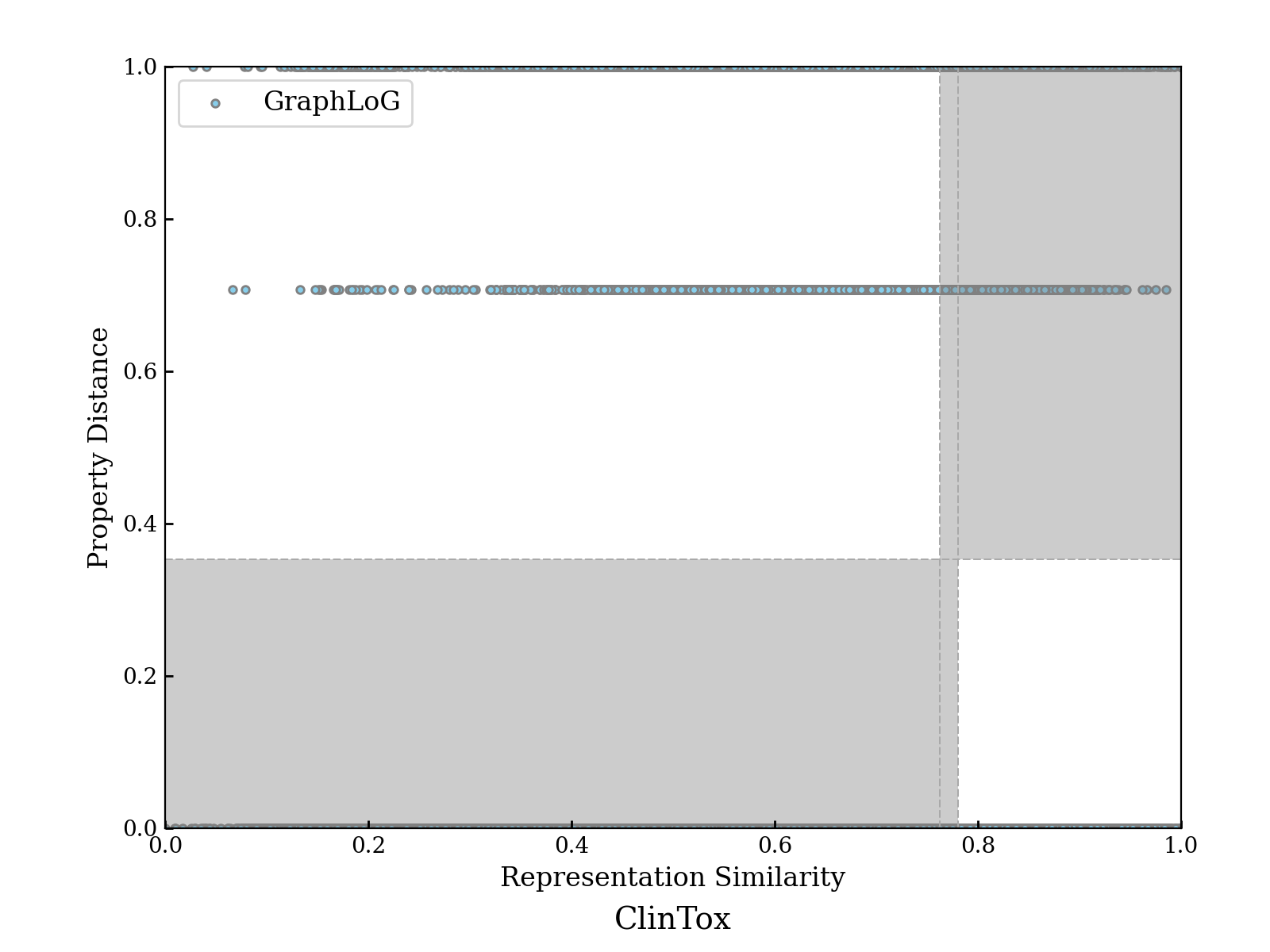}
}
	\subfigure[MAT]{
	\includegraphics[width=3.2cm]{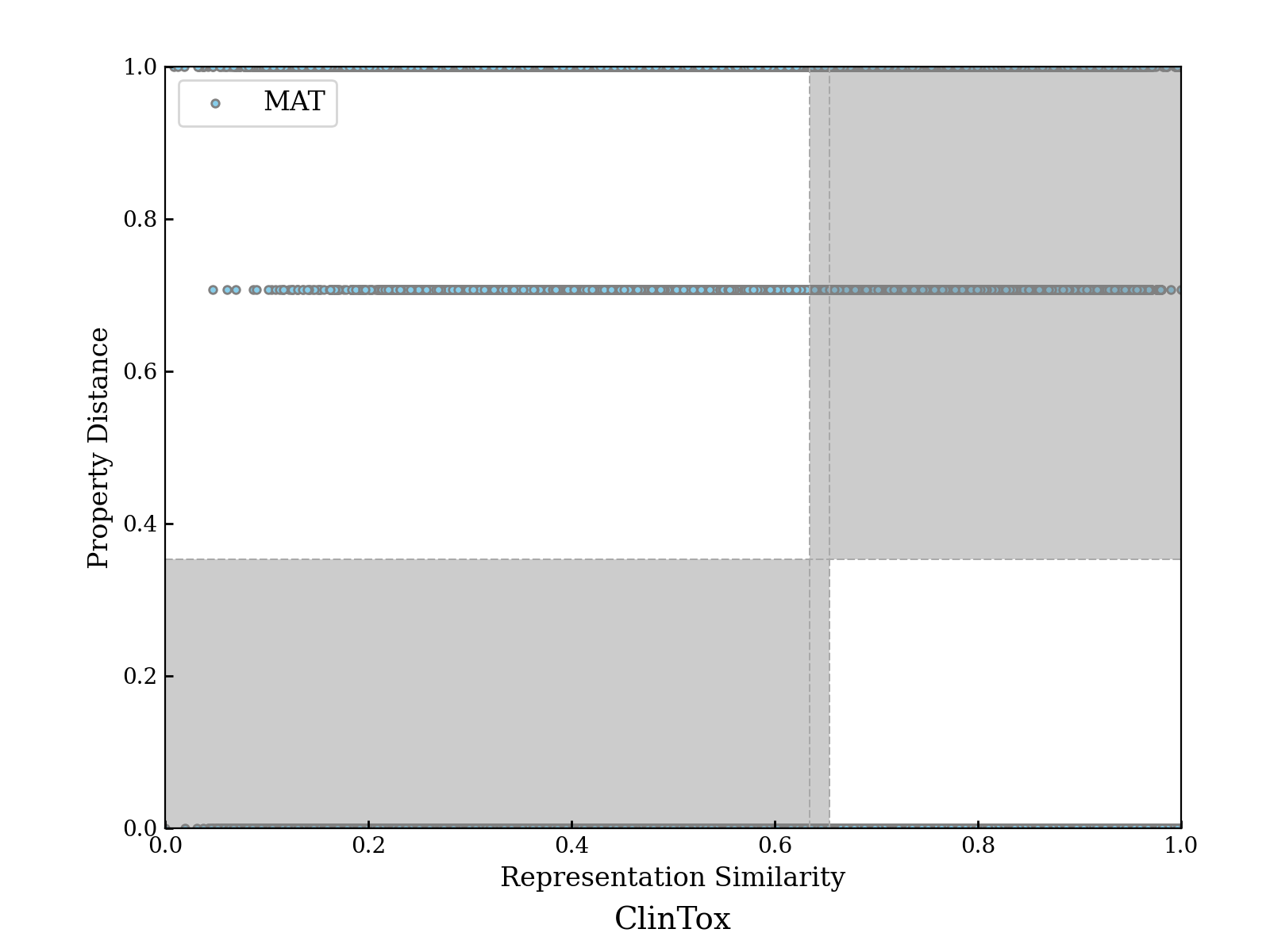}
}
    \subfigure[SMILESTransformer]{
    \includegraphics[width=3.2cm]{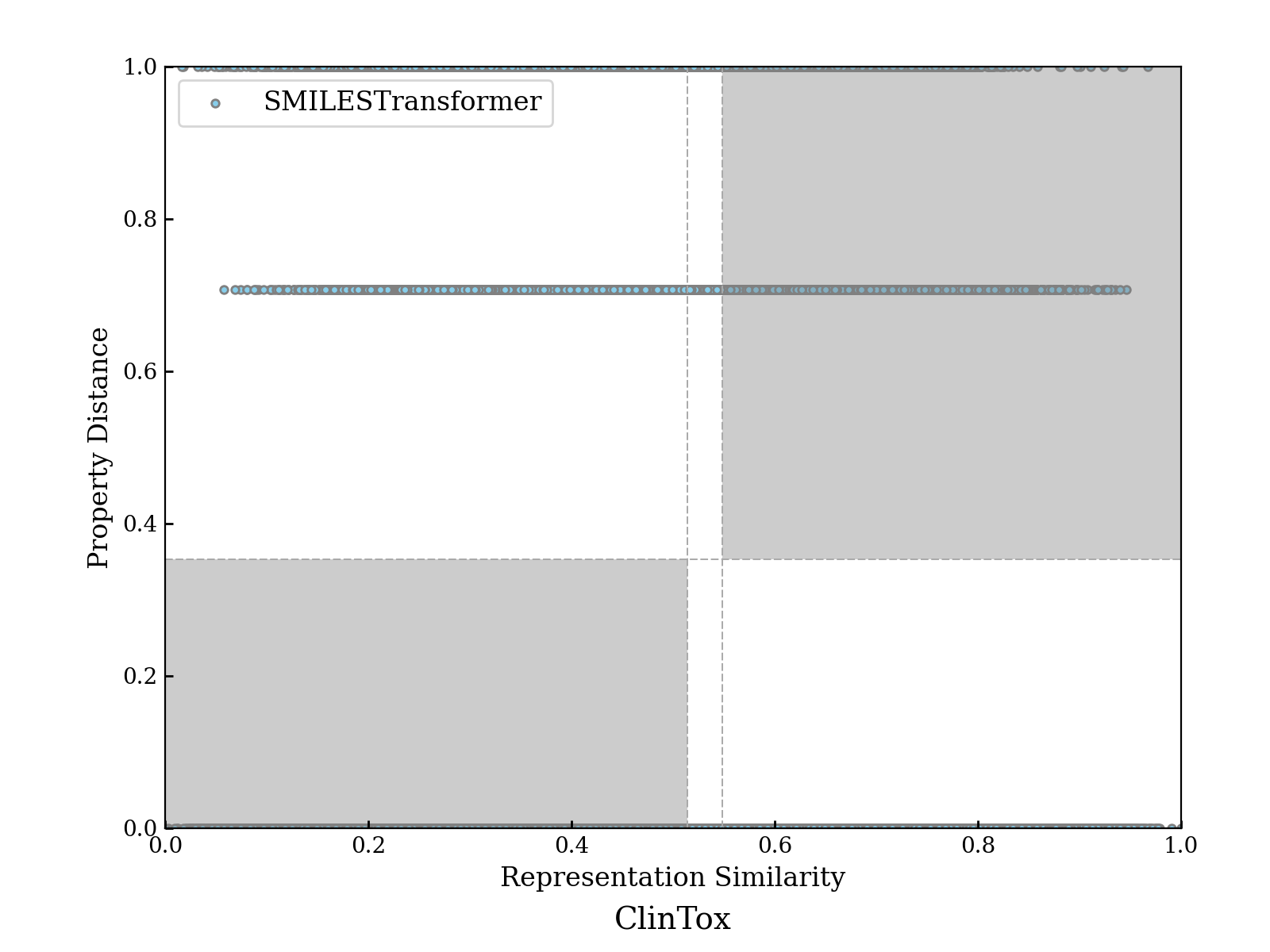}
}
    \subfigure[PretrainGNNs]{
	\includegraphics[width=3.2cm]{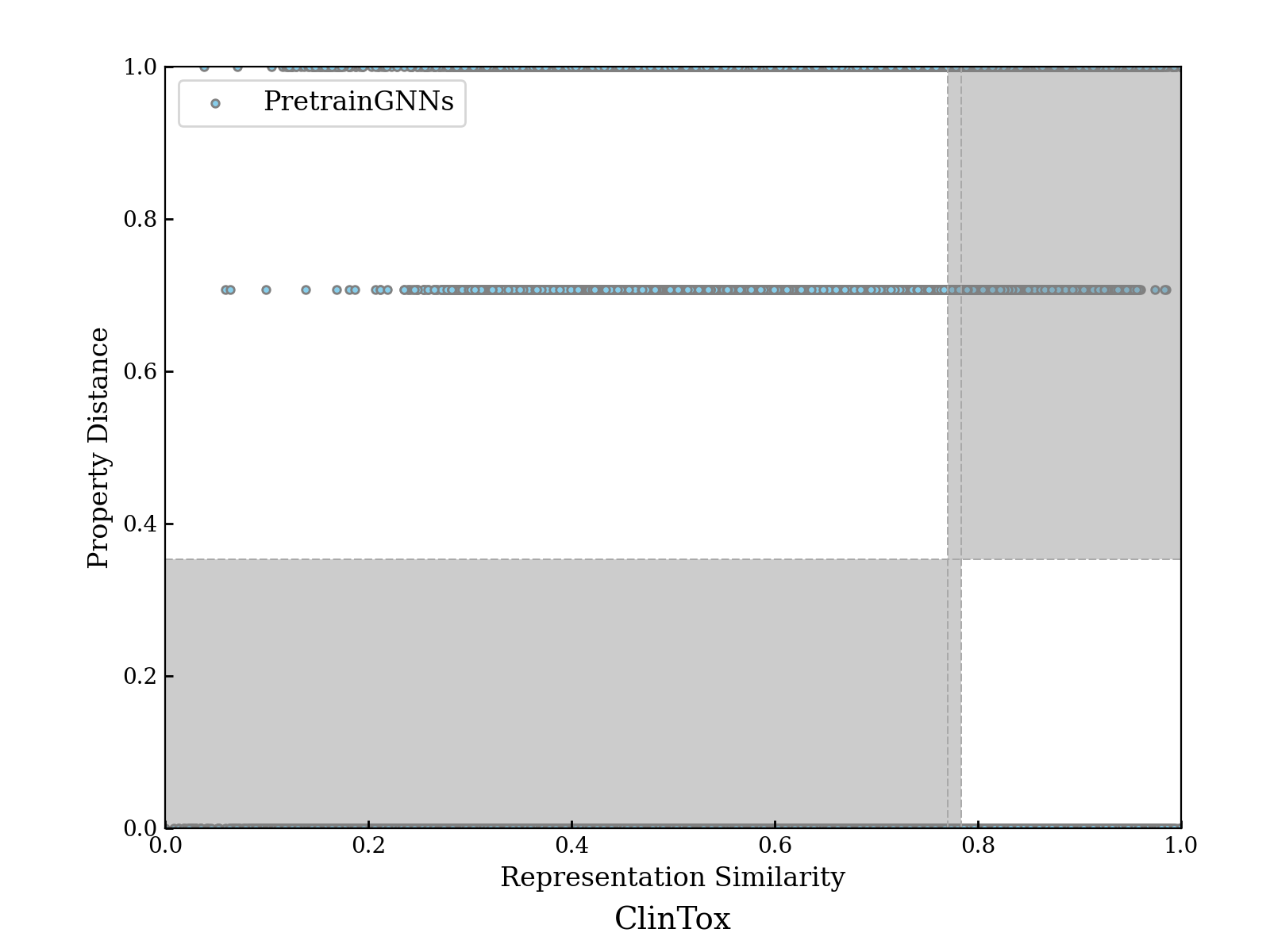}
}
    \subfigure[Pretrain8]{
	\includegraphics[width=3.2cm]{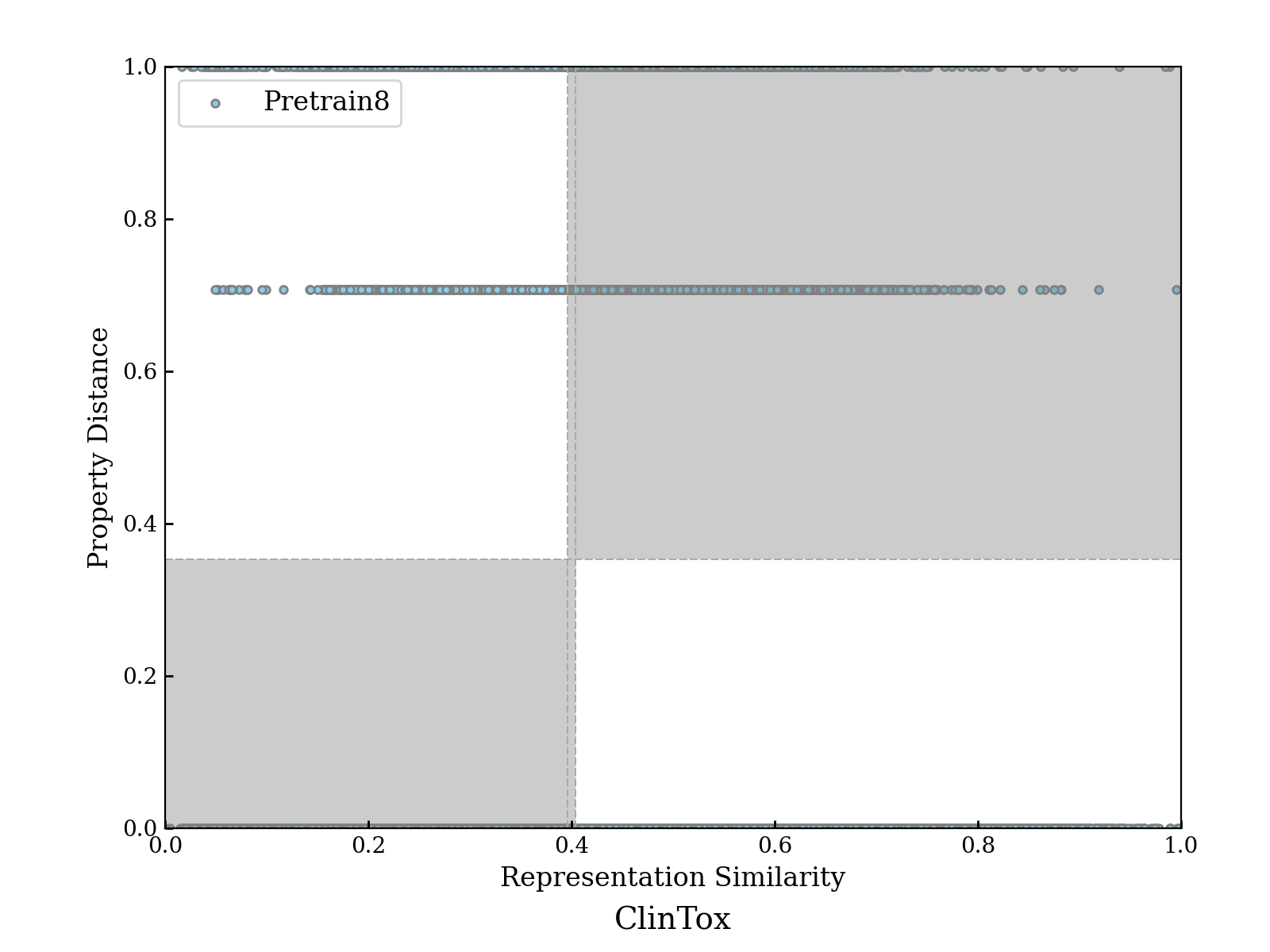}
}

	\caption{RPSMaps with MinMaxEud as metric function on ClinTox Dataset. }
	\label{fig:ClinToxMinMaxEudRPSMap}
\end{figure}

\begin{figure}[thbp]
	\centering
	\subfigure[ECFP]{
	\includegraphics[width=3.2cm]{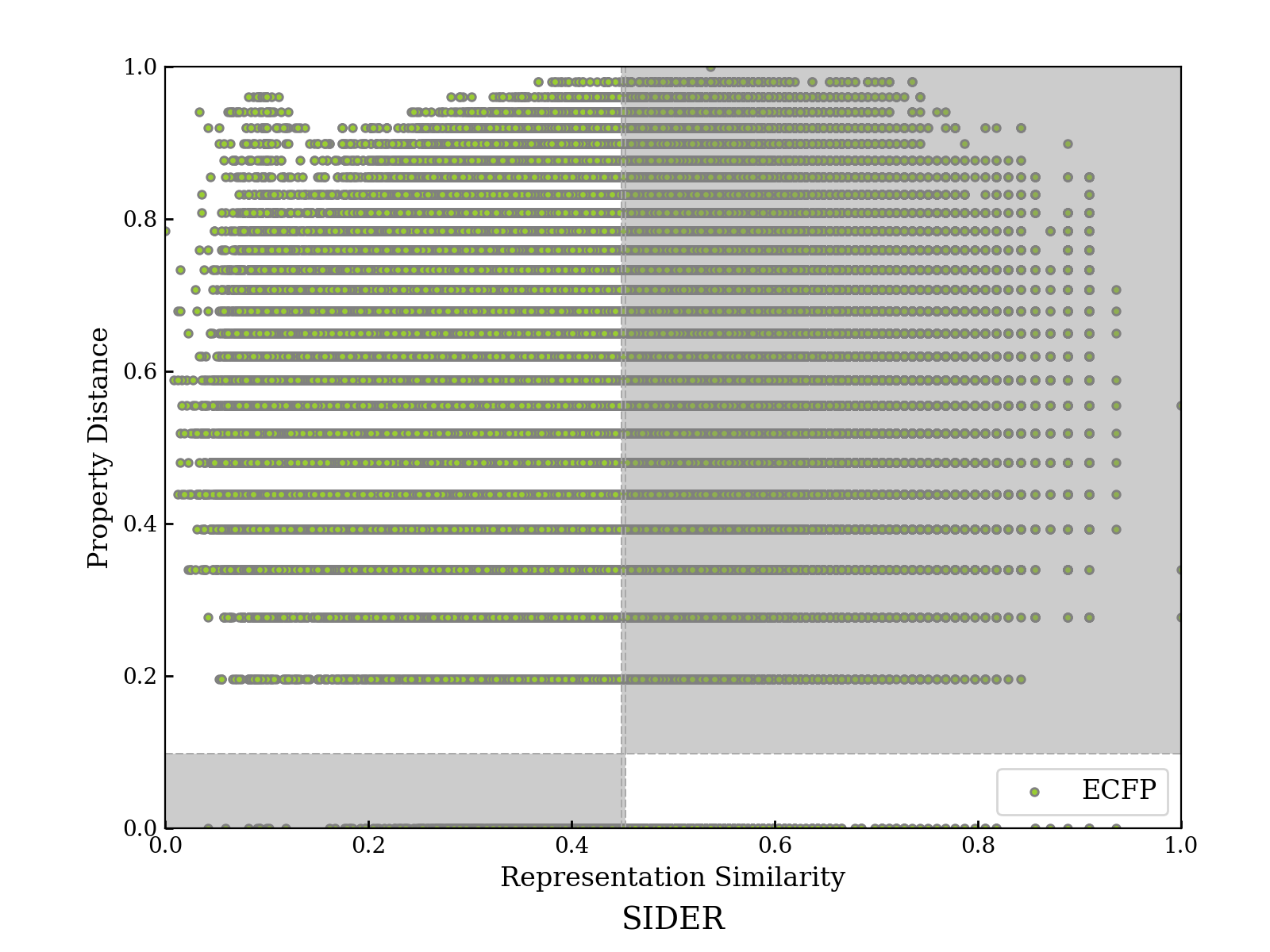}
}
    \subfigure[GROVER]{
    \includegraphics[width=3.2cm]{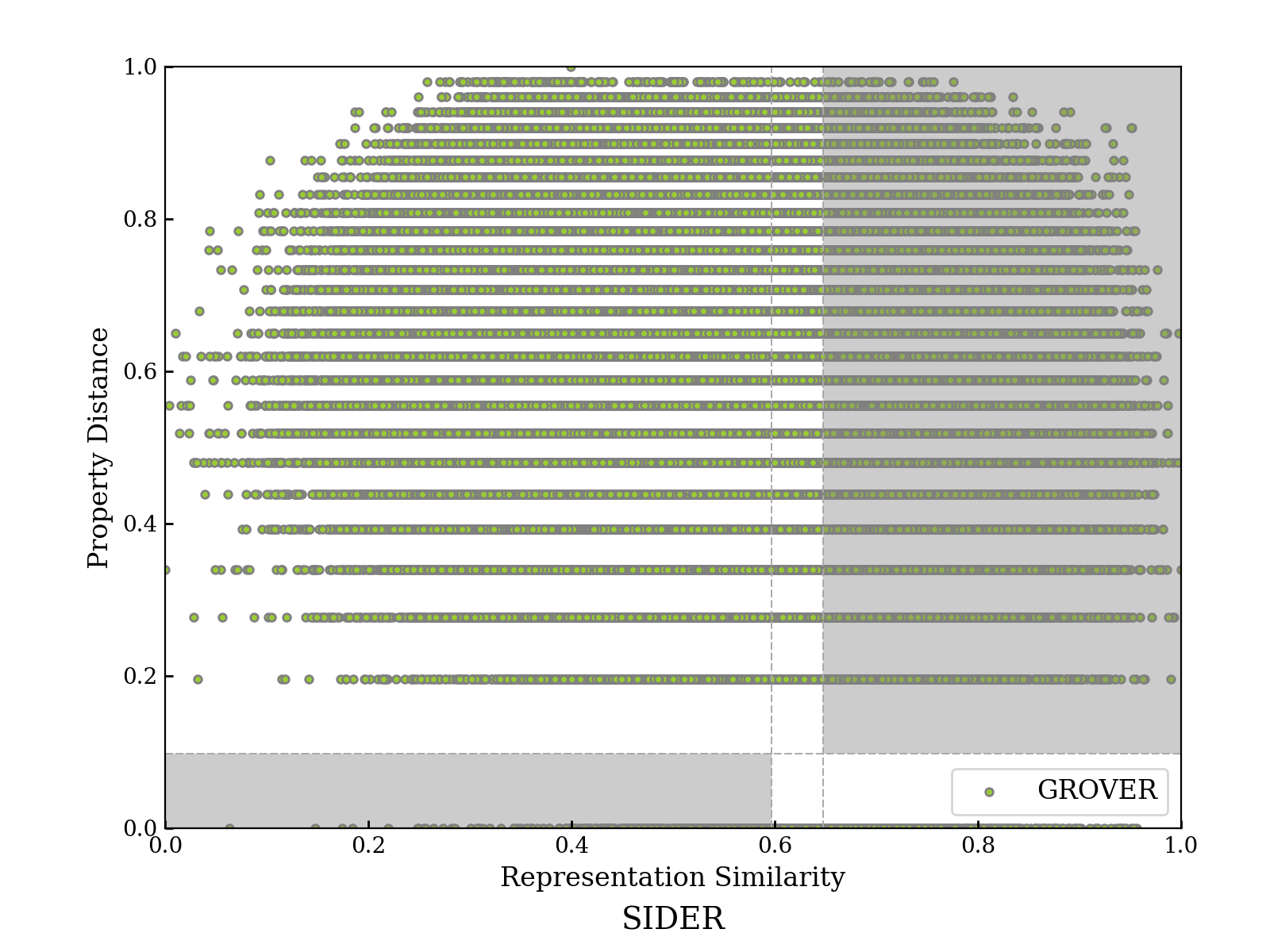}
}
    \subfigure[ChemBert]{
	\includegraphics[width=3.2cm]{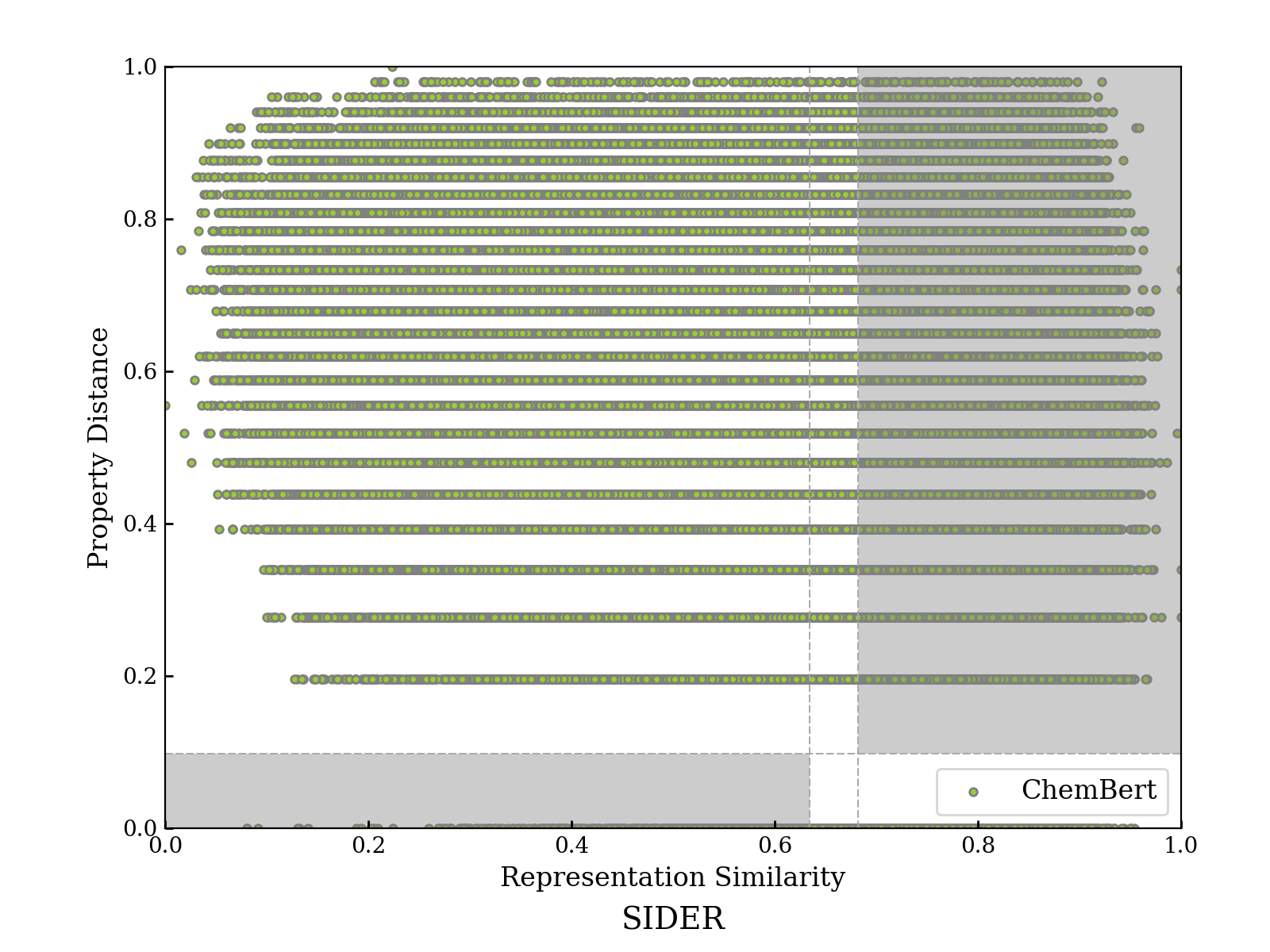}
}
    \subfigure[GraphLoG]{
	\includegraphics[width=3.2cm]{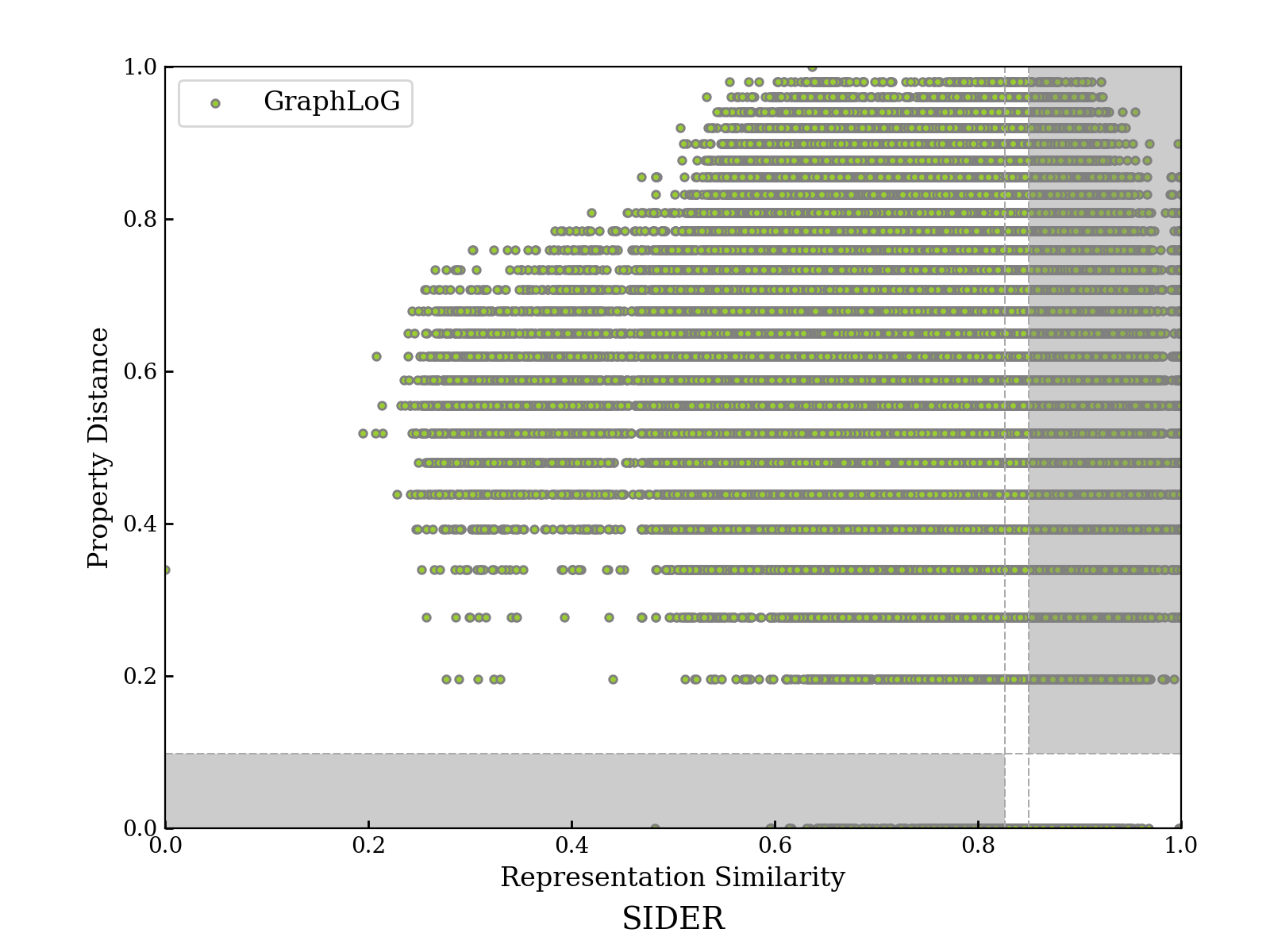}
}
	\subfigure[MAT]{
	\includegraphics[width=3.2cm]{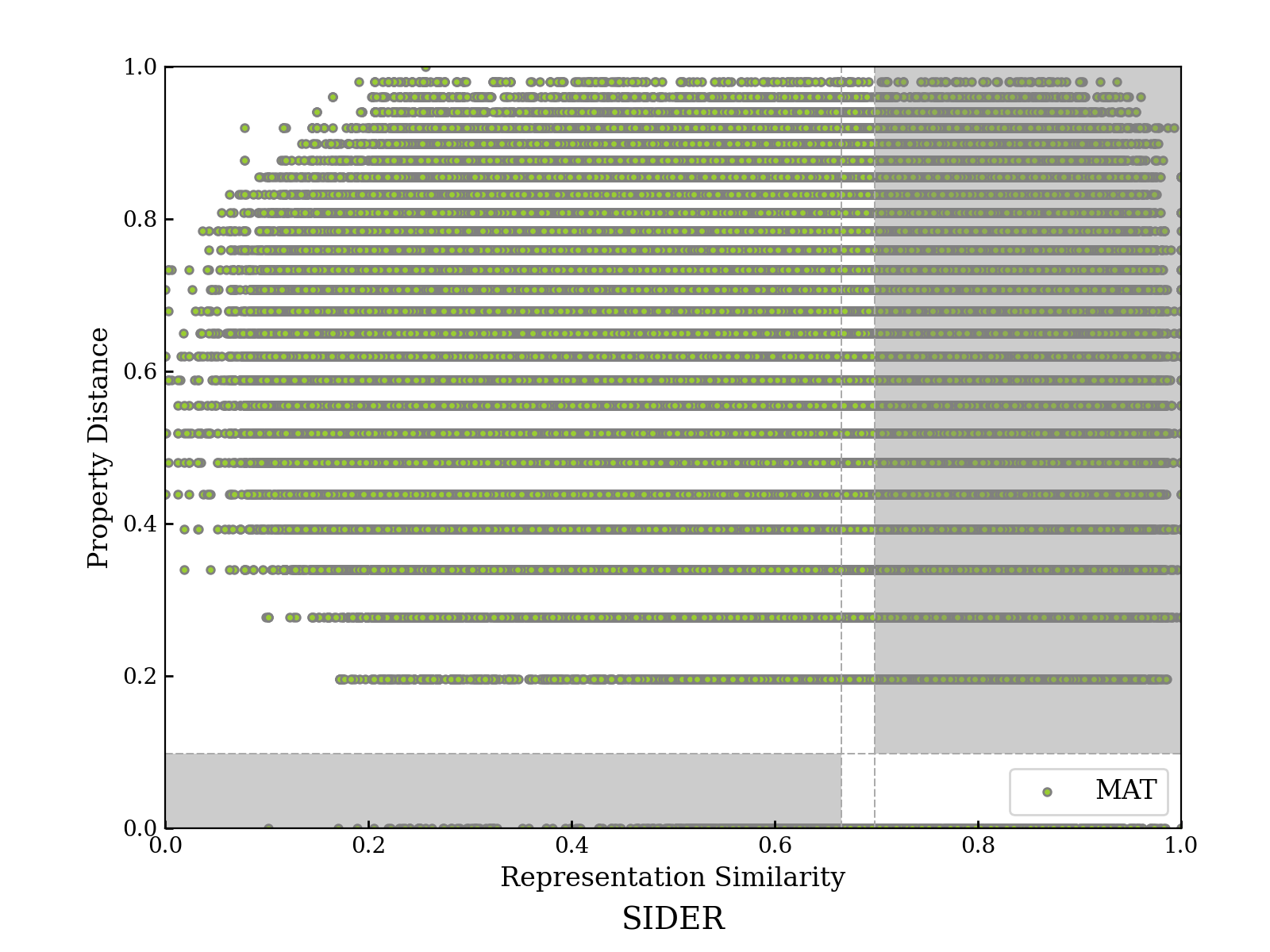}
}
    \subfigure[SMILESTransformer]{
    \includegraphics[width=3.2cm]{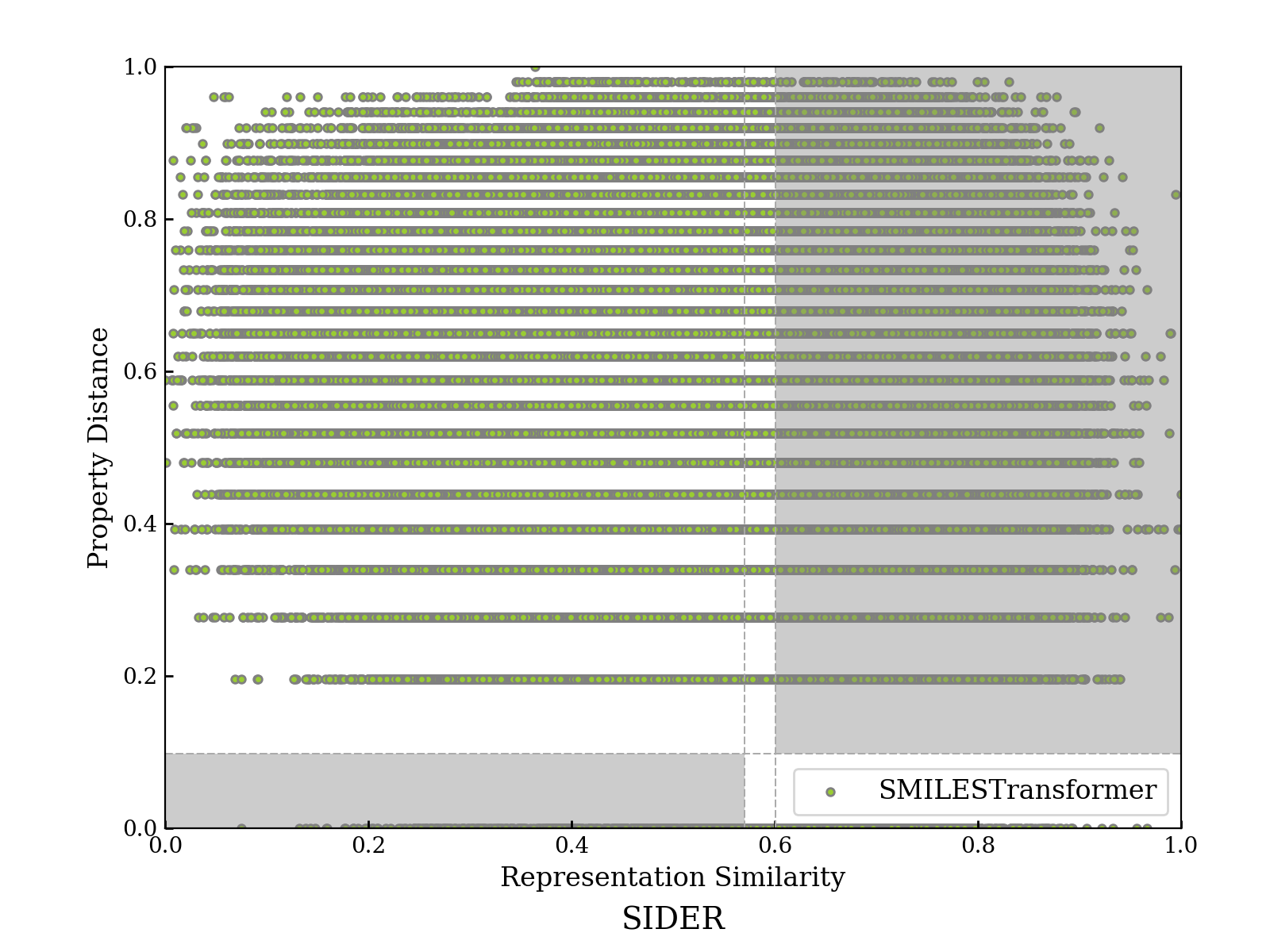}
}
    \subfigure[PretrainGNNs]{
	\includegraphics[width=3.2cm]{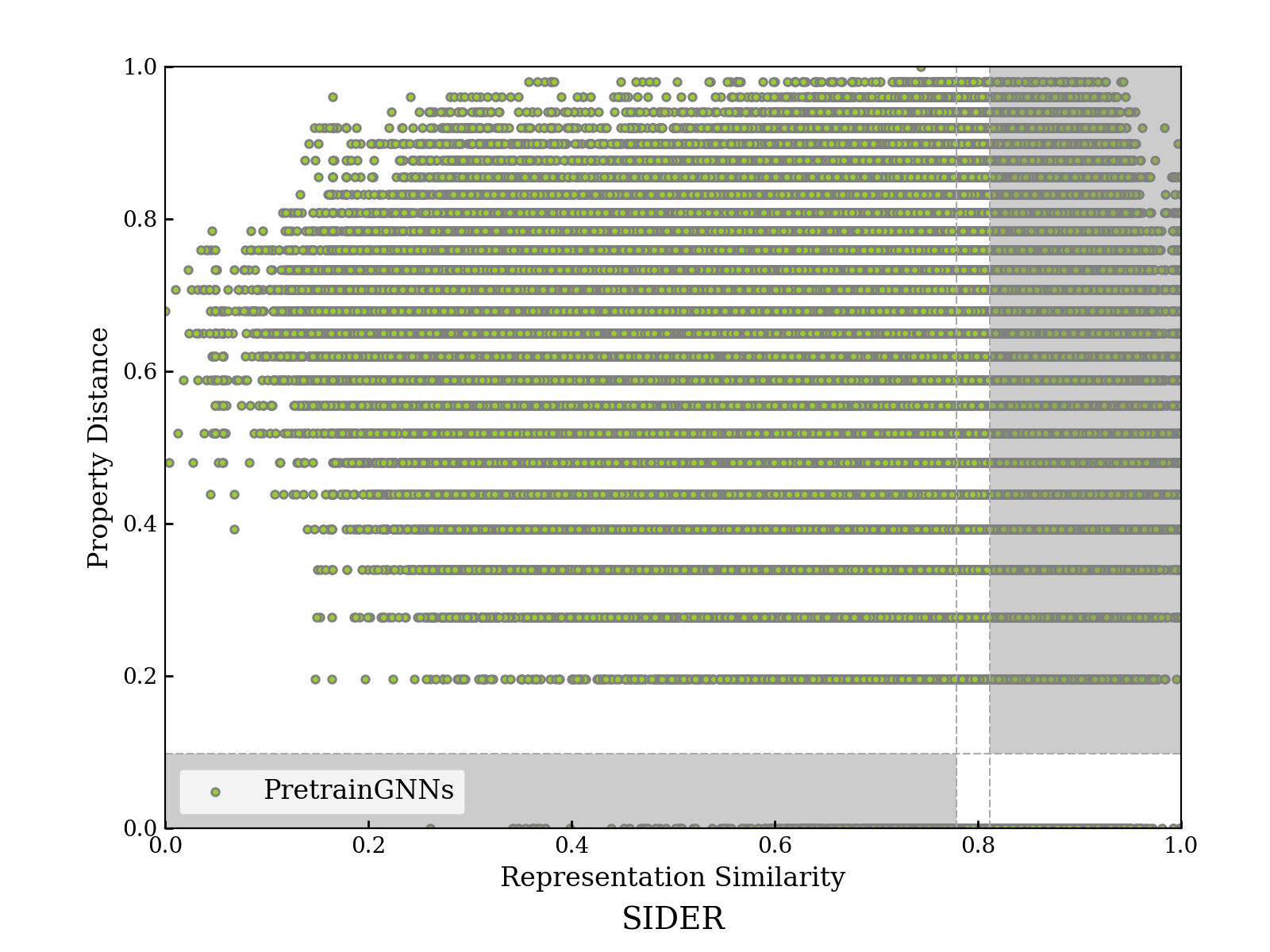}
}
    \subfigure[Pretrain8]{
	\includegraphics[width=3.2cm]{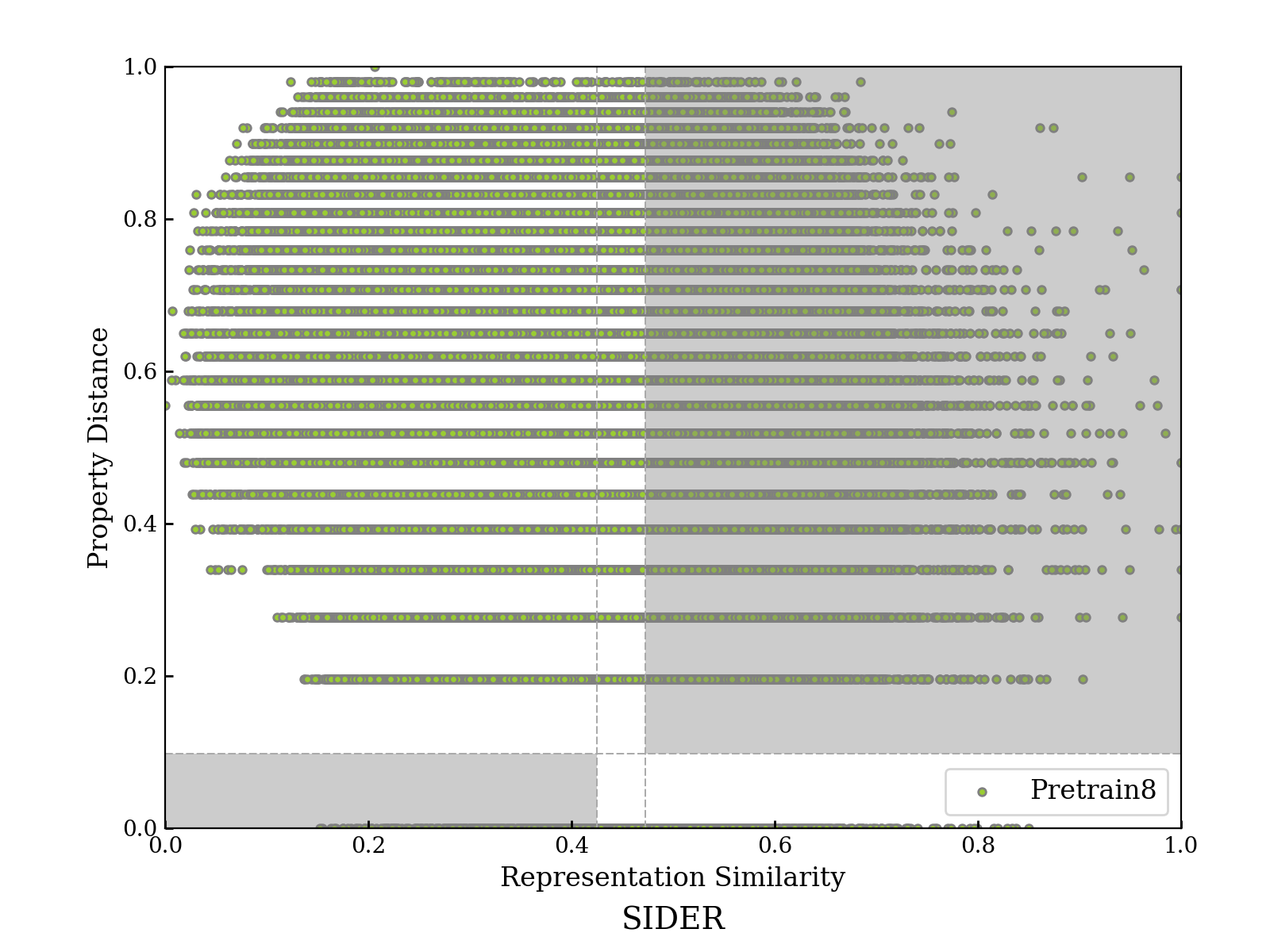}
}

	\caption{RPSMaps with MinMaxEud as metric function on SIDER Dataset. }
	\label{fig:SIDERMinMaxEudRPSMap}
\end{figure}

\clearpage

\section{Distribution of Molecular Similarities with CosineSim as Metric Function}
\begin{figure}[thbp]
	\centering
	\subfigure[ECFP]{
	\includegraphics[width=3.2cm]{CosDistriMapsNew/MorganFP_ESOL_CosineSim_distribution_regression}
	\label{subfig:ECFPESOLCosDistrib}
}
    \subfigure[GROVER]{
    \includegraphics[width=3.2cm]{CosDistriMapsNew/GROVER_ESOL_CosineSim_distribution_regression}
}
    \subfigure[ChemBert]{
	\includegraphics[width=3.2cm]{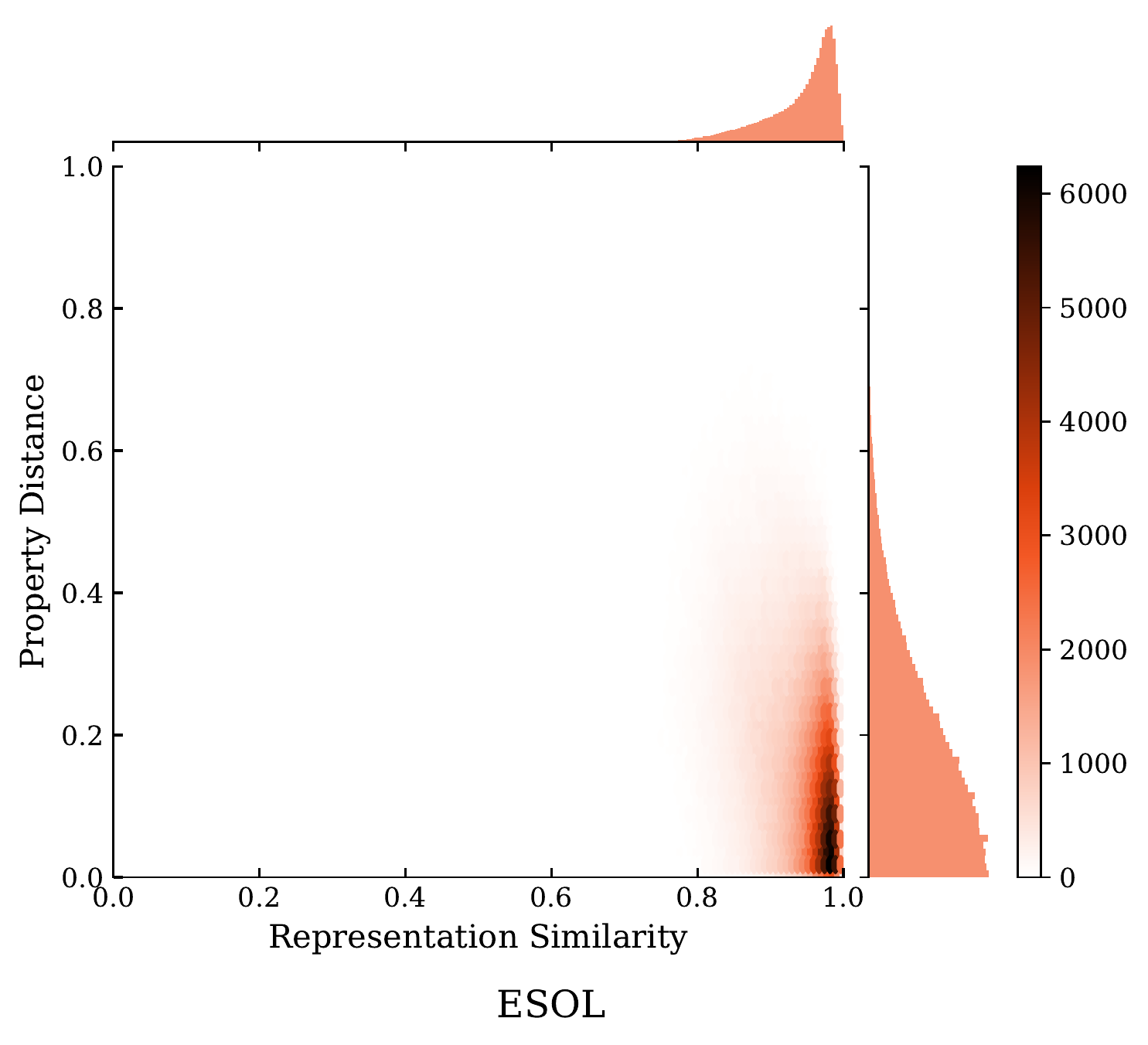}
}
    \subfigure[GraphLoG]{
	\includegraphics[width=3.2cm]{CosDistriMapsNew/GraphLoG_ESOL_CosineSim_distribution_regression}
}
	\subfigure[MAT]{
	\includegraphics[width=3.2cm]{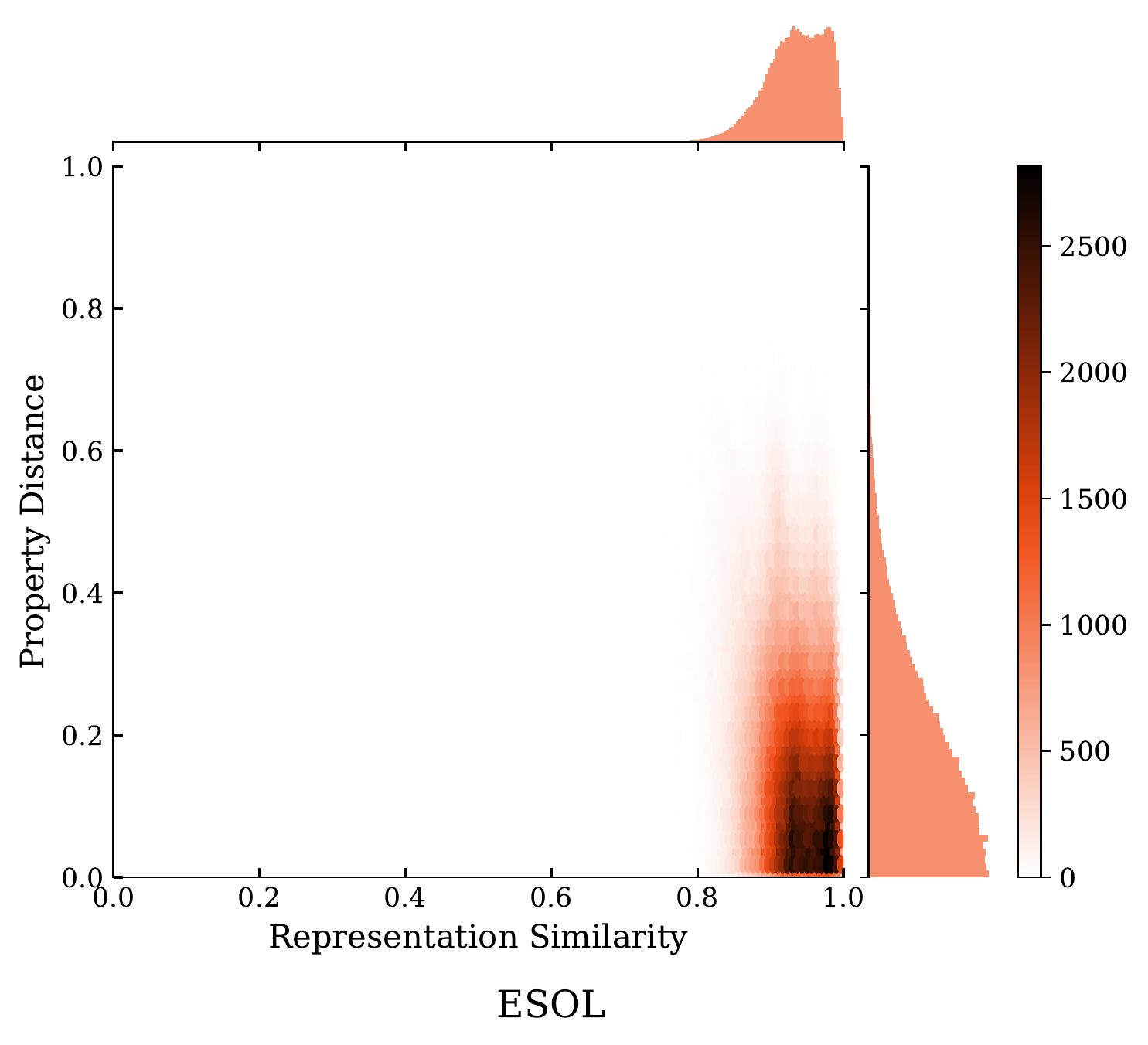}
}
    \subfigure[SMILESTransformer]{
    \includegraphics[width=3.2cm]{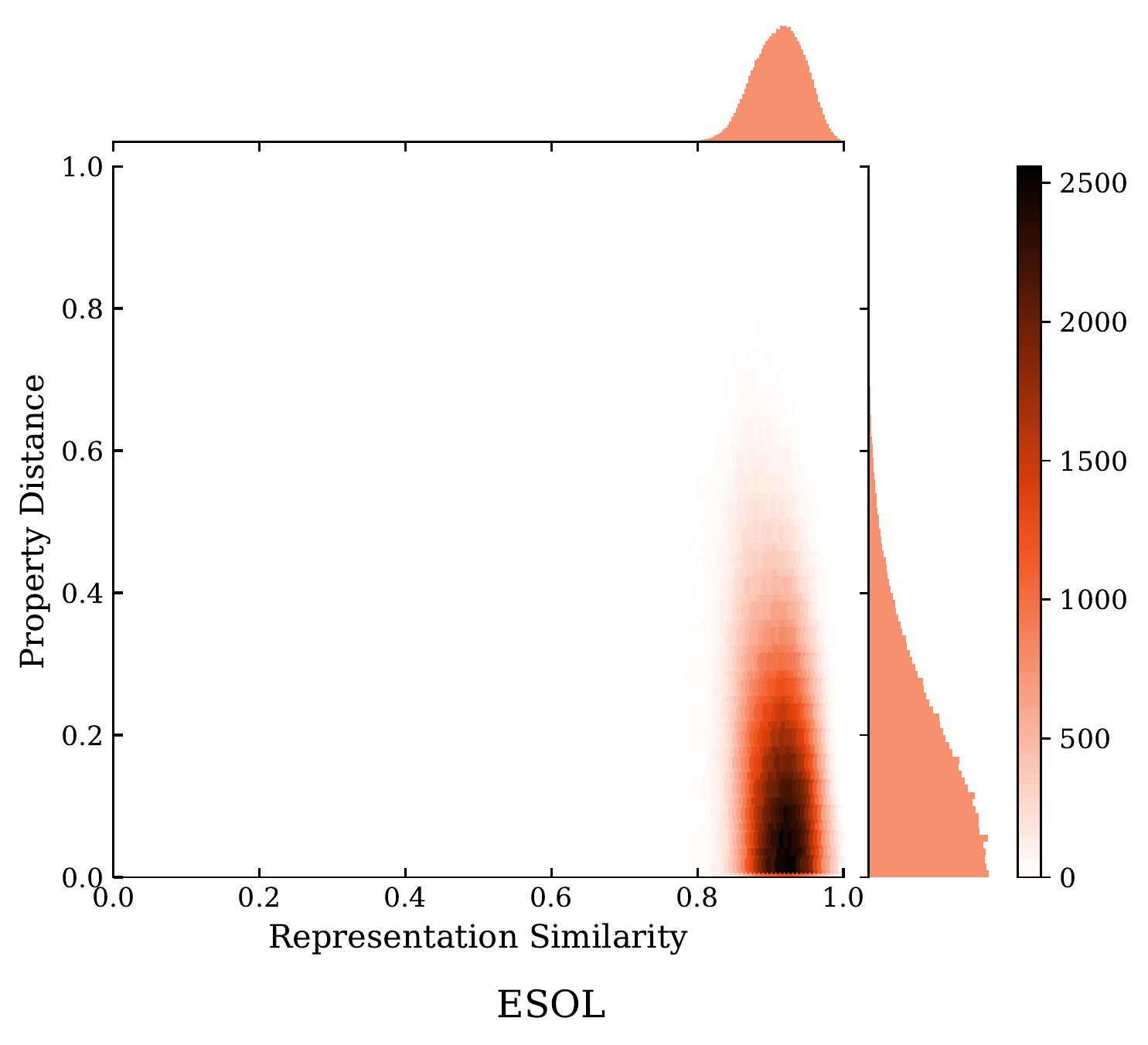}
}
    \subfigure[PretrainGNNs]{
	\includegraphics[width=3.2cm]{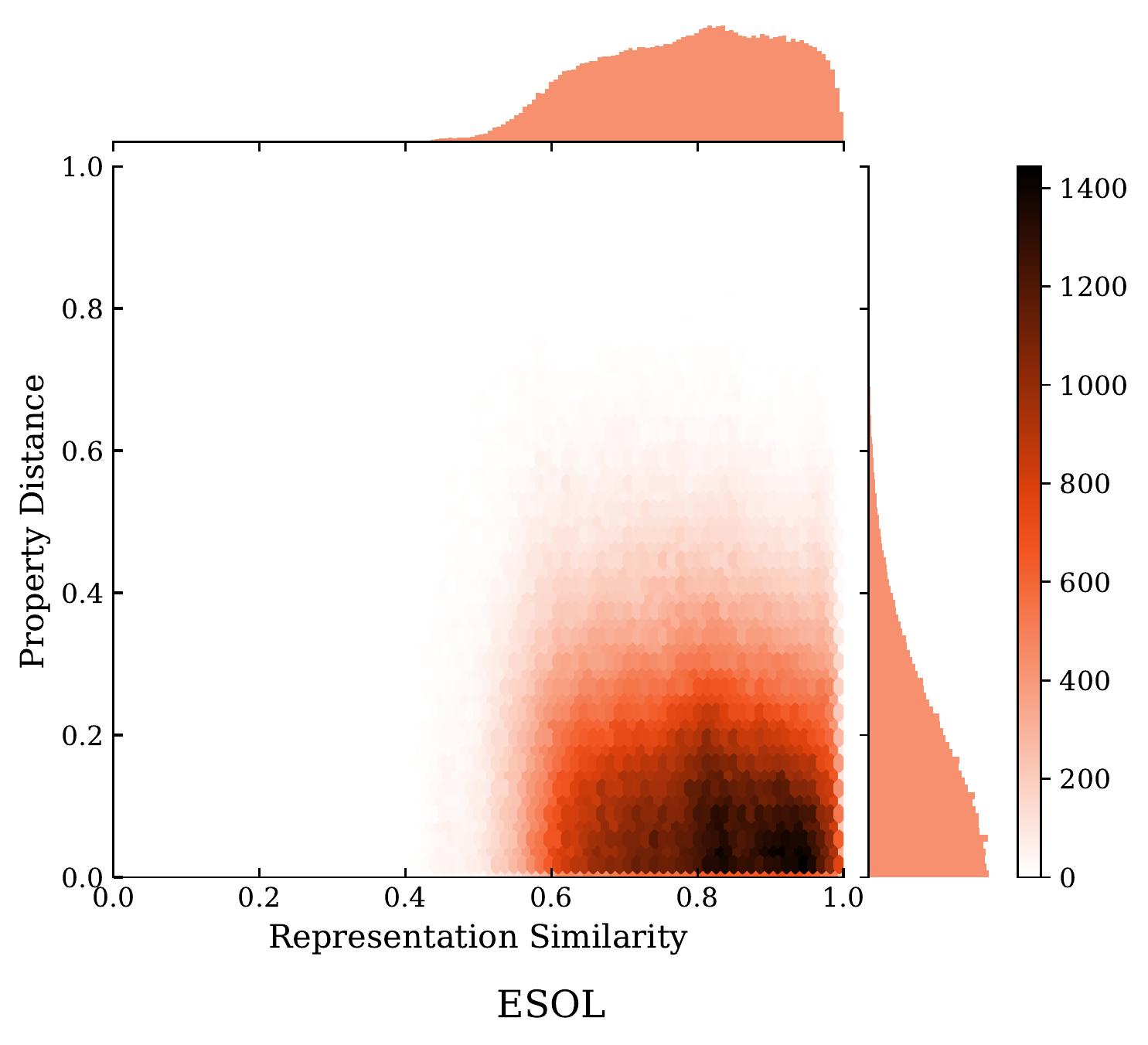}
}
    \subfigure[Pretrain8]{
	\includegraphics[width=3.2cm]{CosDistriMapsNew/Pretrain8_ESOL_CosineSim_distribution_regression}
}

	\caption{Distribution of the molecular similarity on ESOL using CosineSim as metric function.}
	\label{fig:ESOL_Cosine_distribut_all}
\end{figure}

\begin{figure}[thbp]
	\centering
	\subfigure[ECFP]{
	\includegraphics[width=3.2cm]{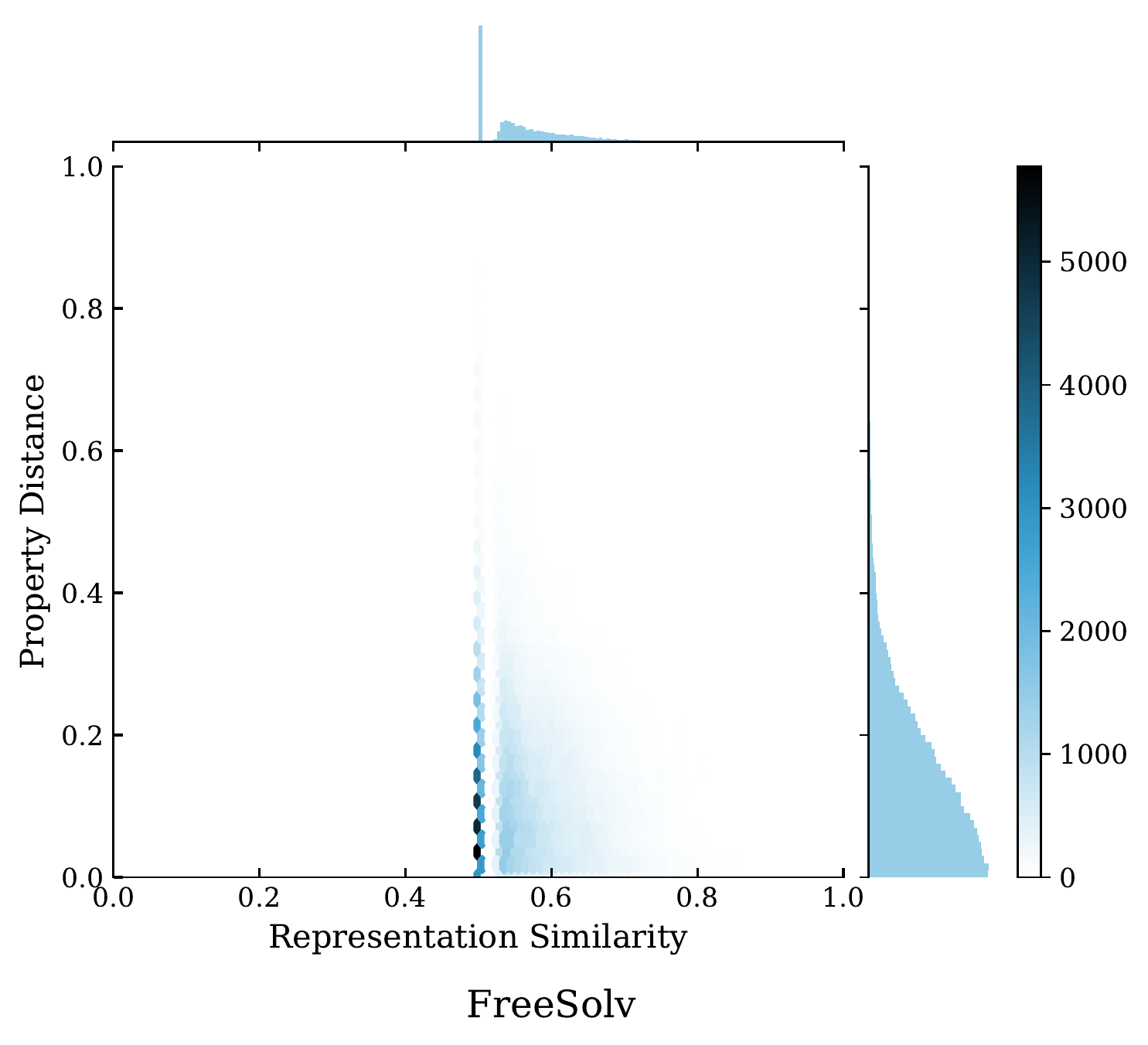}
	\label{subfig:ECFPFreeSolvCosDistrib}

}
    \subfigure[GROVER]{
    \includegraphics[width=3.2cm]{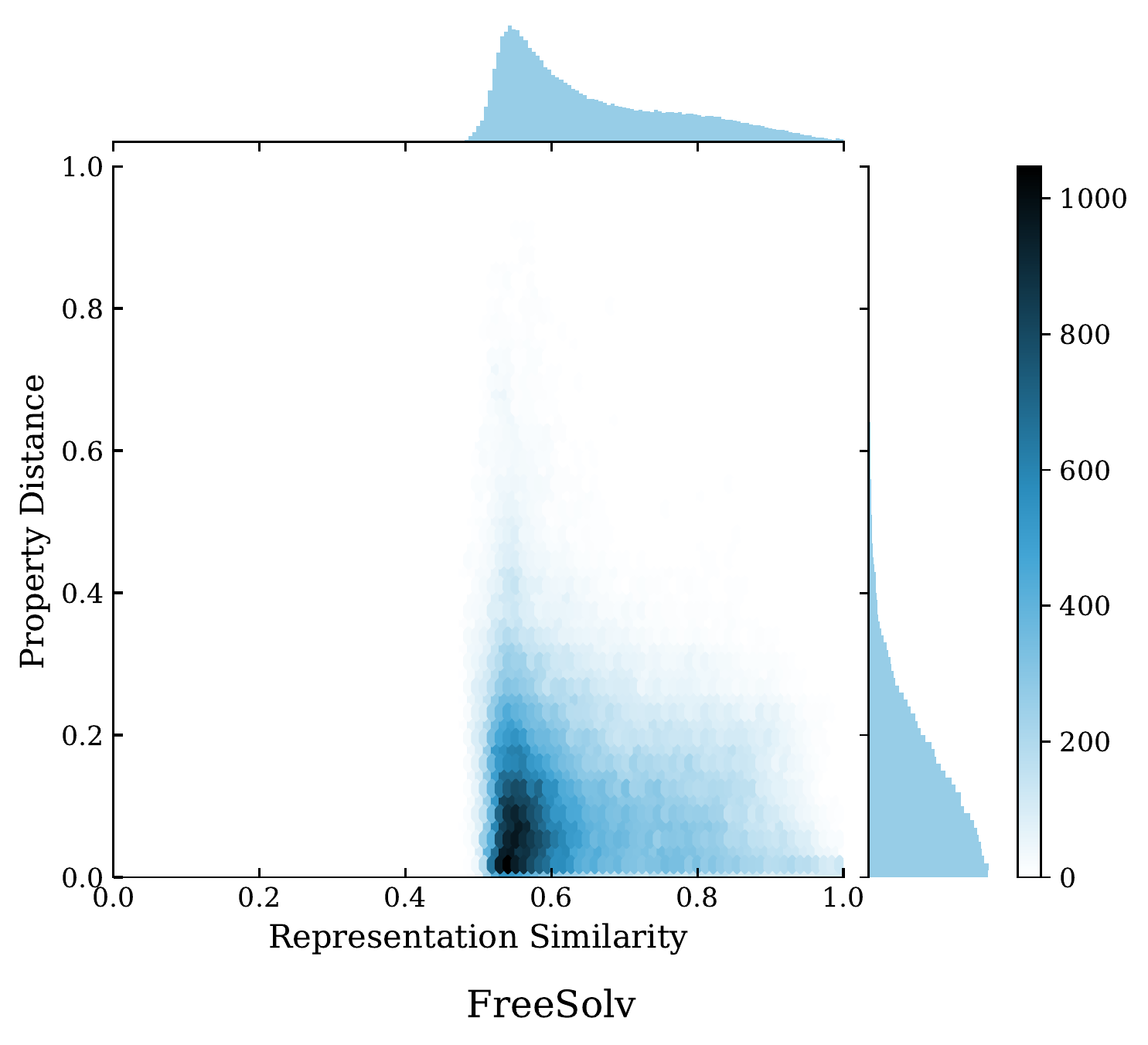}
}
    \subfigure[ChemBert]{
	\includegraphics[width=3.2cm]{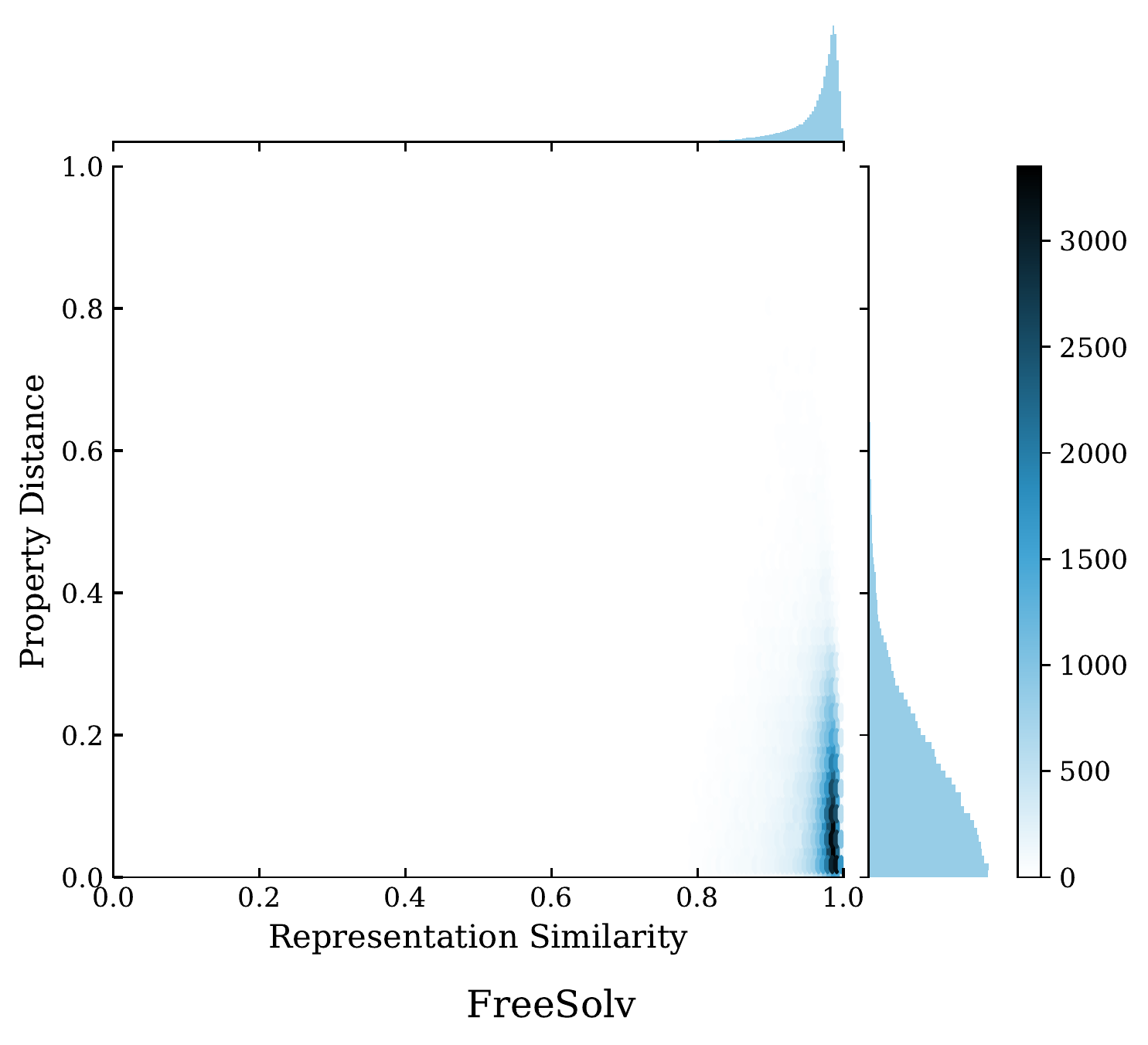}
}
    \subfigure[GraphLoG]{
	\includegraphics[width=3.2cm]{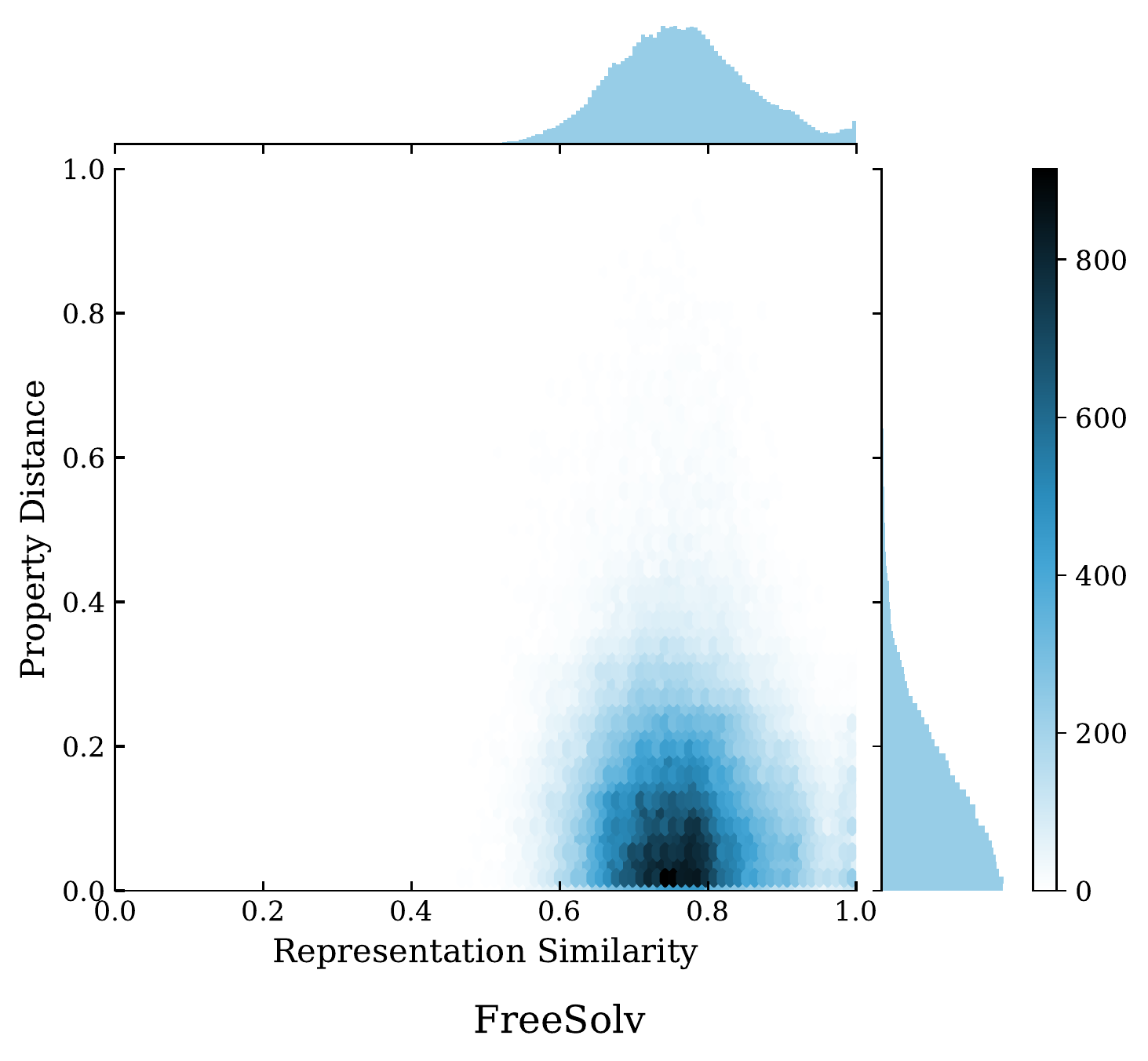}
}
	\subfigure[MAT]{
	\includegraphics[width=3.2cm]{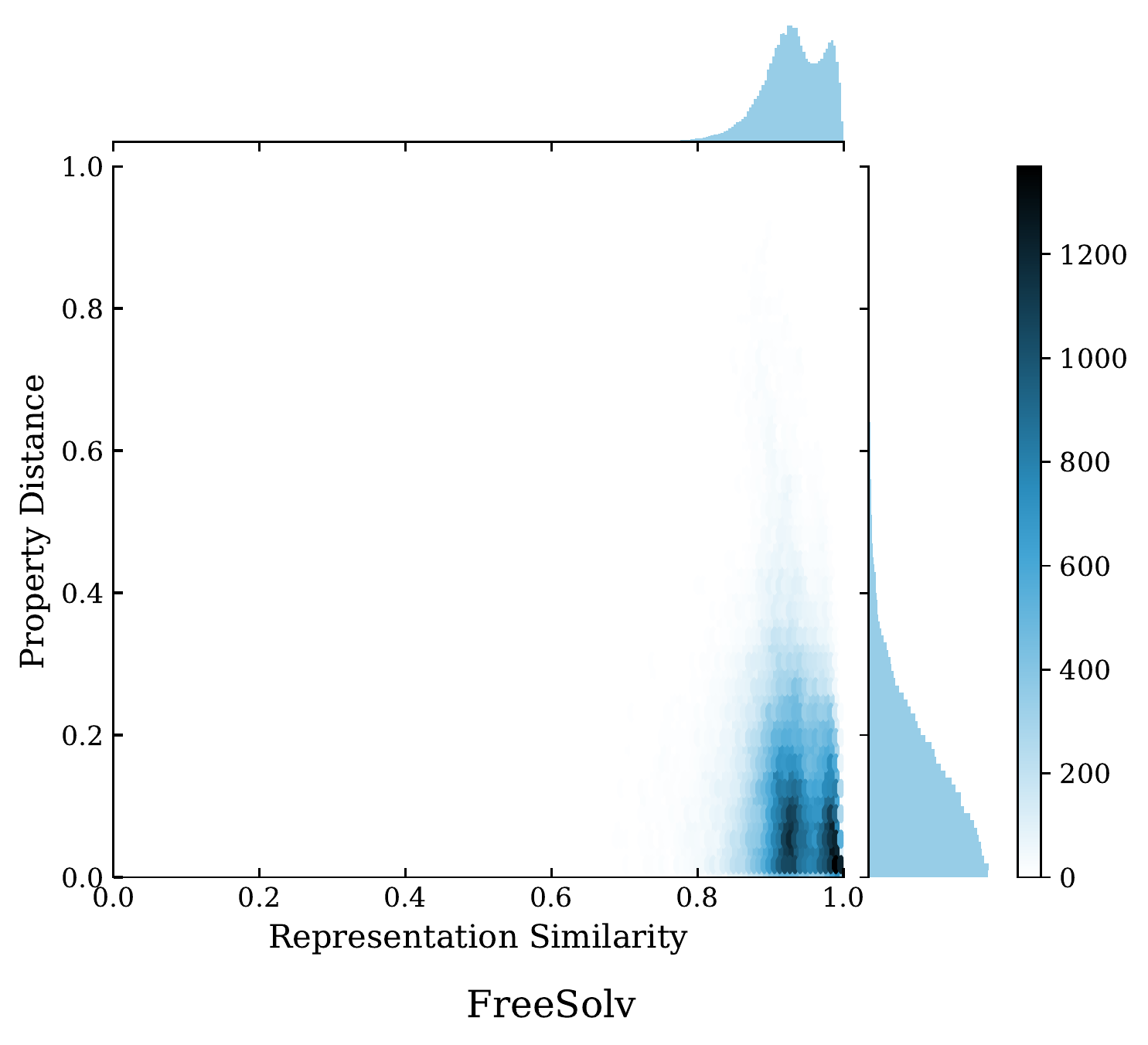}
}
    \subfigure[SMILESTransformer]{
    \includegraphics[width=3.2cm]{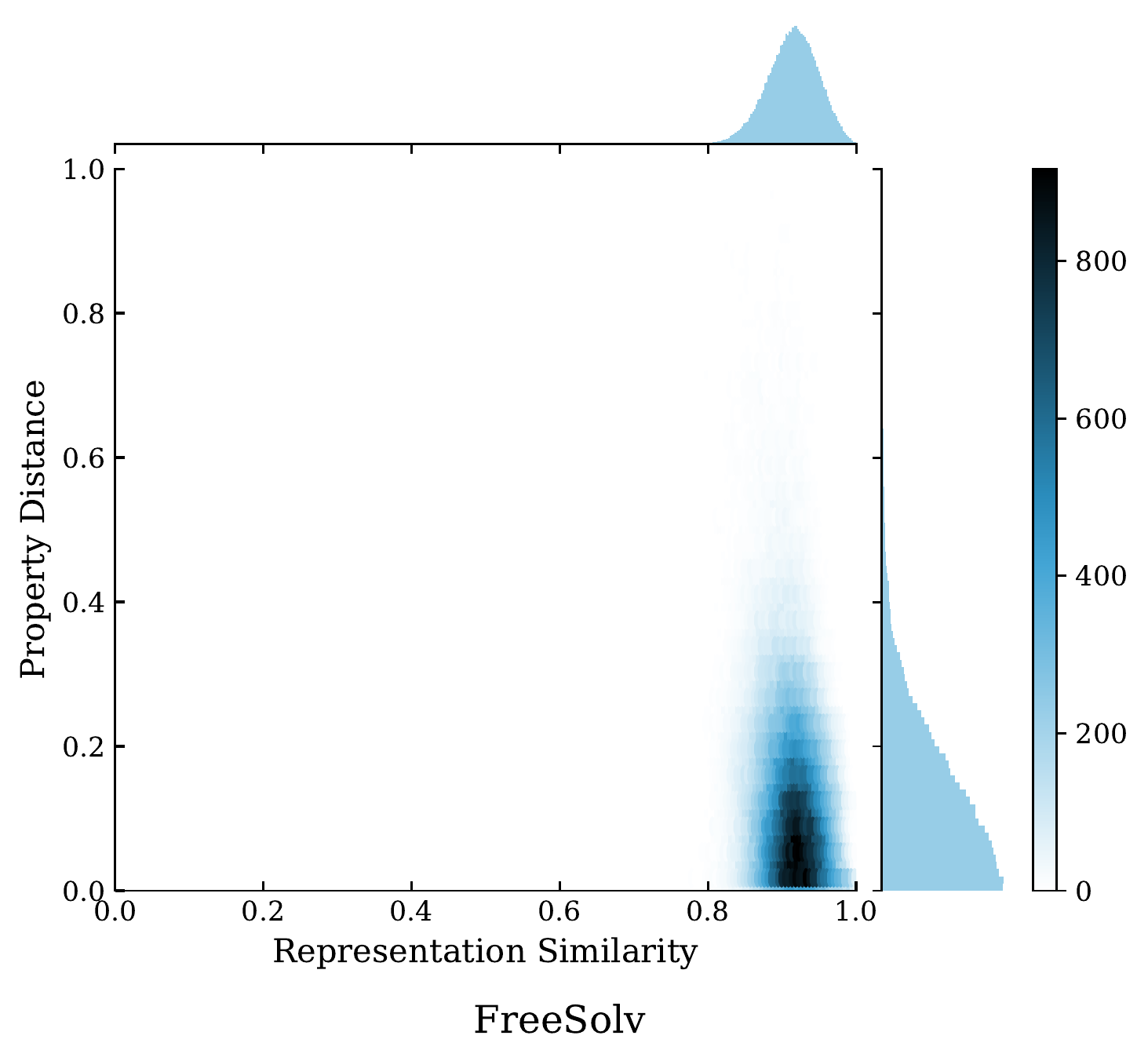}
}
    \subfigure[PretrainGNNs]{
	\includegraphics[width=3.2cm]{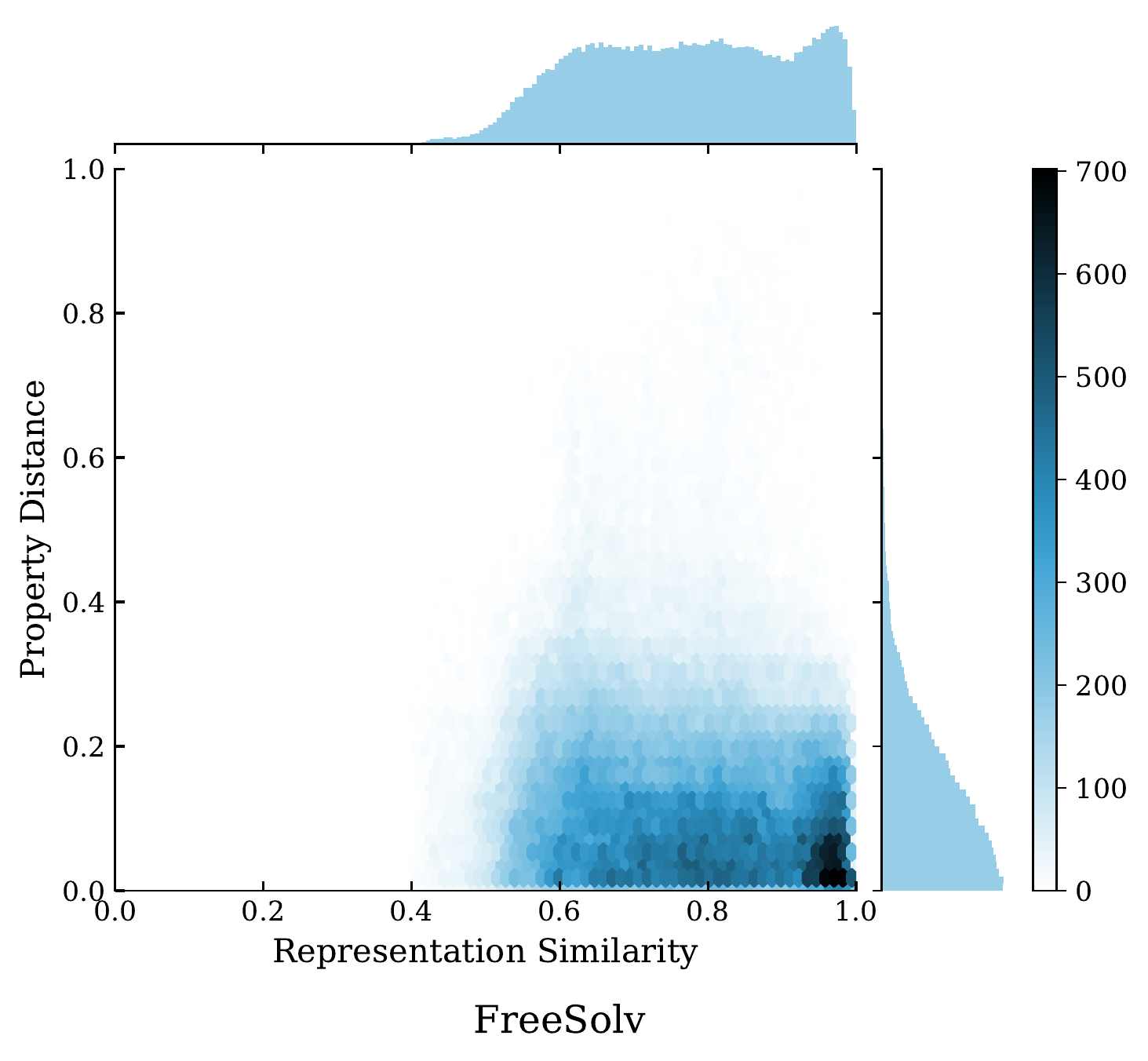}
}
    \subfigure[Pretrain8]{
	\includegraphics[width=3.2cm]{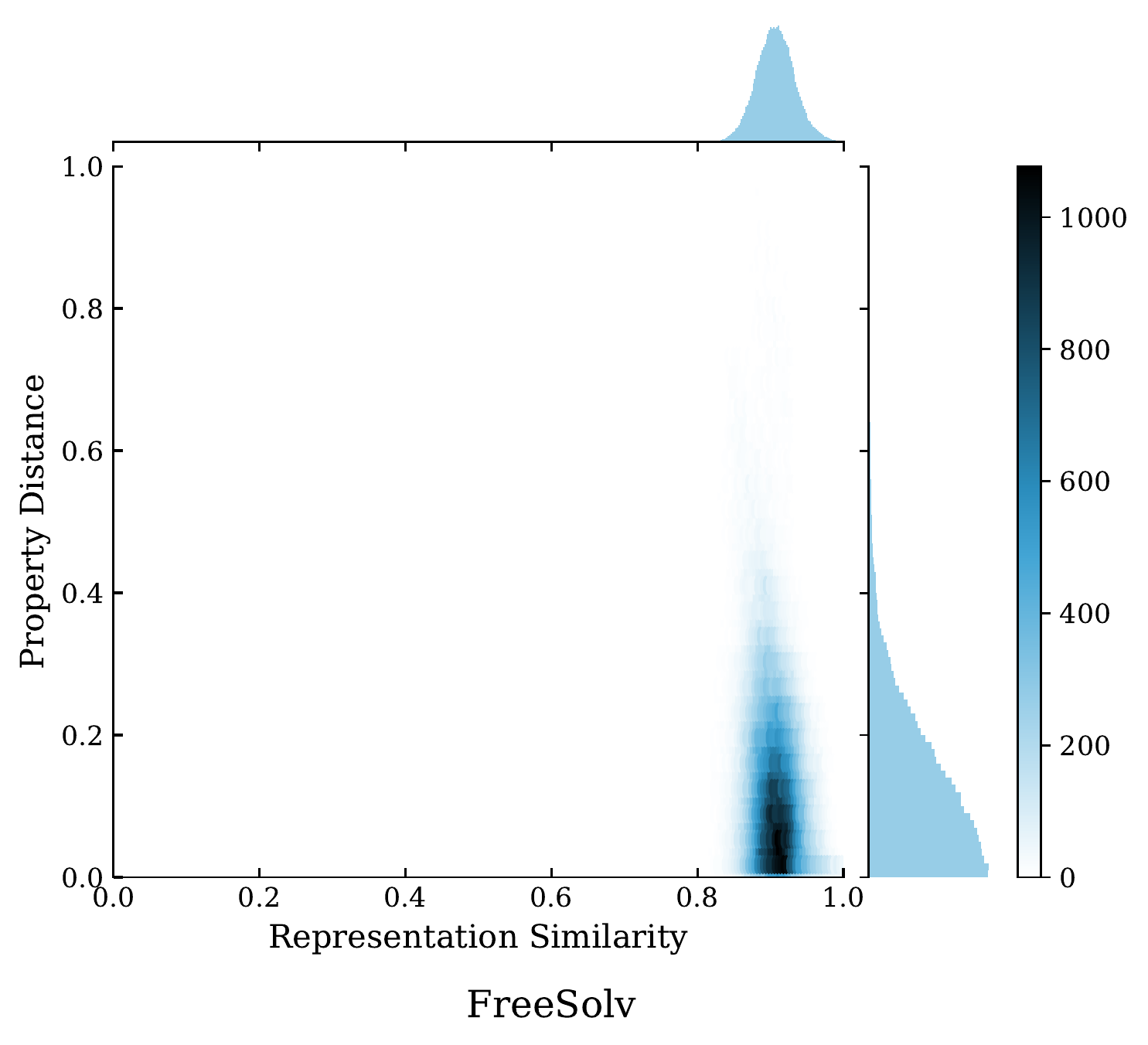}
}

	\caption{Distribution of the molecular similarity on FreeSolv using CosineSim as metric function.}
	\label{fig:FreeSolv_Cosine_distribut}
\end{figure}

As discussed in Appendix~\ref{sec:metrics}, to analysis the information embeded in the angle of two representation vectors, the distribution of molecular similarities with CosineSim as metric function are drawn.
Plots of ESOL and FreeSolv datasets are shown in Fig.~\ref{fig:ESOL_Cosine_distribut_all} and Fig.~\ref{fig:FreeSolv_Cosine_distribut}, respectively.

The basis of the ECFP corresponds to specific molecular substructures, therefore the cosine similarity of ECFPs indicates the proportion of common substructures shared by molecules.
As shown in Fig.~S\ref{subfig:ECFPESOLCosDistrib} and Fig.~S\ref{subfig:ECFPFreeSolvCosDistrib}, most of the molecules in these datasets share no common molecules.
However, this finding cannot be discovered by the distribution of other pre-trained models.
Except for GROVER, distributions of other models are almost normal or with the density concentrating on higher similarity.
It indicates that the relationship between the basis of representation spaces generated by PTMs and the substructures of molecules are not explicit.

\clearpage

\section{Improvement of $C_1$ and $C_4$ of Pre-trained Models over ECFP on ESOL.}

\begin{table}[htbp]
\caption{Improvement of $C_1$ and $C_4$ of pre-trained models over ECFP on ESOL.}
\label{tab:fracofc1c4}
\centering
\begin{tabular}{lcr}
\toprule
Pre-trained Models & $\frac{C_1^{feat}}{C_1^{fp}}$ & $\frac{C_4^{feat}}{C_4^{fp}}$ \\
\midrule
GROVER & 0.982 & 0.941\\
ChemBERT & 0.839 & 0.945\\
GraLoG & 1.028 & 1.019\\
MAT & 0.938 & 0.992 \\
S.T. & 0.978 & 0.841 \\
PreGNNs & 0.964 & 0.997 \\
Pre8 & 1.037 & 0.776\\
\bottomrule
\end{tabular}
\end{table}

\section{Scores for Representation Quality Estimation}
Two scores are proposed to quantitatively evaluate the quality of the representations by different pre-trained models on target tasks.
The $s_{AD}$ scores of ECFP are also provided for reference.
Results are shown in Tab.~\ref{tab:allscores}.
Results of the PTMs with the best score of each dataset are bold.
And the results of ECFP which are better than those of the PTMs are underlined.

\begin{table}[htbp]
\caption{Two proposed scores for representation quality estimation.}
\label{tab:allscores}
\centering
\resizebox{\linewidth}{!}{
\begin{tabular}{lcccccccccr}
\toprule
Tasks & Scores & GROVER & ChemBERT & GraLoG & MAT & S.T. & PreGNNs & Pre8 & ECFP\\
\midrule
BACE & $s_{AD}$ &\textbf{0.067} & 0.086 & 0.071 & 0.191 & 0.073 & 0.073 & 0.076 & \underline{0.051} \\
& $s_{IR}$ & 2.393 & 2.341 & 2.426 & 2.452 & 2.302 & 2.495 & \textbf{2.301} & - \\
\midrule
BBBP & $s_{AD}$ & 0.065 & 0.098 & 0.068 & 0.108 & \textbf{0.037} & 0.104 & 0.044 & \underline{0.030} \\
& $s_{IR}$ & 3.054 & 2.767 & 3.595 & 3.172 & 2.045 & 3.177 & \textbf{2.035} & - \\
\midrule
ClinTox & $s_{AD}$ & 0.107 & 0.104 & 0.077 & 0.144 & \textbf{0.057} & 0.118 & 0.083 & \underline{0.045} \\
& $s_{IR}$ & 3.354 & 2.546 & 3.383 & 3.163 & \textbf{1.951} & 3.069 & 3.081 & -\\
\midrule
ESOL & $s_{AD}$ & 0.019 & 0.017 & 0.021 & 0.019 & 0.016 & 0.021 & \textbf{0.014} & 0.019\\
& $s_{IR}$ & 1.923 & \textbf{1.784} & 2.047 & 1.930 & 1.819 & 1.960 & 1.813 & - \\
\midrule
FreeSolv & $s_{AD}$ & 0.018 & 0.020 & 0.020 & 0.019 & 0.016 & 0.019 & \textbf{0.014} & 0.016\\
& $s_{IR}$ & 2.106 & 2.040 & 2.126 & 2.018 & 1.896 & 2.058 & \textbf{1.849} & - \\
\midrule
Lipo & $s_{AD}$ & 0.016 & 0.024 & 0.015 & 0.025 & 0.015 & \textbf{0.013} & 0.014 & \underline{0.013}\\
& $s_{IR}$ & 1.975 & 2.119 & 1.993 & 2.367 & 1.988 & 1.983 & \textbf{1.935} & -\\
\midrule
QM7 & $s_{AD}$ & 0.012 & \textbf{0.008} & 0.017 & 0.016 & 0.010 & 0.017 & 0.011 & 0.014\\
& $s_{IR}$ & 1.816 & \textbf{1.408} & 2.024 & 1.936 & 1.666 & 1.971 & 1.734 &-\\
\midrule
QM8 & $s_{AD}$ & 0.010 & 0.012 & 0.013 & 0.015 & 0.012 & 0.013 & \textbf{0.010} & \underline{0.010} \\
& $s_{IR}$ & 1.729 & 2.027 & 2.001 & 2.077 & 1.897 & 1.971 & \textbf{1.719} & - \\
\midrule
SIDER & $s_{AD}$ & 0.047 & 0.091 & \textbf{0.025} & 0.092 & 0.082 & 0.052 & 0.043 & 0.081\\
& $s_{IR}$ & 1.248 & 1.526 & 1.252 & 1.595 & 1.547 & 1.437 & \textbf{1.247} & - \\
\midrule
Tox21 & $s_{AD}$ & 0.108 & 0.071 & 0.097 & 0.093 & \textbf{0.054} & 0.075 & 0.062 & \underline{0.031}\\
& $s_{IR}$ & 3.731 & \textbf{2.657} & 3.938 & 3.287 & 2.750 & 3.226 & 3.458 & -\\


\bottomrule
\end{tabular}
}
\end{table}

\clearpage
\section{Evolution of Representations during Finetuning PTMs}

\begin{figure}[thbp]
	\centering
	\subfigure[Epoch 0]{
	\includegraphics[width=4cm]{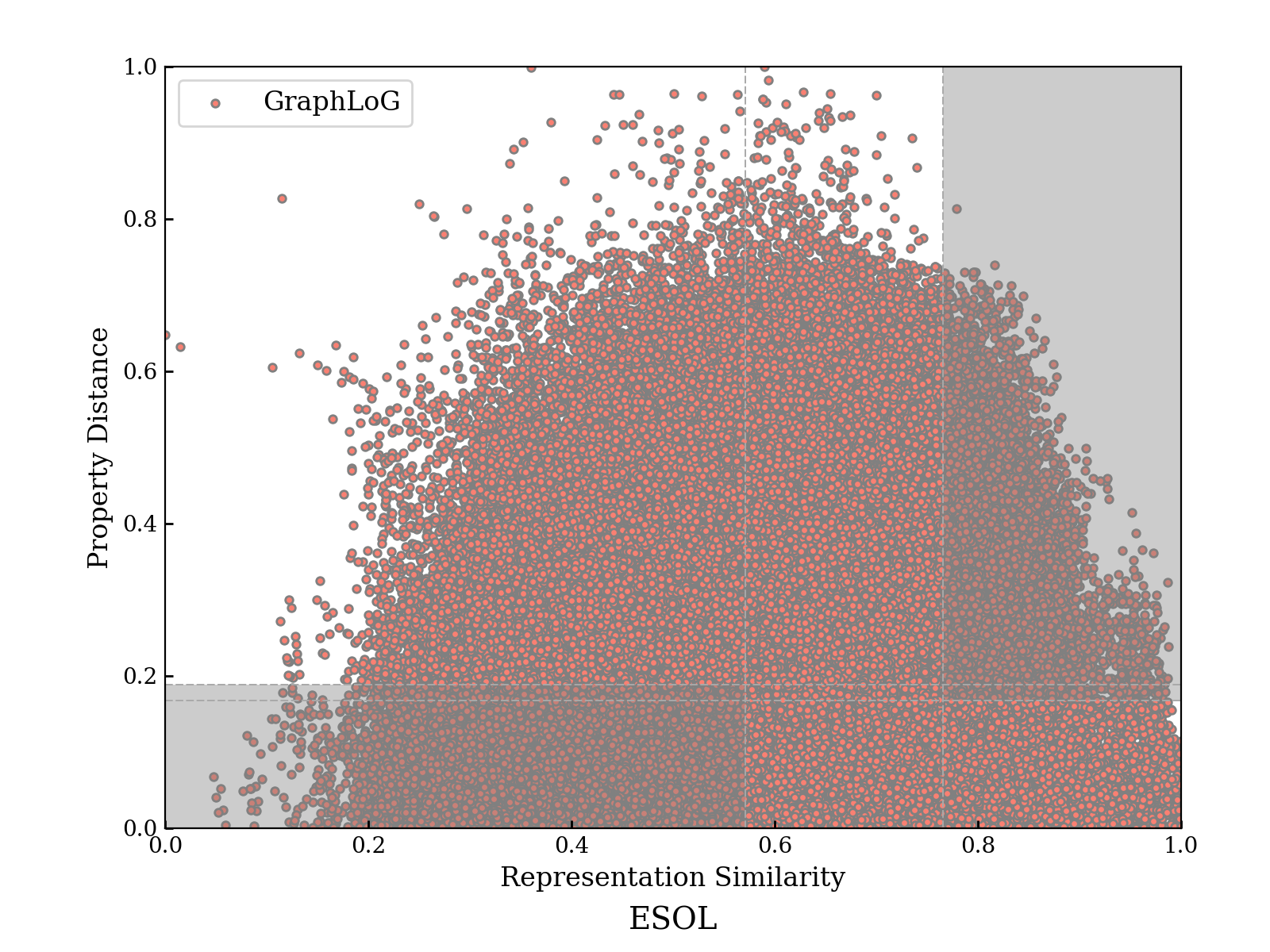}
}
    \subfigure[Epoch 1]{
    \includegraphics[width=4cm]{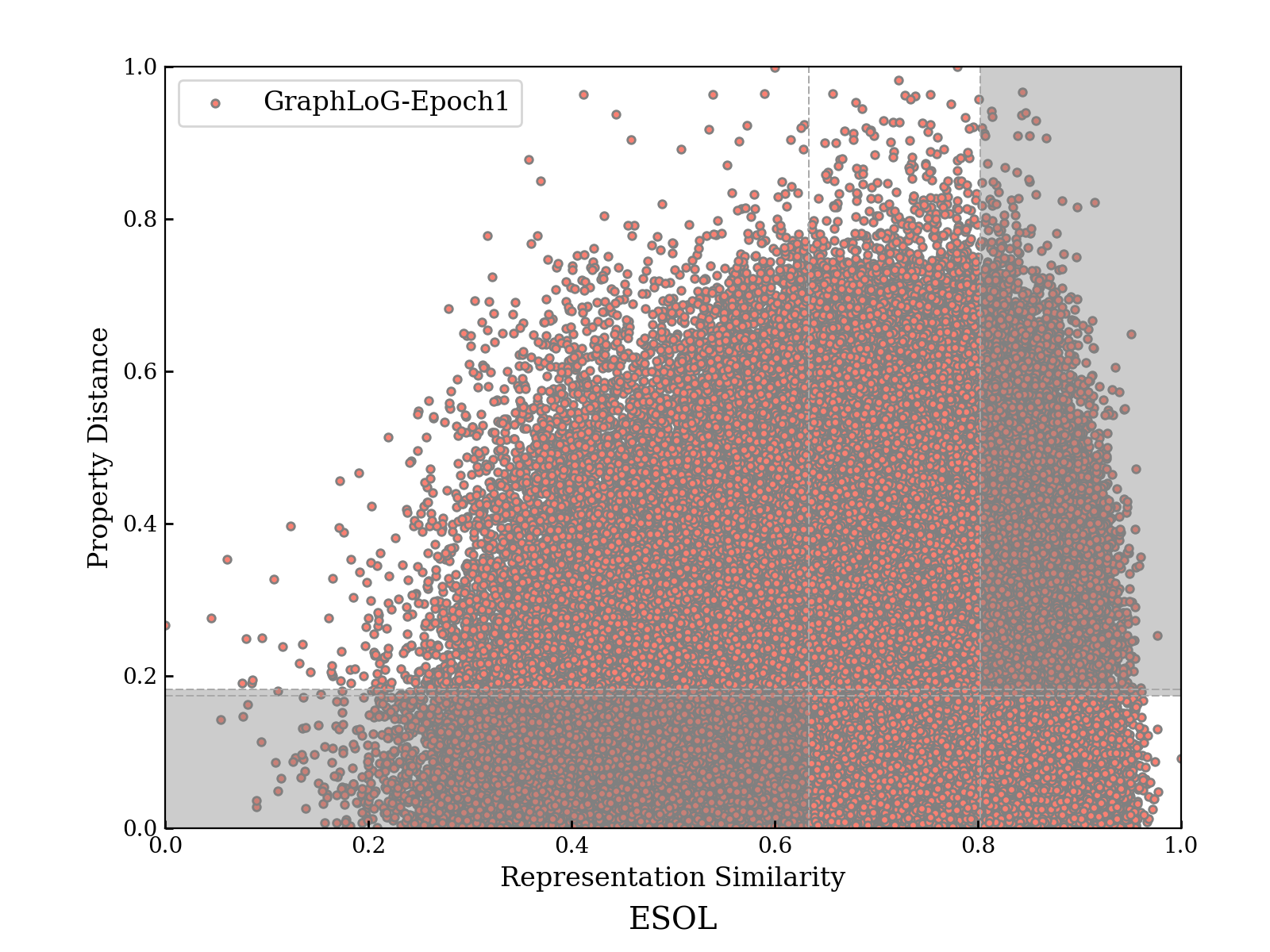}
}
    \subfigure[Epoch 2]{
	\includegraphics[width=4cm]{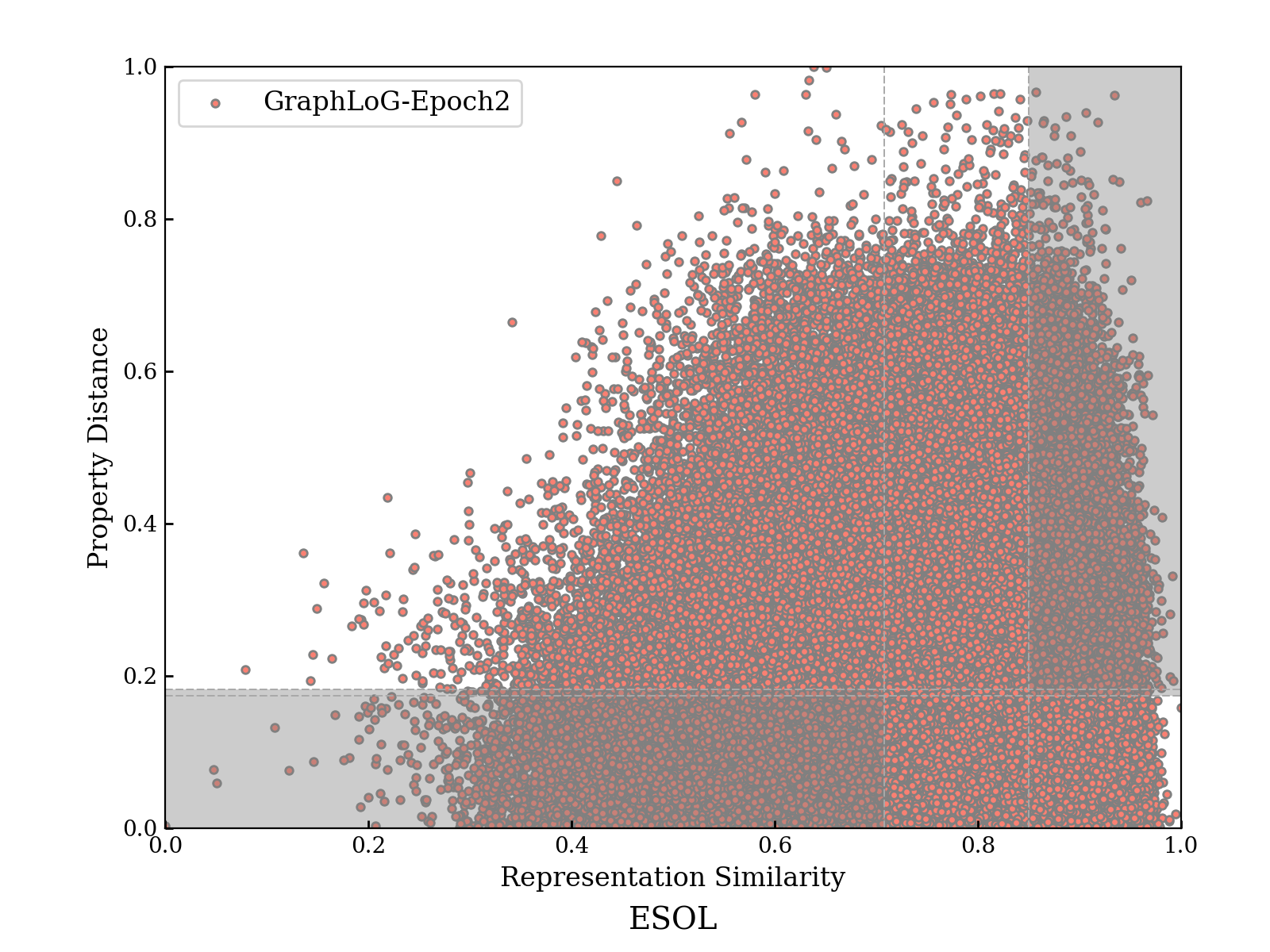}
}
    \subfigure[Epoch 5]{
	\includegraphics[width=4cm]{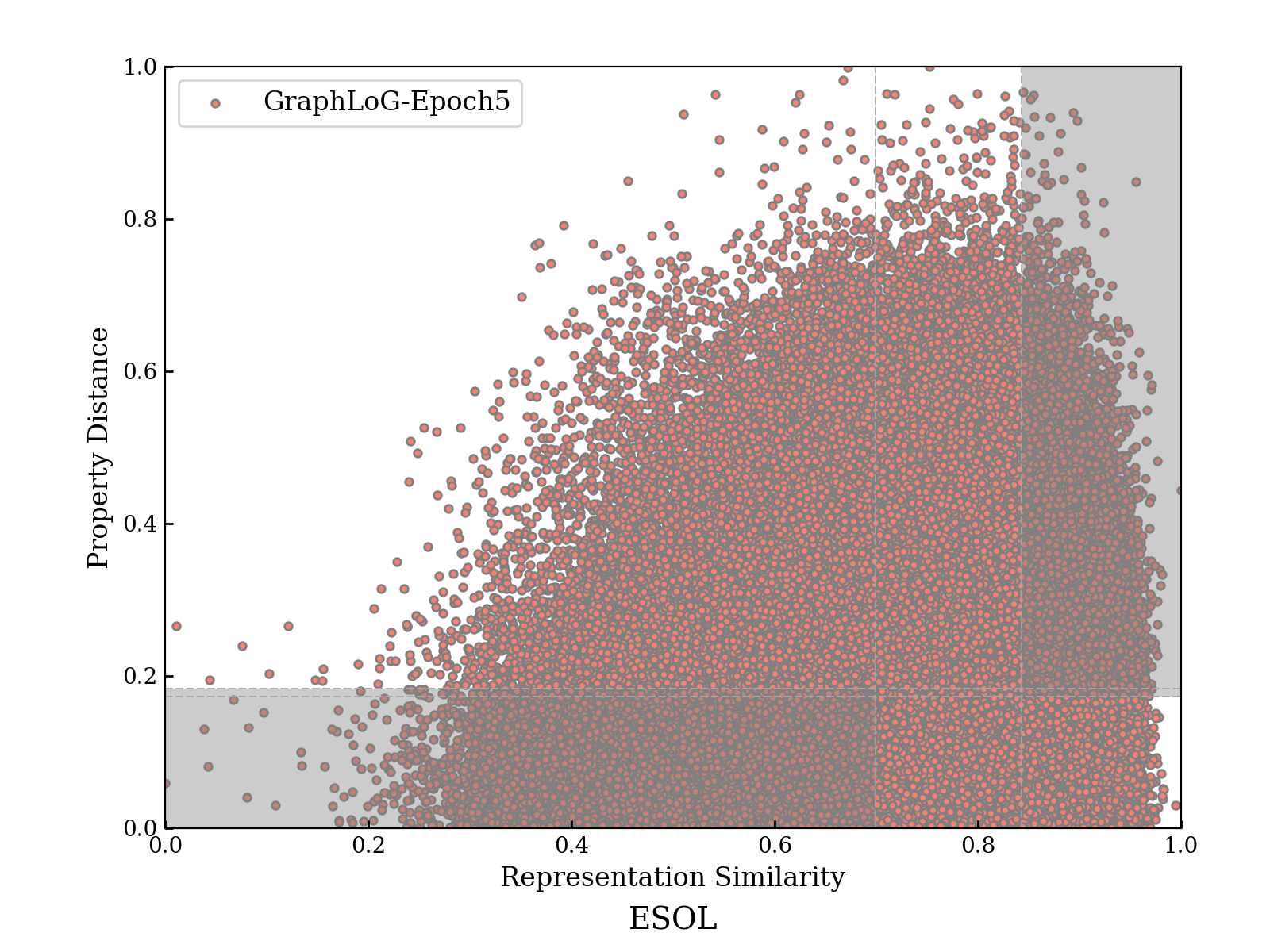}
}
	\subfigure[Epoch 10]{
	\includegraphics[width=4cm]{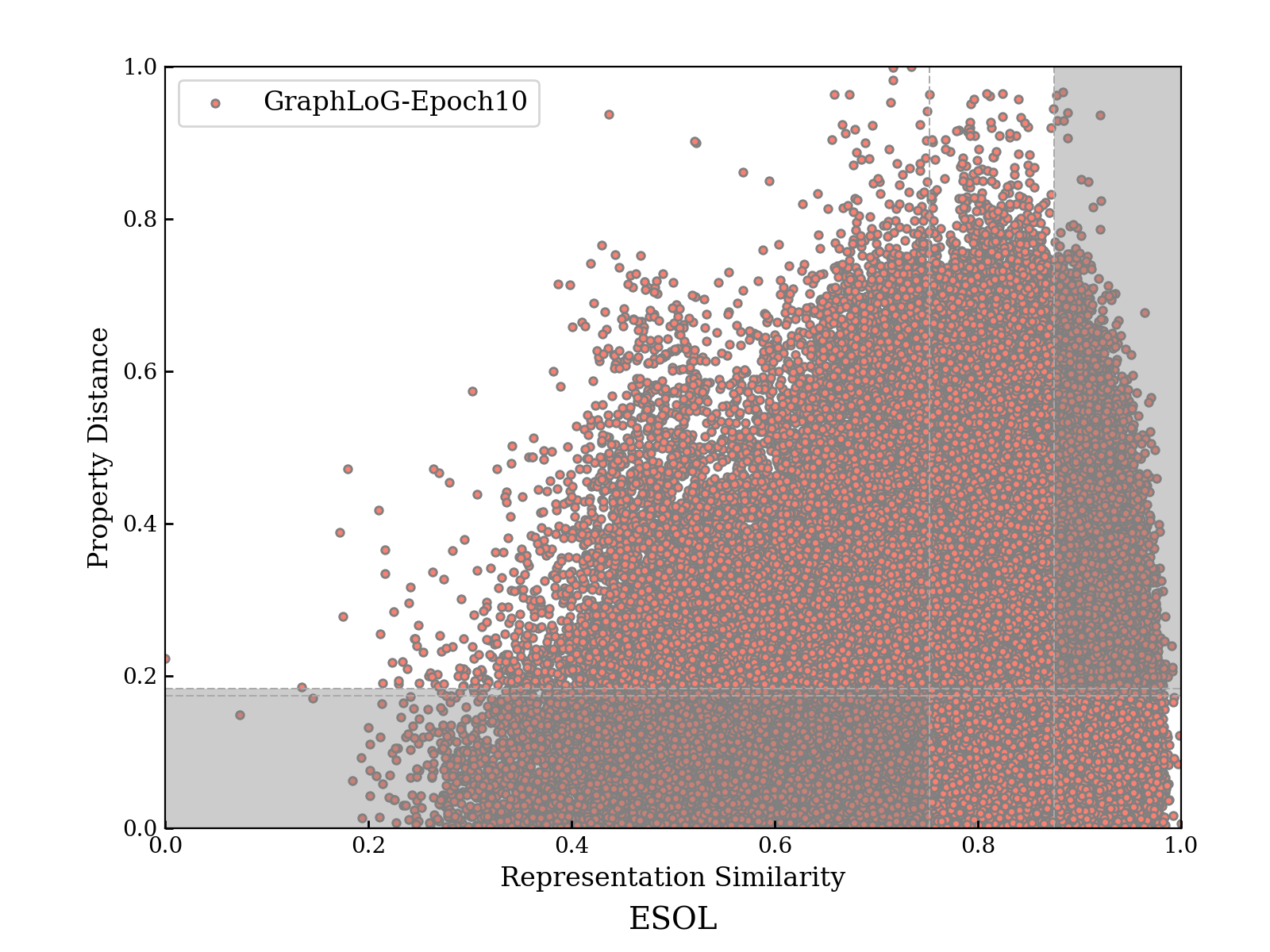}
}
    \subfigure[Epoch 50]{
    \includegraphics[width=4cm]{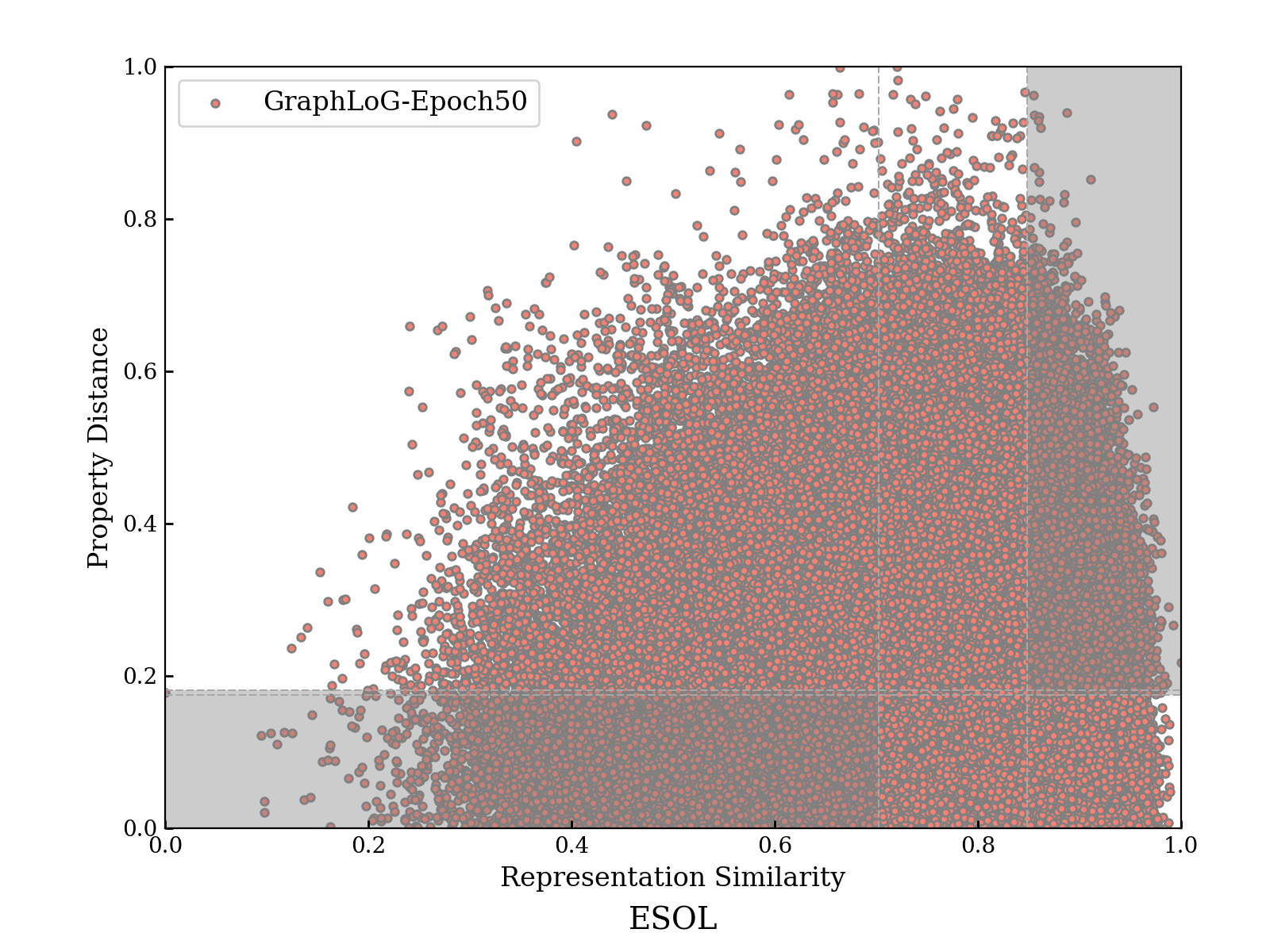}
}
    \subfigure[Epoch 100]{
	\includegraphics[width=4cm]{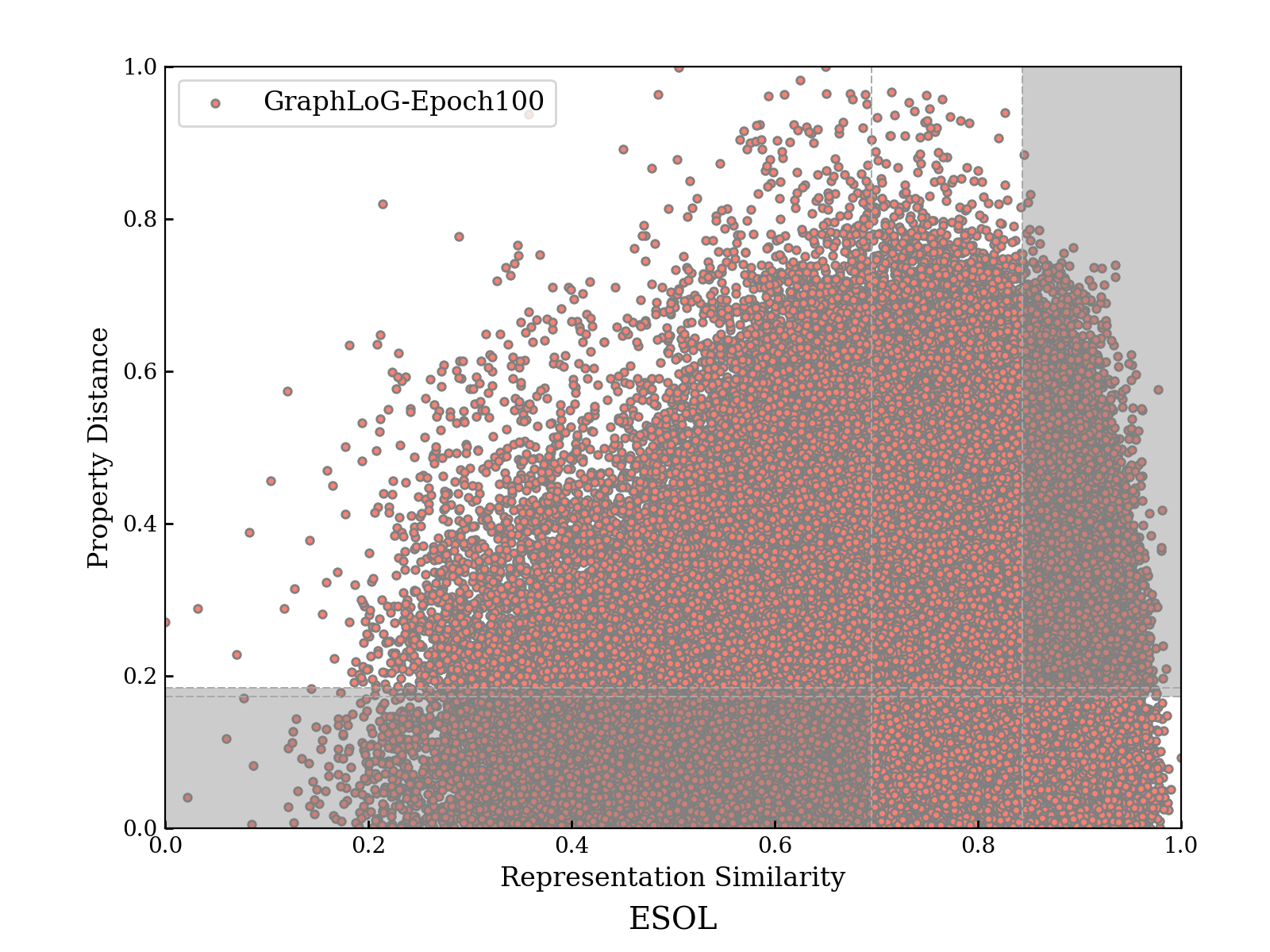}
}

	\caption{RPSMaps of representations generated by finetuned GraphLoG model on ESOL dataset.}
	\label{fig:GraphLoG-Epoch-ESOL}
\end{figure}

Although in this work we concentrate on the representations generated by a \textbf{frozen} PTM.
However, one may be curious about how the representation-property relationship evolves if the PTM is finetuned with a small learning rate on downstream target tasks.
In this section, we have finetuned a GraphLoG model on the ESOL dataset, and the representations generated by GraphLoG after different training epochs are evaluated by the RePRA.
The RPSMaps are shown in Fig.~\ref{fig:GraphLoG-Epoch-ESOL}, and the scores $s_{AD}$ and $s_{IR}$ are presented in Tab.~\ref{tab:GraphLoG-Epoch-ESOL}

It is clearly revealed by the scores $s_{AD}$ and $s_{IR}$ in Tab.~\ref{tab:GraphLoG-Epoch-ESOL} that with the GraphLoG model finetuned on downstream target dataset, the relationship between the generated representations and the target task property is gradually improved.
This finding is interesting.
The RMSE is used as the loss function for downstream target task, which is just a sample-wise loss function and is not designed to regularize the ACs and SH phenomena between representations.
However, by finetuning the PTM with such sample-wise loss function, the pair-wise relationship between representations will be improved implicitly.
It indicates that the extra information provided by the gradient back-propagated from the task layers will indeed guide the PTM to generate better representations.
In addition, it raises a new question: whether a pair-wise loss function with regularization terms for ACs and SH will lead to better representations and benefit the downstream prediction.

\begin{table}[htbp]
\caption{$s_{AD}$ and $s_{IR}$ scores of representations generated by finetuned GraphLoG model on ESOL dataset.}
\label{tab:GraphLoG-Epoch-ESOL}
\centering
\begin{tabular}{lcr}
\toprule
Finetuning Epochs & $s_{AD}$ & $s_{IR}$ \\
\midrule
Epoch 0 & 0.02069 & 2.04692 \\
Epoch 1 & 0.01870 & 1.97499\\
Epoch 2 & 0.01669 & 1.97397\\
Epoch 5 & 0.01687 & 1.96440\\
Epoch 10 & 0.01470 & 1.92687\\
Epoch 50 & 0.01638 & 1.91342\\
Epoch 100 & 0.01677 & 1.93319\\
\bottomrule
\end{tabular}
\end{table}

\clearpage
\section{Influence of the ECFP radius}
\begin{figure}[thbp]
	\centering
	\subfigure[ECFP4]{
	\includegraphics[width=4cm]{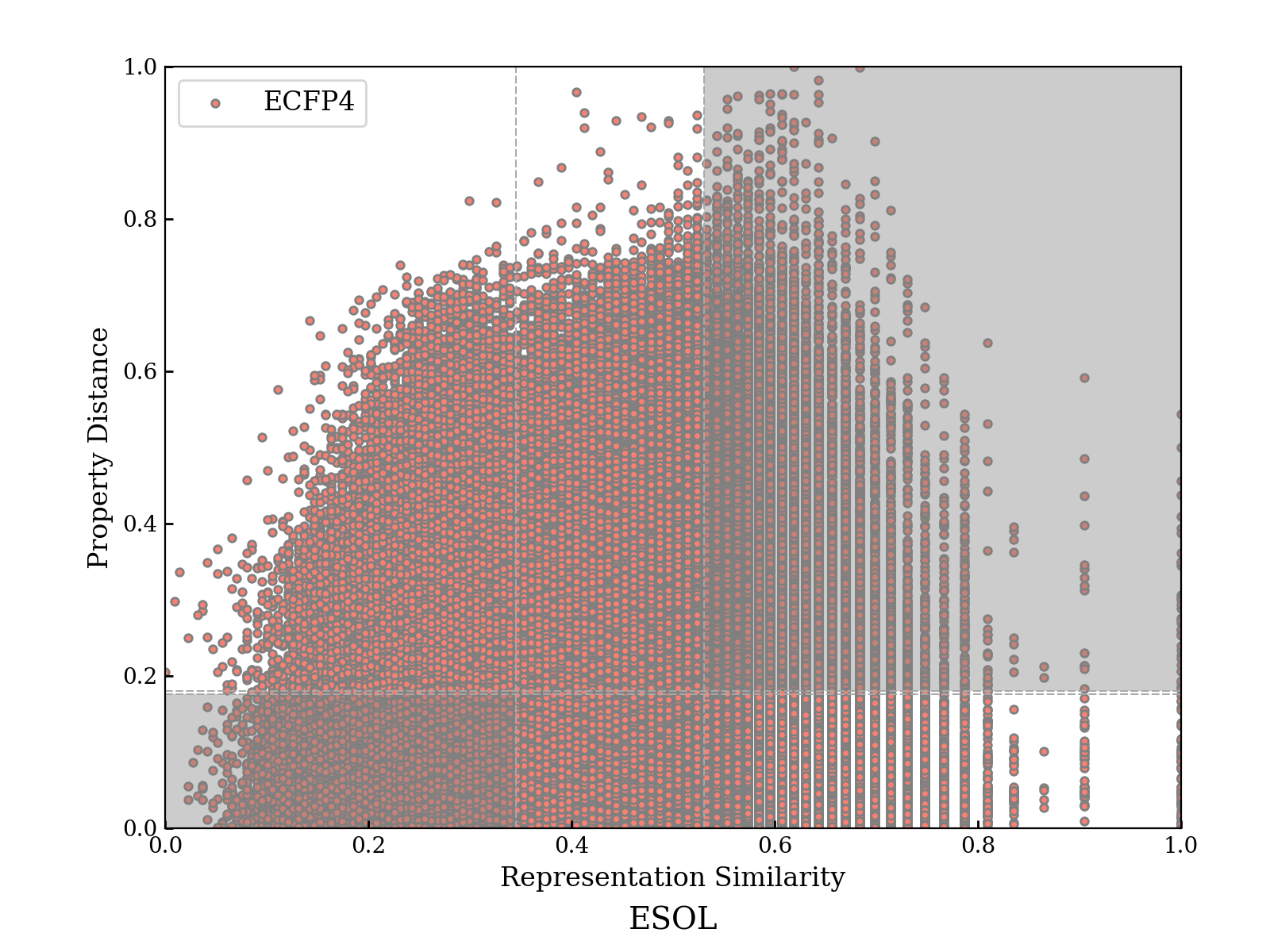}
}
    \subfigure[ECFP6]{
    \includegraphics[width=4cm]{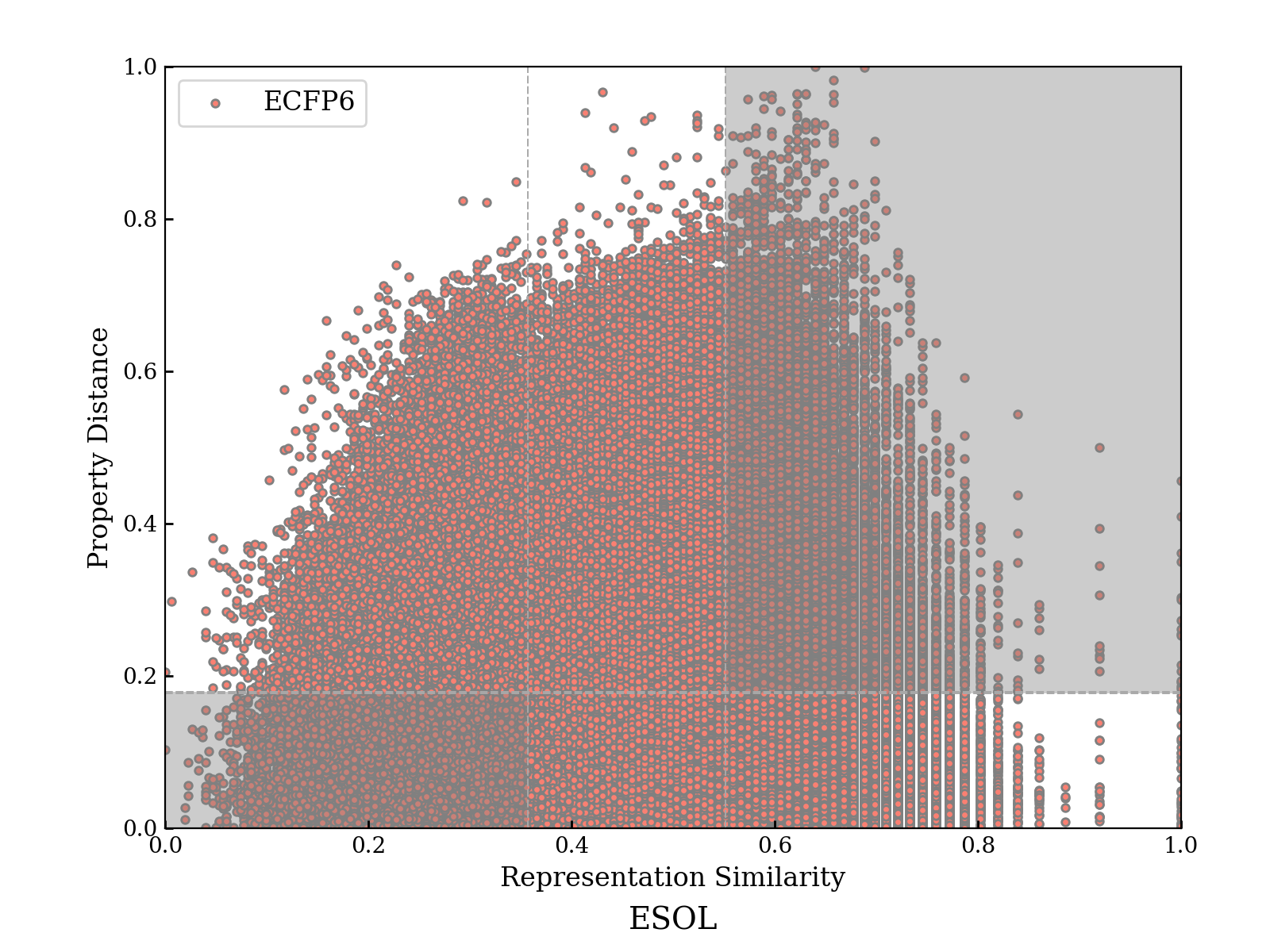}
}
    \subfigure[ECFP8]{
	\includegraphics[width=4cm]{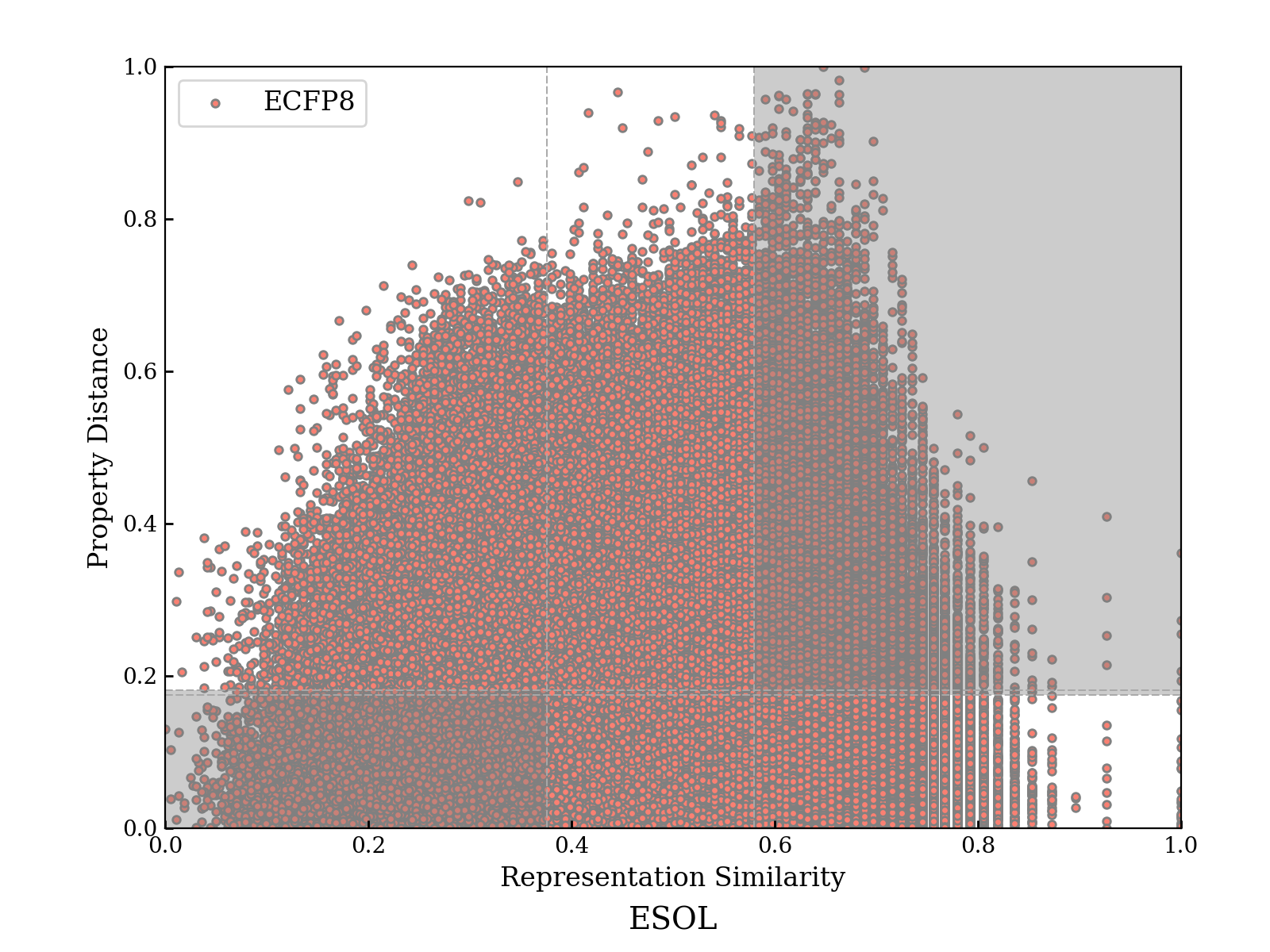}
}

	\caption{RPSMaps of ECFPs with different radius on ESOL dataset.}
	\label{fig:ECFP-radius-ESOL}
\end{figure}

\begin{figure}[thbp]
	\centering
	\subfigure[ECFP4]{
	\includegraphics[width=4cm]{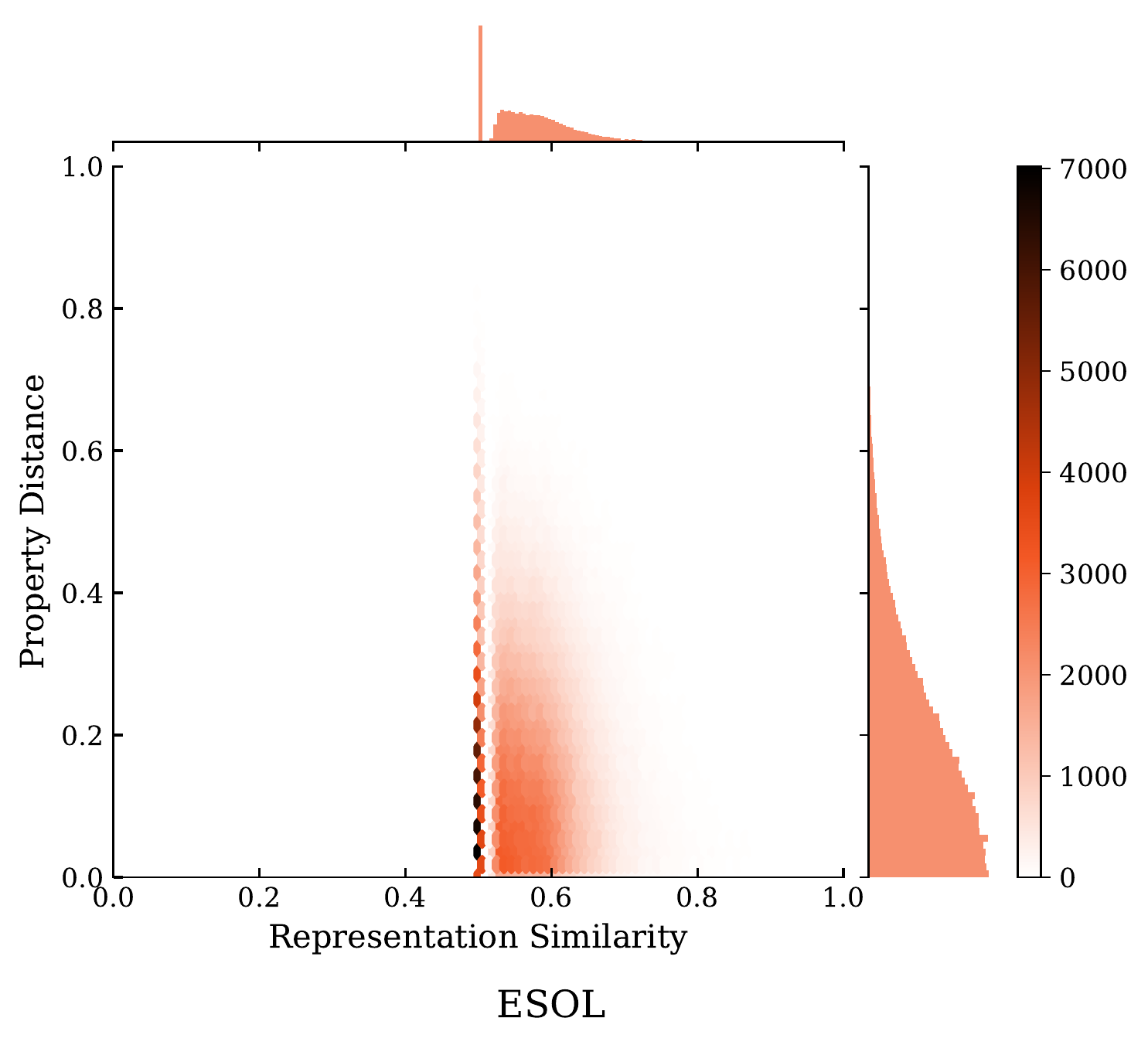}
}
    \subfigure[ECFP6]{
    \includegraphics[width=4cm]{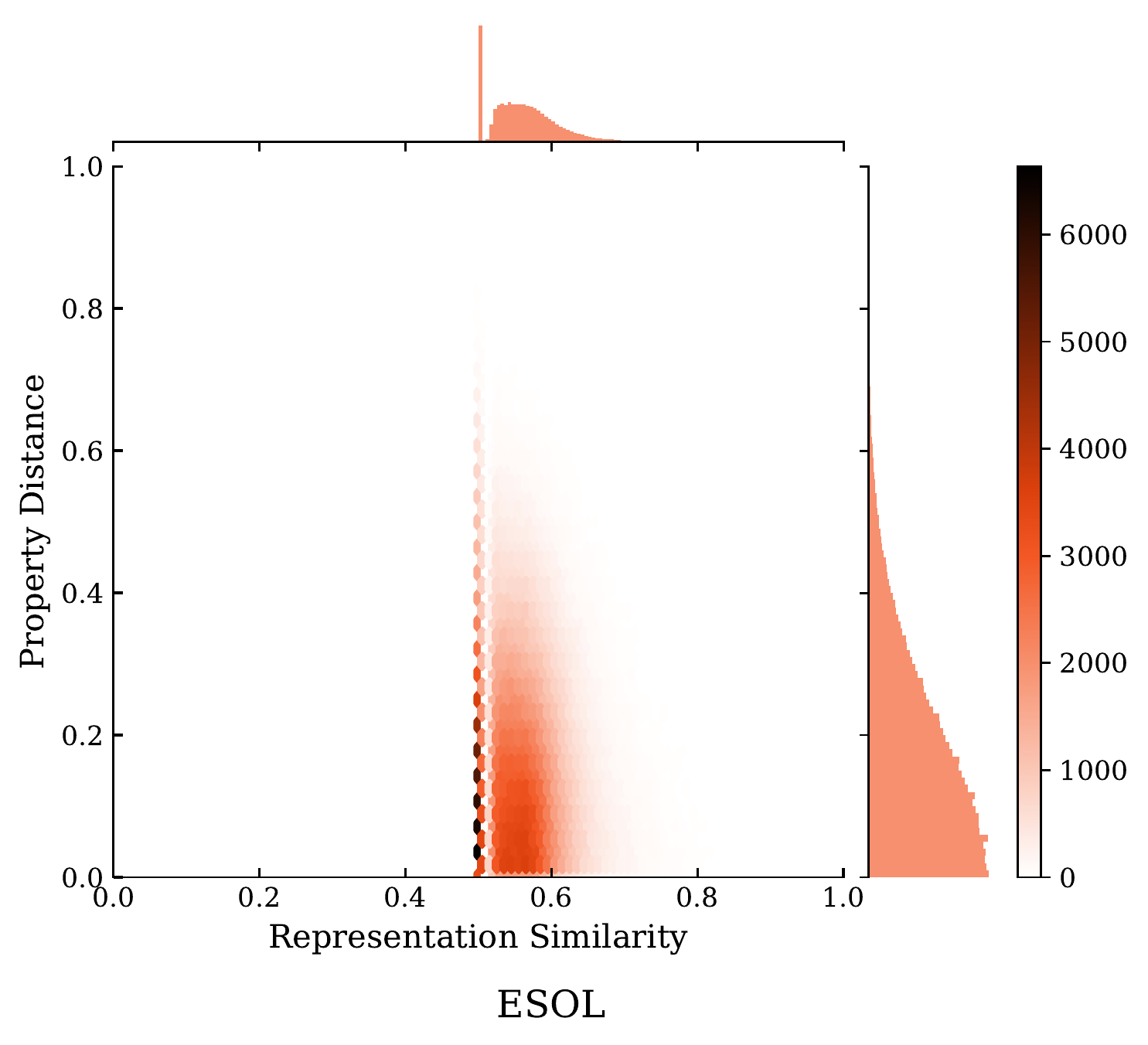}
}
    \subfigure[ECFP8]{
	\includegraphics[width=4cm]{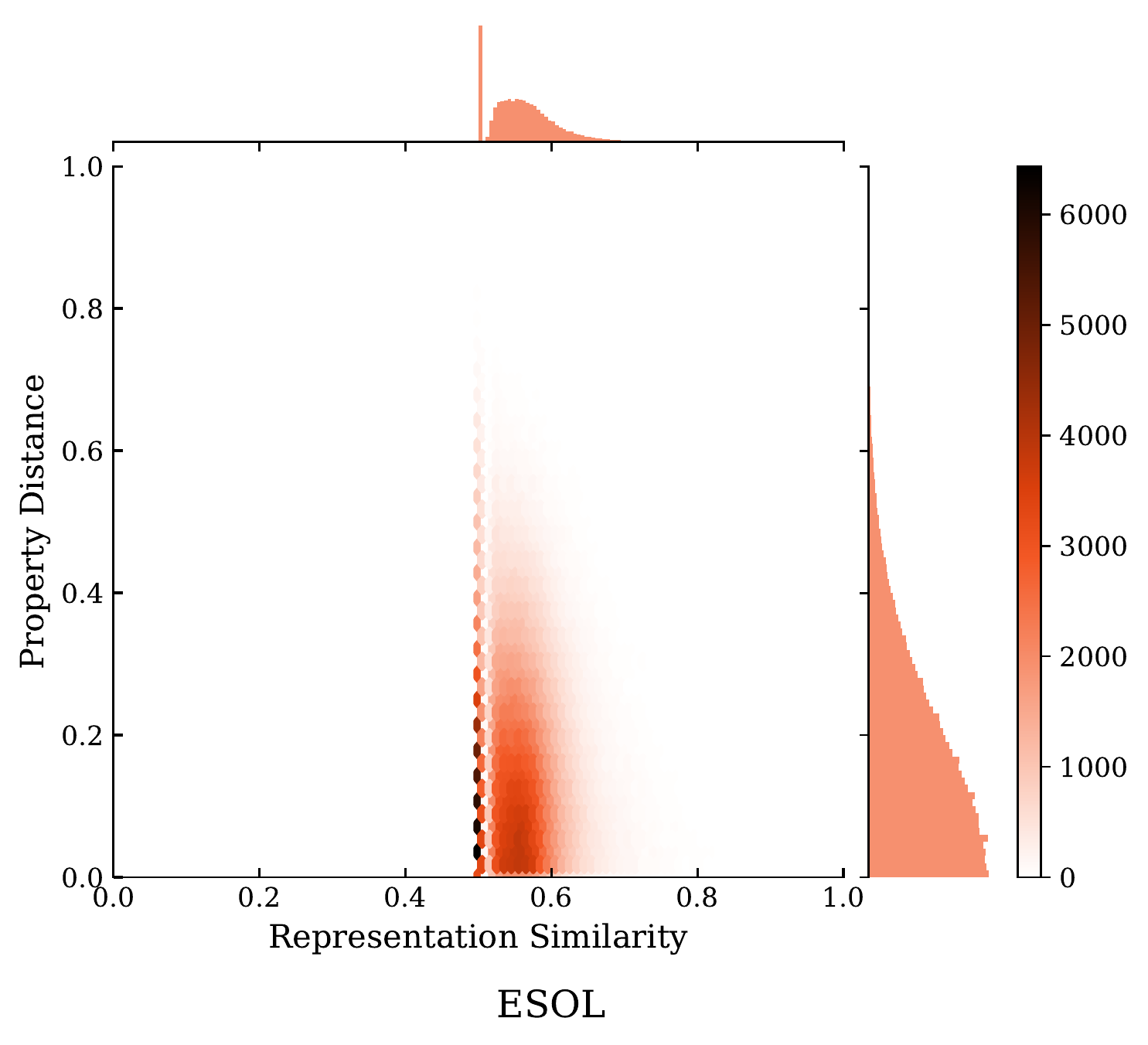}
}

	\caption{Distribution of CosineSim of ECFPs with different radius on ESOL dataset.}
	\label{fig:ECFP-radius-ESOL-Cosine}
\end{figure}

In the experiments, we use ECFP with radius 2 as baseline representations to evaluate the quality of representations produced by PTMs.
In this section, we will test the influence of the ECFP radius to the evaluations.
We set the radius of ECFP to be 3 and 4 (i.e. ECFP6 and ECFP8), and calculate the RPSMaps and scores on ESOL dataset.
The results are shown in Fig.~\ref{fig:ECFP-radius-ESOL}, Fig.~\ref{fig:ECFP-radius-ESOL-Cosine}, Tab.~\ref{tab:ECFP-radius-SAD}, Tab.~\ref{tab:ECFP-radius-SIR} and Tab.~\ref{tab:ECFP-radius-fracofc1c4}.

From the RPSMaps of ECFPs with different radius (Fig.~\ref{fig:ECFP-radius-ESOL}), it can be clearly found that there are less pairs of samples with 1.0 representation similarity, which indicates that the ECFP8 have larger expressive power to distinguish these molecular structures.
And as shown in the distribution of CosineSim (Fig.~\ref{fig:ECFP-radius-ESOL-Cosine}), the cosine similarity of ECFPs with larger radius is distributed more in the lower similarity interval, which can also be attributed to the higher expressive power.
However, as the ECFPs cannot capture the similarity between subgraphs, a larger radius may lead to a higher number of SH.
From Tab.~\ref{tab:ECFP-radius-fracofc1c4} and Tab.~\ref{tab:fracofc1c4}, it is shown that compared with ECFP4, PTMs are showing smaller $\frac{C_1^{feat}}{C_1^{fp}}$ but larger $\frac{C_4^{feat}}{C_4^{fp}}$ when using ECFP8 as baseline representations.
These fraction scores prove that the ECFP8 have more SH than the ECFP4, but less ACs.
And by $s_{AD}$ to integrally take both ACs and SH into account (Tab.~\ref{tab:ECFP-radius-SAD}), the ECFP4 is evaluated to be better than other ECFPs.

\begin{table}[htbp]
\caption{$s_{AD}$ of ECFPs with different radius on ESOL dataset.}
\label{tab:ECFP-radius-SAD}
\centering
\begin{tabular}{lccr}
\toprule
 & ECFP4 & ECFP6 & ECFP8 \\
\midrule
$s_{AD}$ & \textbf{0.019} & 0.020 & 0.021\\
\bottomrule
\end{tabular}
\end{table}

\begin{table}[htbp]
\caption{$s_{IR}$ scores of PTMs using ECFP with different radius as baselines on ESOL dataset.}
\label{tab:ECFP-radius-SIR}
\centering
\begin{tabular}{lccccccr}
\toprule
radius & GROVER & ChemBERT & GraLoG & MAT & S.T. & PreGNNs & Pre8 \\
\midrule
ECFP4 & 1.923 & \textbf{1.784} & 2.047 & 1.930 & 1.819 & 1.960 & 1.813\\
ECFP6 & 1.964 & 1.835 & 2.093 & 1.979 & 1.848 & 2.008 & \textbf{1.832}\\
ECFP8 & 1.960 & \textbf{1.829} & 2.089 & 1.974 & 1.846 & 2.003 & 1.831\\
\bottomrule
\end{tabular}
\end{table}

\begin{table}[htbp]
\caption{Improvement of $C_1$ and $C_4$ of pre-trained models over ECFP8 on ESOL.}
\label{tab:ECFP-radius-fracofc1c4}
\centering
\begin{tabular}{lcr}
\toprule
Pre-trained Models & $\frac{C_1^{feat}}{C_1^{fp}}$ & $\frac{C_4^{feat}}{C_4^{fp}}$ \\
\midrule
GROVER & 0.928 & 1.031\\
ChemBERT & 0.793 & 1.036\\
GraLoG & 0.972 & 1.116\\
MAT & 0.887 & 1.087 \\
S.T. & 0.925 & 0.921 \\
PreGNNs & 0.911 & 1.092 \\
Pre8 & 0.980 & 0.850\\
\bottomrule
\end{tabular}
\end{table}

\end{document}